\documentclass[preprint]{elsarticle}
\usepackage{amssymb,amsmath}
\usepackage{graphicx}
\usepackage{array}
\usepackage{ifthen}

\usepackage{figwidth}

\newlength{\mywidth}
\newcommand{\slashed}[1]{\settowidth{\mywidth}{#1}\hspace*{0.5\mywidth}\makebox[0ex][c]{$#1$}\makebox[0ex][c]{$\slash$}\hspace*{0.5\mywidth}}

\newcommand{\eq}[1]{Eq.~(#1)}
\newcommand{\reffig}[1]{Fig.~\ref{#1}}

\newcommand{\reftable}[1]{Table~\ref{#1}}
\newcommand{\units}[1]{\, \mathrm{#1}}
\newcommand{\Tr}{\mathrm{Tr}}

\begin{document}

\ifthenelse{\equal{\Qclass}{revtex4}}{
\title{The topological susceptibility from grand canonical simulations in the interacting instanton liquid model: chiral phase transition and axion mass.}

\author{Olivier Wantz}
\email[Electronic address: ]{O.Wantz@damtp.cam.ac.uk}
\affiliation{Department of Applied Mathematics and Theoretical Physics,
Centre for Mathematical Sciences,\\ University of Cambridge,
Wilberforce Road, Cambridge CB3 0WA, United Kingdom}
\author{E.P.S. Shellard}
\email[Electronic address: ]{E.P.S.Shellard@damtp.cam.ac.uk}
\affiliation{Department of Applied Mathematics and Theoretical Physics,
Centre for Mathematical Sciences,\\ University of Cambridge,
Wilberforce Road, Cambridge CB3 0WA, United Kingdom}
}{}
\ifthenelse{\equal{\Qclass}{elsarticle}}{
\title{The topological susceptibility from grand canonical simulations in the interacting instanton liquid model: chiral phase transition and axion mass.}

\author[damtp]{Olivier Wantz}
\ead{O.Wantz@damtp.cam.ac.uk}
\author[damtp]{E.P.S. Shellard}
\ead{E.P.S.Shellard@damtp.cam.ac.uk}
\address[damtp]{Department of Applied Mathematics and Theoretical Physics,
Centre for Mathematical Sciences,\\ University of Cambridge,
Wilberforce Road, Cambridge CB3 0WA, United Kingdom}
}{}

\bibliographystyle{plain}

\begin{abstract}
This is the last in a series of papers on the topological susceptibility in the interacting instanton liquid model (IILM). We will derive improved finite temperature interactions to study the thermodynamic limit of grand canonical Monte Carlo simulations in the quenched and unquenched case with light, physical quark masses. In particular, we will be interested in chiral symmetry breaking. The paper culminates by giving, for the first time, a well-motivated temperature-dependent axion mass. Especially, this work finally provides a computation of the axion mass in the low temperature regime, $m^2_a f^2_a = 1.46\;10^{-3}\Lambda^4 \frac{1+0.50\,T/\Lambda}{1+\left(3.53\, T/\Lambda\right)^{7.48}}$. It connects smoothly to the high temperature dilute gas approximation; the latter is improved by including quark threshold effects. To compare with earlier studies, we also provide the usual power-law $m^2_a = \frac{\alpha_a \Lambda^4}{f_a^2 (T/\Lambda)^n}$, where $\Lambda=400\units{MeV}$, $n=6.68$ and $\alpha=1.68 \,10^{-7}$.
\end{abstract}

\maketitle

\section{Introduction}
\label{sec:introduction:T}

The purpose of this paper is to continue the study of the interacting instanton liquid model (IILM) at finite temperature, with an ancillary goal being to improve our understanding of the temperature dependence of the axion mass. For the first time, we will be able to give a well-motivated axion mass that covers all temperatures down to $T=0$.

In \cite{wantz:iilm:1} we set up the formalism underlying the IILM and developed a numerical framework to compute the interactions given by an arbitrary background ansatz. Simulations at zero temperature were performed with the so-called ratio ansatz to determine the parameters that enter the model: the lambda parameter, $\Lambda$, and the quark masses, $m_q$.

In this paper we will investigate the IILM at finite temperature based on the caloron solution of Harrington and Shepard \cite{harrington:shepard:caloron}. Using as input the physical parameters we determined at zero temperature, we will study chiral symmetry restoration, based on the ideas of instanton--anti-instanton molecule formation \cite{ilgenfritz:shuryak:chiral:symmetry:restoration:iilm}, \cite{ilgenfritz:shuryak:quark:correlations:chiral:transition}, \cite{schaefer:shuryak:verbaarschot:chiral:phase:transition:molecules}, and determine the topological susceptibility.

\begin{figure}[tbp]
\begin{center}
 \includegraphics[width=\figwidth,clip=true,trim=0mm 0mm 15mm 10mm]{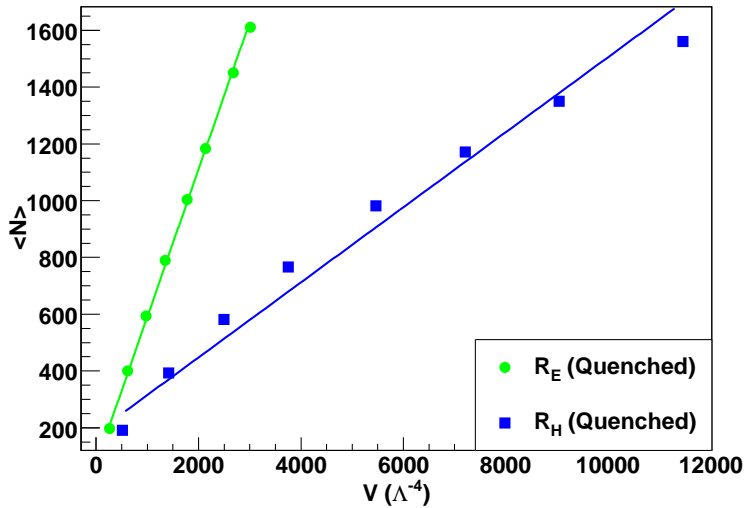}
\end{center}
 \caption{The $R_H$ ansatz does not lead to a well-defined thermodynamic limit at finite temperature, whereas the $R_E$ ansatz does exhibit the correct linear scaling of an extensive quantity. Here we display results from quenched simulations, but the same problems persist with dynamical quarks. The instanton number $N$ shows the strongest response to the unphysical interactions.}\label{fig:thermodynamic:limit}
\end{figure}

In order to deal with light, physical quark masses, we study the thermodynamic limit. As mentioned in the previous paper \cite{wantz:iilm:1}, we found that the interactions derived previously in \cite{shuryak:verbaarschot:interactions:finite:T} are deficient in this respect. Although the old and new interactions agree rather well at zero temperature, this is no longer the case at $T\neq 0$ because of an unphysical behaviour for the instanton--instanton interaction that only decays as $O(1/R_s)$, with $R_s$ the spatial instanton separation, see \eq{3.11} in \cite{shuryak:verbaarschot:interactions:finite:T}. This long-range interaction prohibits a thermodynamic limit as it is not integrable, see \reffig{fig:thermodynamic:limit}. In their paper, the authors do discuss this long-range interaction and report they found the $O(1/R_s)$ dyon--dyon behaviour for a wide range of intermediate separations. It might well be that for the simulation boxes used in a subsequent numerical investigation (see \cite{schaefer:shuryak:iilm}) the interactions are still well described by this ansatz, i.e.\ that the spatial extent of the box is bounded by these intermediate separations. However, for studying the large volume behaviour this ansatz is not appropriate. Removing this particular part of the interactions, we were able to retrieve a well-defined thermodynamic limit.

Apart from this deficiency, it seems obvious that at finite temperature it will be much harder to find a good parametrisation for the action of the background ansatz because the underlying $O(4)$ symmetry is broken and the constituent gauge fields are more complicated. This fact was already pointed out in \cite{shuryak:verbaarschot:interactions:finite:T}. We improve the existing interactions by extending the formalism set up in \cite{wantz:iilm:1} to finite temperature; as we will see, the only difficulties are of a technical nature.

With this paper we will achieve our initial aim of computing the temperature dependent axion mass. So far the mass was computed within a dilute gas approximation which breaks down at low temperatures. The connection between the high temperature regime, where the dilute gas becomes ever more accurate, and the zero temperature result, obtained through chiral perturbation theory, has been performed in a rather crude manner up to date: either by unsmoothed matching \cite{turner:axion:cosmology,bae:huh:kim:axion} or by an ad hoc interpolation prescription \cite{turner:kolb:cosmology}. Our determination of the axion mass will for the first time give a well-motivated interpolation between the zero and finite temperature regimes. Comparison with lattice data will allow for a critical evaluation of the systematic uncertainties. In particular, considerations regarding the anthropic axion with large decay constant \cite{linde:axion,turner:wilczek:axion,tegmark:aguirre:rees:wilczek:constants} are potentially very sensitive to the non-perturbative effects of the QCD phase transition, when their mass becomes sizable.

In section \ref{sec:interactions:T} we will re-derive the finite temperature interactions for the ratio ansatz. We will then discuss, in section \ref{sec:numerical:implementation:T}, the new elements that finite temperature introduces in the numerical framework. After these technical preliminaries we will have a short investigation of the topological susceptibility in the quenched sector in section \ref{sec:quenched:T}, before we discuss the main numerical results of the unquenched IILM regarding chiral symmetry restoration, section \ref{sec:chiral}, and the topological susceptibility and axion mass, section \ref{sec:top:susc:axion}.

\section{Interactions in the IILM at finite temperature}
\label{sec:interactions:T}

In terms of the 't Hooft potential $1+\Pi$, the Harrington--Shepard caloron \cite{harrington:shepard:caloron}, an infinite sequence of singular gauge BPST instantons \cite{bpst:instanton} along the Euclidean time direction, is given by
\begin{equation}
 A^a_\mu = - O_i^{ab} \zeta^b_{\mu\nu} \frac{\partial_\nu \Pi(x,\{y,\rho\})}{1+\Pi(x,\{y,\rho\})}\,,
\end{equation}
where $O$ is the colour matrix in the adjoint representation of the embedding $SU(2) \to SU(3)$, and $\zeta^b_{\mu\nu} = \bar{\eta}^b_{\mu\nu}$ ($\zeta^b_{\mu\nu} = \eta^b_{\mu\nu}$) for instantons (anti-instantons); $\eta$ are the 't Hooft symbols. The 't Hooft potential has the following form
\begin{equation}
 \Pi(x,\{y,\rho\}) = \frac{\pi \rho^2}{\beta r} \frac{\sinh\frac{2\pi r}{\beta}}{\cosh\frac{2\pi r}{\beta}-\cos\frac{2\pi t}{\beta}}\,,
\end{equation}
with $r^2 = (\vec{x}-\vec{y})^2$ and $t = x^4-y^4$; the collective coordinates are: $y$ the centre, $\rho$ the size and $O$ the colour orientation. At finite temperature bosonic quantities such as $\Pi$ are periodic in the Euclidean time direction with period $\beta=1/T$. Note that $\Pi$ approaches the zero temperature instanton potential in the singular gauge for $\beta \to \infty$.

We will use as background the ratio ansatz and, as for the zero temperature case, only consider two-body interactions. The gauge field is then given by
\begin{equation}
 A^a_\mu = - \frac{\sum_i O_i^{ab} \zeta^b_{\mu\nu} \partial_\nu \Pi_i(x,\{y_i,\rho_i\})}{1+\sum_i \Pi_i(x,\{y_i,\rho_i\})}\,,
\end{equation}
with $O=O_1^t O_2$. This pair interaction has been derived in \cite{shuryak:verbaarschot:interactions:finite:T}, and we will refer to it as $R_H$; the corresponding forces derived in this paper will be denoted by $R_E$. However, in studying the volume dependence of various quantities, we noticed that these interactions did not allow for a thermodynamic limit. The problem can be traced back to the $\log$ term in the instanton--instanton interaction, (3.11) in \cite{shuryak:verbaarschot:interactions:finite:T}, only decaying like $O(1/R_s)$ for large $R_s$, where $R_s$ is the spatial separation of the pair. Note that this term is attributed to the dyon--dyon interaction for intermediate separations, $\beta \ll R_s \ll \rho^2/\beta$, in the high temperature limit.

The ratio ansatz has the same functional form in terms of $\Pi$ as for zero temperature, and so we can use our previous result \cite{wantz:iilm:1} to write 
\begin{multline}
 F^a_{\mu\nu} F^a_{\mu\nu} = I + (\Tr O^tO + (\bar{\eta} O \eta)_{\mu\nu\mu\nu}) J + (\bar{\eta} O \eta)_{\rho\mu\rho\nu} I_{\mu\nu} \\
+ (\bar{\eta} O \eta)_{\mu\rho\nu\sigma} I_{\mu\rho\nu\sigma} + (\eta O^tO \eta)_{\mu\rho\nu\sigma} J_{\mu\rho\nu\sigma} + (\bar{\eta} O \eta)_{\alpha\mu\alpha\rho} (\bar{\eta} O \eta)_{\beta\nu\beta\sigma} K_{\mu\rho\nu\sigma}\,. \label{eq:glue_interaction}
\end{multline}
The different terms are given in appendix \ref{app:interaction:T:gluonic}. Due to charge renormalisation the action, $S[A]=\frac{1}{4g^2}\int F^a_{\mu\nu} F^a_{\mu\nu}$, acquires a quantum contribution in the form of the running coupling constant. The classical interaction is given by
\begin{equation}
 S^g_{12}/S_0 \equiv V_{12} \equiv (S[A]/S_0 - 2)\,,
\end{equation}
where $S_0=8\pi/g^2$ is the single instanton action. The quantum effects substitute $g$ in $S_0$ for the running coupling constant; the RG scale is estimated by the geometric mean $\sqrt{\rho_1 \rho_2}$, as proposed in \cite{shuryak:instantons:liquid:II,schaefer:shuryak:iilm}.

\begin{figure}[tbp]
\begin{center}
 \includegraphics[width=\figwidth,clip=true,trim=0mm 0mm 15mm 10mm]{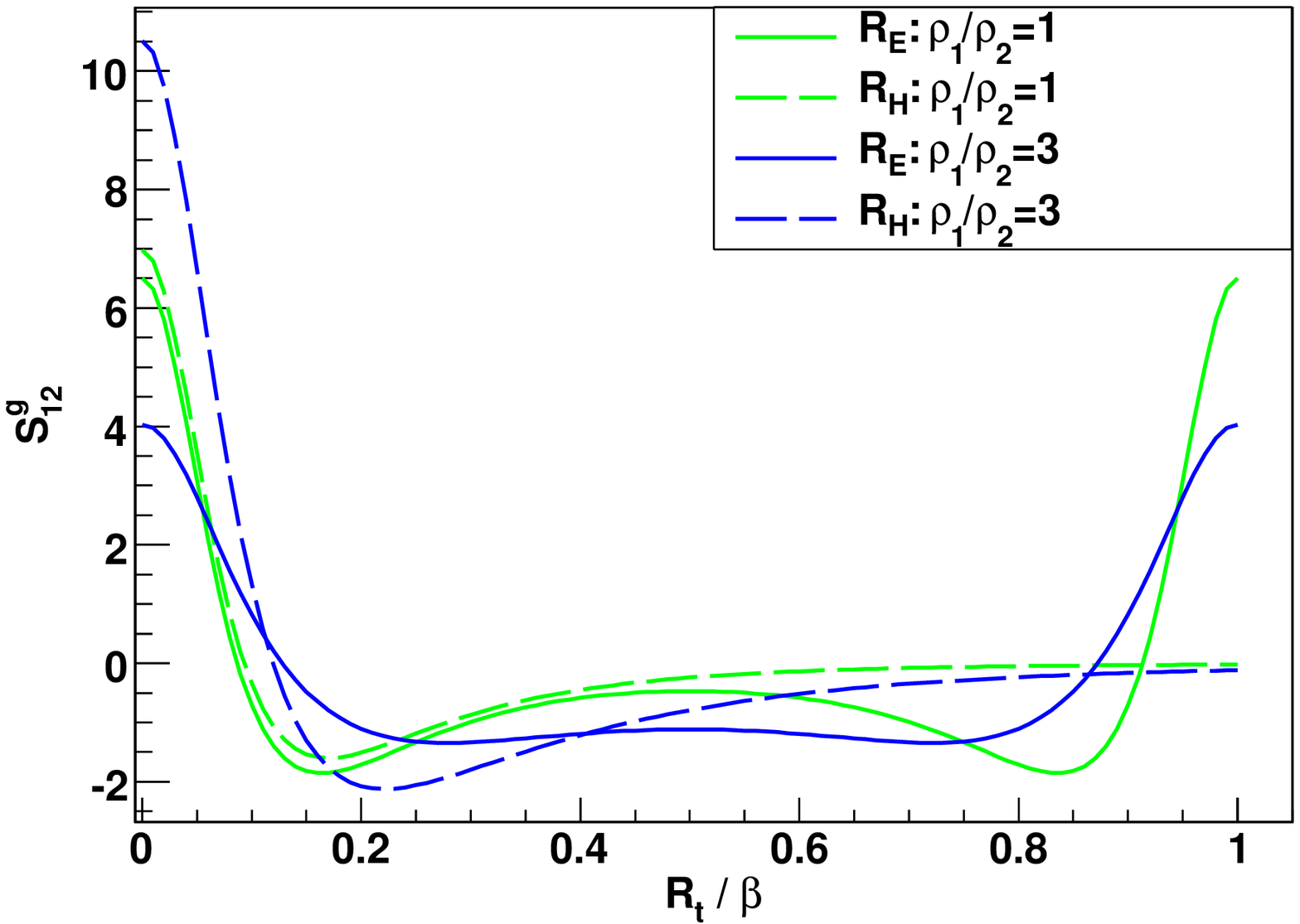}
 \includegraphics[width=\figwidth,clip=true,trim=0mm 0mm 15mm 10mm]{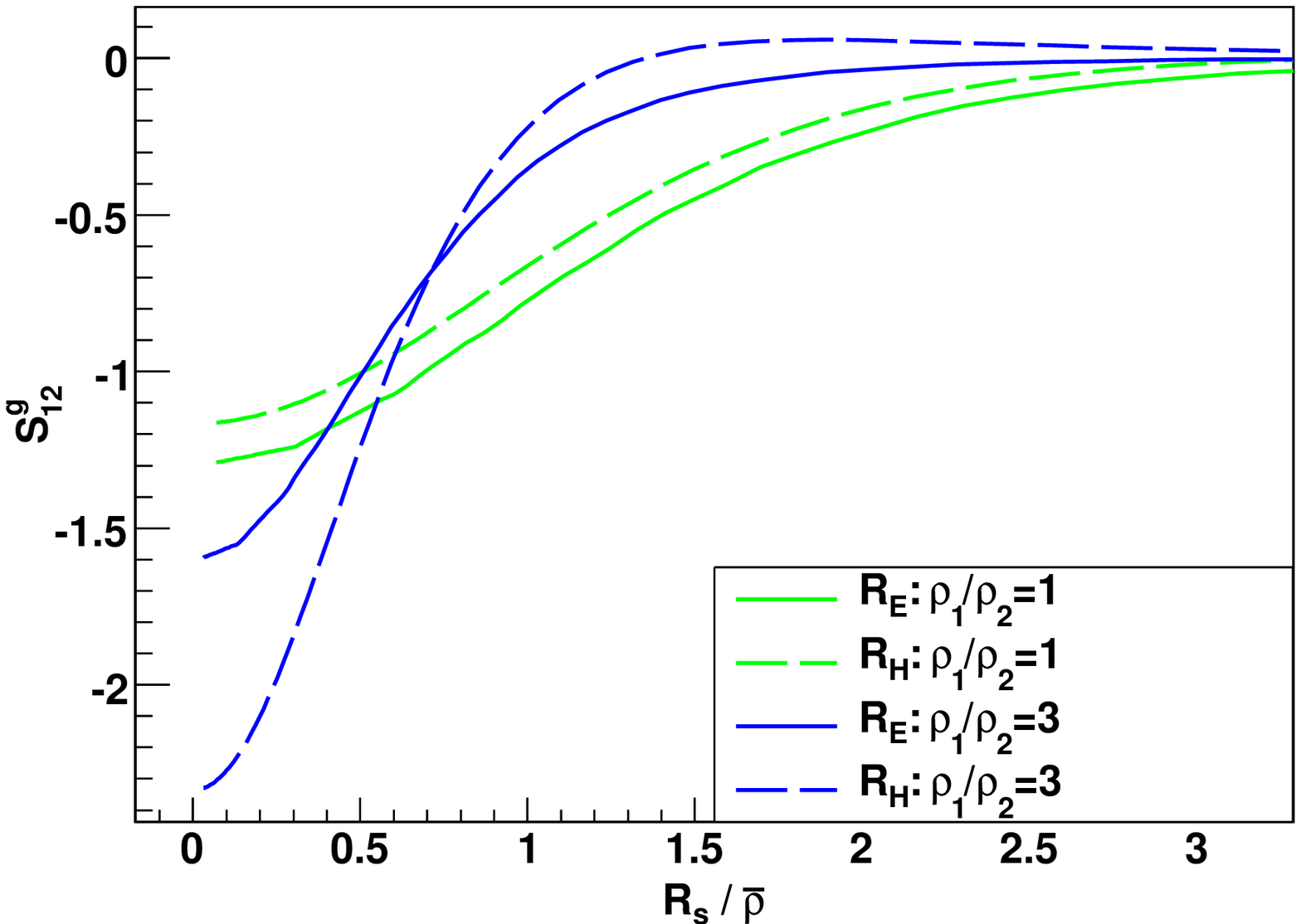}
\end{center}
 \caption{The major difference follows from the fact that $R_H$ interactions are not periodic in $R_t$. This is clearly a deficiency of the analytical formulas because it follows directly from (\ref{eq:glue_interaction}) that the interactions should have period $\beta$. For spatial separations,  the main differences occur for unequal size parameters, e.g.\ $\rho_1/\rho_2=3$ in this case. The reason is that the dependence on the sizes is more complicated than the functional form, $\sqrt{\rho_1\rho_2}$, used in $R_H$. (We have set $\bar{\rho}=\sqrt{\rho^2_1+\rho^2_2}$.)}\label{fig:interaction:T:gluonic}
\end{figure}

First, we will look at the dependence of the interactions on $R_t$, the instanton separation in the (imaginary) time direction. Compared to the zero temperature case, the differences are substantial even for pairs with equal sizes. The major difference comes from the fact that the $R_H$ interactions are not periodic. For unequal sizes, the difference is even more pronounced as was the case at zero temperature, see \reffig{fig:interaction:T:gluonic}. The reason is again that the functional form on the instanton sizes in $R_H$ is not general enough.

\begin{figure}[tbp]
\begin{center}
 \includegraphics[width=\figwidth,clip=true,trim=0mm 0mm 15mm 10mm]{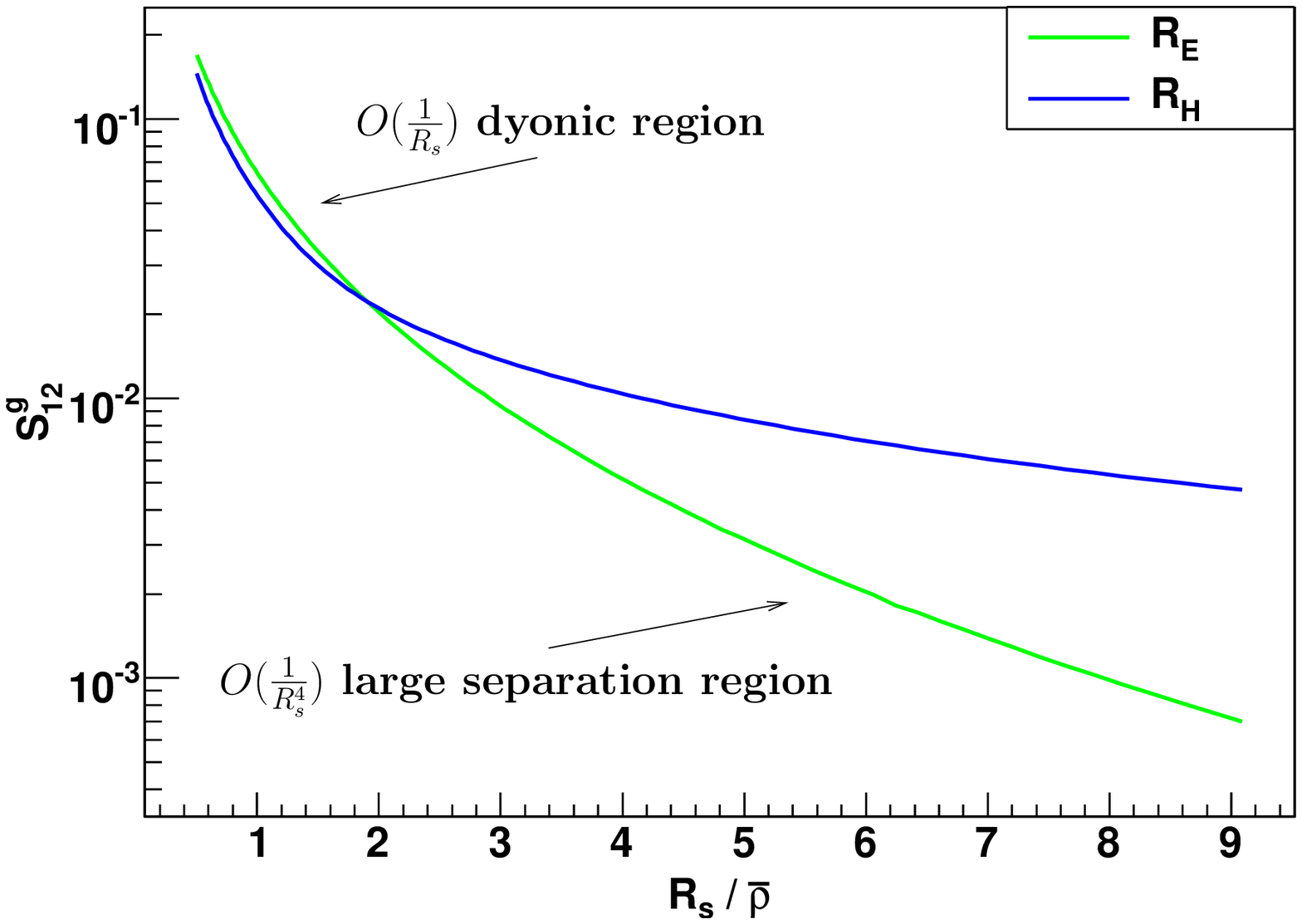}
\end{center}
 \caption{The instanton--instanton interaction behaves very differently for $R_H$ and $R_E$. In the high temperature, large size limit we do see the dyon--dyon behaviour but for larger separations, $R_s \gg \rho^2/\beta$, the interaction decays much faster, $O(R_s^{-4})$. This fall-off behaviour is integrable and we can study the thermodynamic limit, in contrast to the purely dyonic interaction. (We have set $\bar{\rho}=\sqrt{\rho^2_1+\rho^2_2}$.)}\label{fig:interaction:T:gluonic:II}
\end{figure}

The interaction between instantons and anti-instantons does not lead to as close a match between $R_E$ and $R_H$ as in the $T=0$ case, but is still qualitatively similar for not too large separations. In the instanton--instanton case, however, we have found significant differences, see \reffig{fig:interaction:T:gluonic:II}: we do find the dyon--dyon behaviour for intermediate distances but for separations $R_s \gg \rho^2/\beta$ the functional dependence changes into an integrable $O(R_s^{-4})$ for $R_E$ whereas the non-integrable dyon--dyon interaction persists in the $R_H$ ansatz. That the large separation fall-off should be different from the dyonic regime is clear from the discussion in \cite{gross:pisarski:yaffe:instantons:finite:T} where the authors show that the caloron field develops a dipole-like character in the far-field region. Note that this leads to a three-dimensional dipole--dipole interaction for an instanton--anti-instanton pair, but that the $R_H$ ansatz retains the zero temperature (four-dimensional) dipole--dipole interaction\footnote{instanton--instanton pairs decay slightly faster, at $O(1/R_s^4)$, however, and their interaction is not of dipole type; note that this happens at $T=0$ too, where instanton--anti-instanton pairs exhibit a dipole--dipole interaction but instanton--instanton pairs don't \cite{diakonov:instanton:variational}.}.

The fermionic interaction follows from the quark wave function overlaps
\begin{equation}
 (\slashed{D}+m)_{ij}=\langle \xi_i|\slashed{D}+m|\xi_j \rangle=\slashed{D}_{ij}+m\delta_{ij}\,. \label{eq:quark:overlap}
\end{equation}
Even though the set of eigenfunctions $\{\xi_i\}$ is generally not an orthonormal basis, the extra contributions are neglected; effectively, we treat $\{\xi_i\}$ as being orthonormal which leads to the diagonal mass term. The fermionic zero mode at finite temperature is given by \cite{grossman:dirac:zeromode}, \cite{gross:pisarski:yaffe:instantons:finite:T}, \cite{shuryak:verbaarschot:interactions:finite:T},
\begin{eqnarray}
 \xi_I &=& \frac{1}{2\pi \rho_I} \sqrt{1+\Pi_I} \slashed{\mbox{$\partial$}} \frac{\chi_I}{1+\Pi_I} 
\left(
\begin{array}{c}
 U_I \varphi\\
 0
\end{array}
\right)\,,\\
\xi_A &=& \frac{1}{2\pi \rho_A} \sqrt{1+\Pi_A} \slashed{\mbox{$\partial$}} \frac{\chi_A}{1+\Pi_A} 
\left(
\begin{array}{c}
 0\\
 U_A \varphi
\end{array}
\right)\,,\\
\chi &=& \Pi \; \frac{\cos \frac{\pi t}{\beta}}{\cosh \frac{\pi r}{\beta}}\,,
\end{eqnarray}
with $\varphi_{\alpha a}=\epsilon_{\alpha a}$, normalised according to $\epsilon_{12}=1$, and $U_i$ the collective coordinates for the colour embedding in the fundamental representation.

\begin{figure}[tbp]
\begin{center}
 \includegraphics[width=\figwidth,clip=true,trim=0mm 0mm 15mm 10mm]{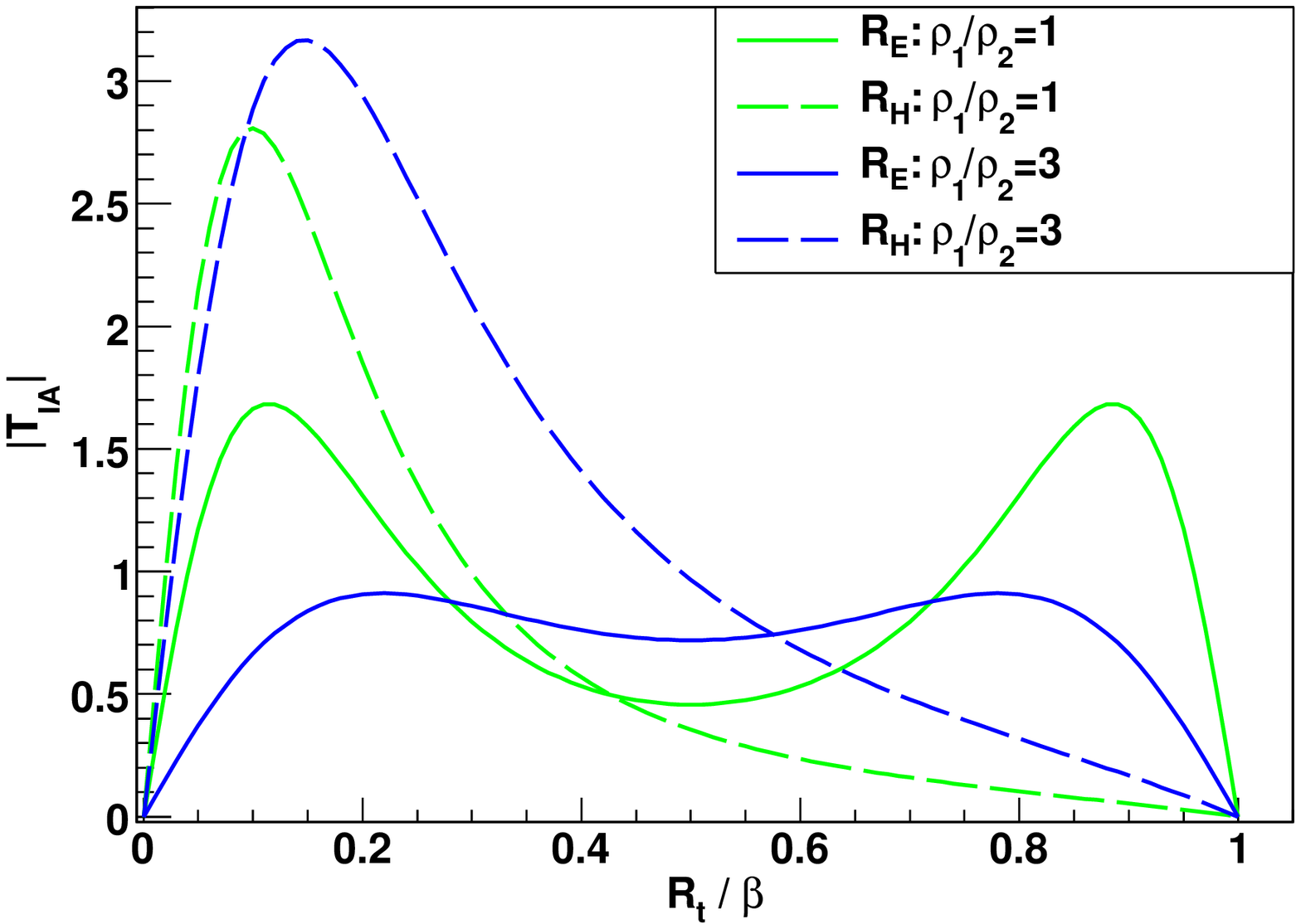}
 \includegraphics[width=\figwidth,clip=true,trim=0mm 0mm 15mm 10mm]{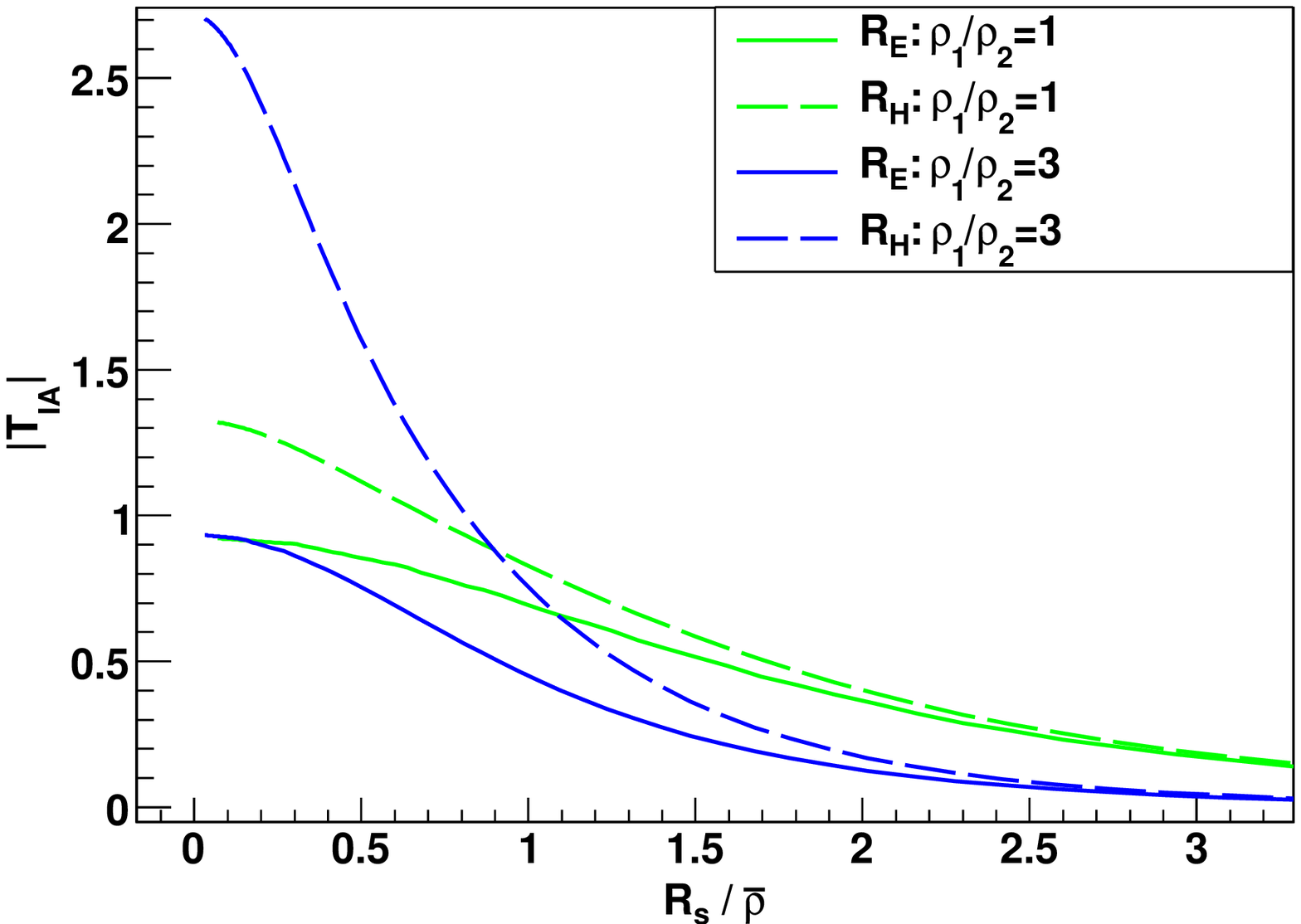}
\end{center}
 \caption{The large discrepancy between $R_E$ and $R_H$ for both temporal and spatial separations is due to the fact that $R_E$ uses the full ratio ansatz in the Dirac operator whereas for $R_H$ a sum ansatz is used. Periodicity, which follows directly from (\ref{eq:quark:overlap}), is lacking for overlaps in the $R_H$ ansatz but is realised in the $R_E$ ansatz. (We have set $\bar{\rho}=\sqrt{\rho^2_1+\rho^2_2}$.)}\label{fig:overlap:T:quark}
\end{figure}

The overlaps $T_{IA}= \int \xi_I^\dagger i \slashed{D} \xi_A$ have a slightly more complicated form than their $T=0$ counterparts and are given by
\begin{multline}
T_{IA} = \int_{\mathbb{R}^3 \times S^1} \frac{1}{4\pi^2\rho_I \rho_A} \left(\frac{1}{2}\Tr (U \tau^{+}_{\beta}) I_\beta - \frac{i}{2} \Tr (U \tau^{+}_\beta \tau_a) \bar{\eta}^a_{\mu\alpha} J_{\beta\mu\alpha} \right.\\
\left. + \frac{i}{2} \Tr (U \tau_a \tau^{+}_\beta) \eta^a_{\mu\alpha} K_{\beta\mu\alpha}\right)\,. \label{eq:quark_interaction}
\end{multline}
The different contributions can be found in appendix \ref{app:interaction:T:quark}.

The difference between the $R_E$ and $R_H$ interactions is quite large for temporal and spatial separations, see \reffig{fig:overlap:T:quark}. However, it is not straightforward to compare both ans\"atze because the $R_H$ overlaps are computed on a different background \cite{shuryak:verbaarschot:interactions:finite:T}, the sum ansatz, whereas we use the full ratio ansatz in $R_E$. As in the gluonic case, the quark overlaps are not periodic for the $R_H$ ansatz as opposed to those of $R_E$. As always, unequal sizes increase the differences between the $R_H$ and $R_E$ ans\"atze even more. For large separations, when the ratio ansatz becomes indistinguishable from the sum ansatz, and for equal sizes, we find very good agreement between $R_H$ and $R_E$.

The total interaction, after normalising to the dilute gas approximation, is given by 
\begin{equation}
 S_\mathrm{int} = \sum_{\mathrm{pairs}\,(i,j)} S_0(\sqrt{\rho_i \rho_j}) V_{ij} - \sum_{n=1}^{N_f}
 \left\{
 \begin{array}{cl}
 \ln\det(\mathbb{I} + \frac{TT^\dagger}{m^2_n}) &, Q<0 \\
 \ln\det(\mathbb{I} + \frac{T^\dagger T}{m^2_n}) &, Q>0
 \end{array}
\right. \,, \label{eq:interaction:total}
\end{equation}
where $Q=N_I-N_A$ is the topological charge and $N_f$ the number of quark flavours. For details see \cite{wantz:iilm:1}.

\begin{figure}[tbp]
\begin{center}
 \includegraphics[width=\figwidth,clip=true,trim=0mm 0mm 15mm 10mm]{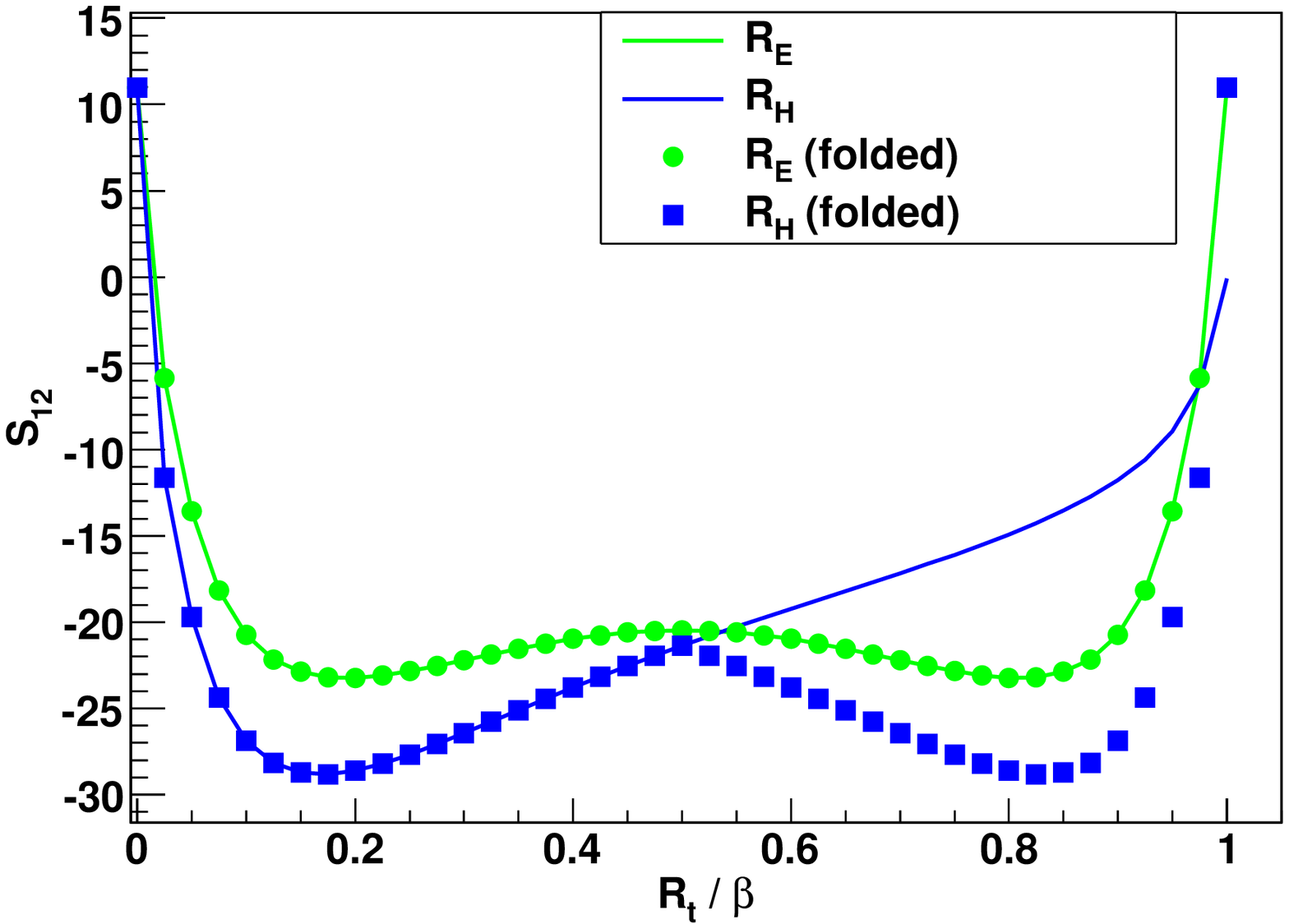}
\end{center}
 \caption{Folding back the temporal separation into the fundamental interaction region, $R_t \in [-\beta/2,\beta/2]$, we explicitly retrieve periodic interactions for $R_H$. The $R_E$ ansatz is intrinsically periodic.}\label{fig:interaction:T:full:t}
\end{figure}

Even though the analytic formulas of the $R_H$ ansatz are not periodic, this is not really a major shortcoming because periodicity can be realised by folding back the instantons into the fundamental interaction region $R_t \in [-\beta/2,\beta/2]$. We demonstrate this for the total interaction in \reffig{fig:interaction:T:full:t}. We can clearly see that the $R_E$ ansatz is intrinsically periodic.

\section{Numerical Implementation}
\label{sec:numerical:implementation:T}

\subsection{Interpolation and asymptotic matching}
\label{sec:numerical:implementation:T:interpolation:matching}

As for zero temperature, we will need to set up a grid for the numerical evaluation of the two-body interactions. The look-up tables will depend on four variables: the spatial separation $R_s$, the temporal separation $R_t$ and the two sizes $\rho_1$ and $\rho_2$. The colour degrees of freedom $O$, or equivalently $U$, have been completely factored out and can be treated exactly.

As we have seen in section \ref{sec:introduction:T}, the dyon--dyon regime is characterised by a fairly slow fall-off. However, not all bosonic interactions in appendix \ref{app:interaction:T:gluonic:exact} decay this slowly; rather they can be grouped according to polynomial and exponential decay. The fermionic overlaps fall-off exponentially, but slightly more slowly than their bosonic counterparts. Therefore, we found it advantageous to define three different grids:
\begin{itemize}
 \item For the polynomial grid, the maximal separation $R^\mathrm{max}_s$ will depend on $\tilde{\rho} \equiv \pi \rho^2/\beta$, the natural size parameter for separations beyond $\beta$. Note that for very small sizes $R^\mathrm{max}_s~<~\beta$ the prescription for $R^\mathrm{max}_s$ should switch over to the $T=0$ case; this is implemented by setting $R^\mathrm{max}_s = \max(\alpha^p_0 \rho,\alpha^p_T \pi \rho^2/\beta)$.
 \item The exponential decay sets in at $R_s \approx \beta$. To accommodate very small instantons we set again $R^\mathrm{max}_s = \min(\alpha^{e^g}_0 \rho,\alpha^{e^g}_T \beta)$.
 \item The grid for quark overlaps is set in an analogous fashion, i.e.\ $R^\mathrm{max}_s = \min(\alpha^{e^q}_0 \rho,\alpha^{e^q}_T \beta)$.
\end{itemize}
The constants $\alpha^j_i$ are fine-tuned so as to achieve fast and stable numerical integrations and a good matching to the analytic expressions used for separations beyond $R^\mathrm{max}_s$. In the temporal direction the grid is bounded by $|R_t|=\beta/2$; care needs to be taken again for small instantons and we set $|R^\mathrm{max}_t|=\min(R^\mathrm{max}_s,\beta/2)$.

\begin{figure}[tbp]
\begin{center}
 \includegraphics[width=0.9\figwidth]{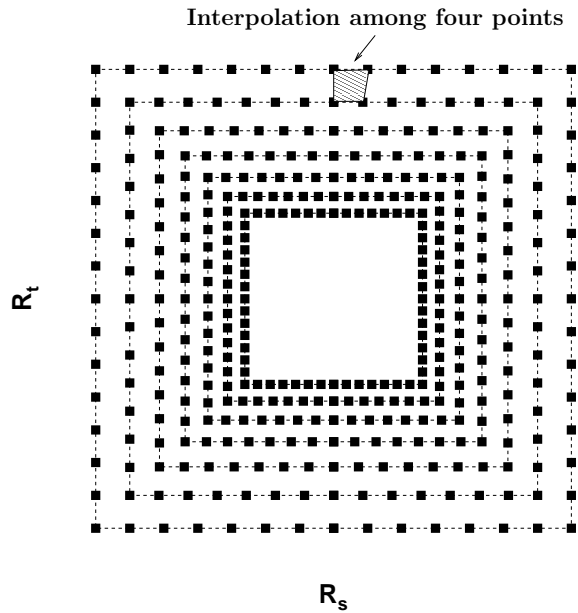}
\end{center}
\caption{It turns out that the interactions are more or less identical on the $R_s-R_t$ plane, when expressed in units of $\rho_1^2+\rho_2^2$. We therefore choose the grid to be irregular, i.e.\ the points on the $R_s-R_t$ plane are different for different values of the sizes. Since instantons are localised field configurations, the separation between grid points grows geometrically in the $R_s$ direction. In the $\rho$ directions the mesh is regular, but still `logarithmic'. The inner and outer regions are matched to asymptotic expansions: for small separations the matching is performed radially from the origin because in this region the interactions are roughly $O(4)$ symmetric; for large separations we match along the $R_s$ direction with constant $R_t$ because the interactions become $t$-independent. (Note that for graphical reasons not all $870$ grid points are displayed and that we actually display $\log R_s$. Also the actual grid only contains the $R_s>0$ points.)}
\label{fig:grid}
\end{figure}

The size distribution is supported on $30$ grid points and for each pair $(\rho_i,\rho_j)$ the $R_s-R_t$ plane consists of $30 \cdot 29$ nodes; this leads to a total of roughly 1 million interpolation points. The grid is `logarithmic' in the size and $R_s$ direction. A typical $R_s-R_t$ plane is shown in \reffig{fig:grid}.

The minimal and maximal sizes supported by the grid are fixed as in the $T=0$ case, i.e.\ they are chosen so small, respectively large, that they correspond to a very low, respectively high, quantile of the normalised instanton density. As we will see in section \ref{sec:numerical:implementation:T:mc}, at finite temperature the instanton density becomes temperature dependent, reflecting the screening of coherent background fields. Instead of computing a grid for every temperature we run simulations at, we can exploit the following scaling transformations: under coordinate rescaling, $x_\mu \to \alpha x_\mu$,
\begin{eqnarray}
 V(\rho_i,R_j,\beta) &=& V(\rho_i/\beta,R_j/\beta)\,, \\
 T(\rho_i,R_j,\beta) &=& \frac{1}{\beta}T(\rho_i/\beta,R_j/\beta)\,.
\end{eqnarray}
These can be used to transform the grid and interactions to different temperatures. We just need to make sure to choose the original sizes large enough to accommodate the low temperature behaviour, i.e.\ $\rho_{\max}=\Lambda$. We defined the grid and the interactions at $\beta=1$ with $\rho \in [0.01,1.6]$.

As for the zero temperature case, the matching consists in deriving asymptotic formulas $f_\mathrm{asy}$ that are patched on to the numerically integrated interactions according to 
\begin{equation}
f(R) = f_\mathrm{asy}(R) \frac{f_\mathrm{ex}(R_m)}{f_\mathrm{asy}(R_m)}\,,
\label{eq:asymptotic:matching}
\end{equation}
where $R_m$ is the matching point. For details see \cite{wantz:iilm:1}. At $T\neq 0$, however, we match differently for small and large separations. In the former case the interactions will be approximately $O(4)$ symmetric and the matching is performed in an $O(2)$ symmetric way on the $R_s-R_t$ plane, i.e.\ along a ray connecting the origin with $(R_s,R_t)$. For large separations the interactions become $t$-independent and we match along a line of constant $R_t$.

\begin{figure}[tbp]
\begin{center}
 \includegraphics[width=0.8\figwidth]{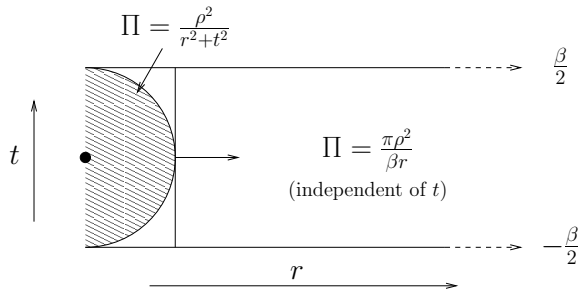}
\end{center}
\caption{In order to derive simple analytical expressions in the limit of small and large pair separations, we approximate the 't Hooft potential to have a simple functional form in different integration regions. This procedure works well for instantons with sizes that do not exceed $\beta$. As we will see the plasma screening effects limit instanton sizes to be rather small compared to $\beta$.}
\label{fig:integration:split}
\end{figure}

To derive the large separation asymptotic formulas, we consider two different integration regions, see \reffig{fig:integration:split}. Therefore, we effectively approximate integrals by
\begin{equation}
 \int_{S^1 \times \mathbb{R}^3} \approx \int_{S^3 \times [0,\beta/2]} + \beta \int_{S^2 \times [\beta/2,\infty]}\,,
\end{equation}
where we exploit the respective spherical symmetry of the 't Hooft potentials. In both regions we only have to deal with rational functions which can be easily integrated exactly. Apart from this extra complication the strategy is the same as in the zero temperature case and is illustrated in \reffig{fig:interaction:T:separation:large}. The results are given in appendix \ref{app:interaction:T:gluonic:large} and \ref{app:interaction:T:quark:large}.

\begin{figure}[tbp]
\begin{center}
 \includegraphics[width=0.8\figwidth]{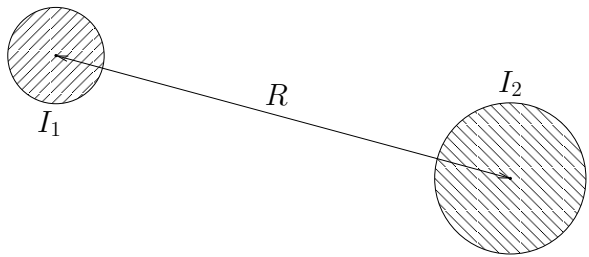}
\end{center}
 \caption{The instantons $I_1$ and $I_2$ are so far apart that within the shaded region that give the dominant contribution to the field strength of each, the other's field strength is roughly constant and fixed at $x_\mu-R_\mu \approx -R_\mu$. We can then safely extend the integration region to be all of $S^1 \times \mathbb{R}^3$, with a negligible error due to the rather strong localisation of the individual instantons. The functional form of the relevant 't Hooft potential changes from $\Pi=\rho^2/r^2_{4d}$ to $\Pi=\pi\rho^2/\beta r_{3d}$ at $r_{4d} = \beta/2$, see \reffig{fig:integration:split}.}\label{fig:interaction:T:separation:large}
\end{figure}

\begin{figure}[tbp]
\begin{center}
 \includegraphics[width=\figwidth]{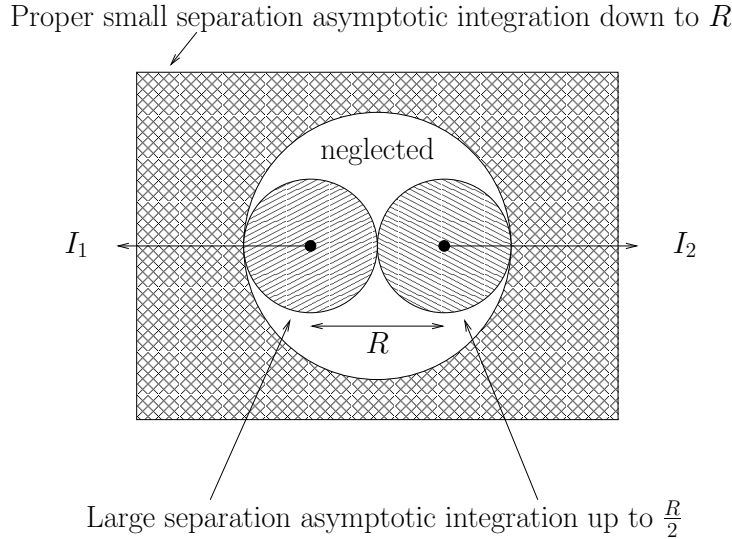}
\end{center}
 \caption{The instantons $I_1$ and $I_2$ are strongly overlapping. We approximate the integral by first integrating over $I_1$ keeping $I_2$ fixed at $R_\mu$ as in the large separation case, see \reffig{fig:interaction:T:separation:large}, but with upper limit $R/2$; to this we add the analogous contribution from $I_2$. The possibly singular behaviour is picked up by integrating from infinity down to $R$ and approximating the arguments by $x_\mu-R_\mu/2 \approx x_\mu$ and $x_\mu+R_\mu/2 \approx x_\mu$, respectively.}\label{fig:interaction:T:separation:small}
\end{figure}

The main objective for small separations is to capture adequately the singularity structure, for which it is sufficient to restrict ourselves to the $T=0$ region. To capture contributions that do not blow up, we add the large separation contributions truncated at the pair separation $(R_t,R_s)$. The results are given in appendix \ref{app:interaction:T:gluonic:small} and \ref{app:interaction:T:quark:small}. Apart from considering different functional forms for the 't Hooft potential this is the same procedure as for the $T=0$ asymptotic interactions and is illustrated in \reffig{fig:interaction:T:separation:small}.

\begin{figure}[tbp]
\begin{center}
 \includegraphics[width=\figwidth,clip=true,trim=0mm 0mm 15mm 10mm]{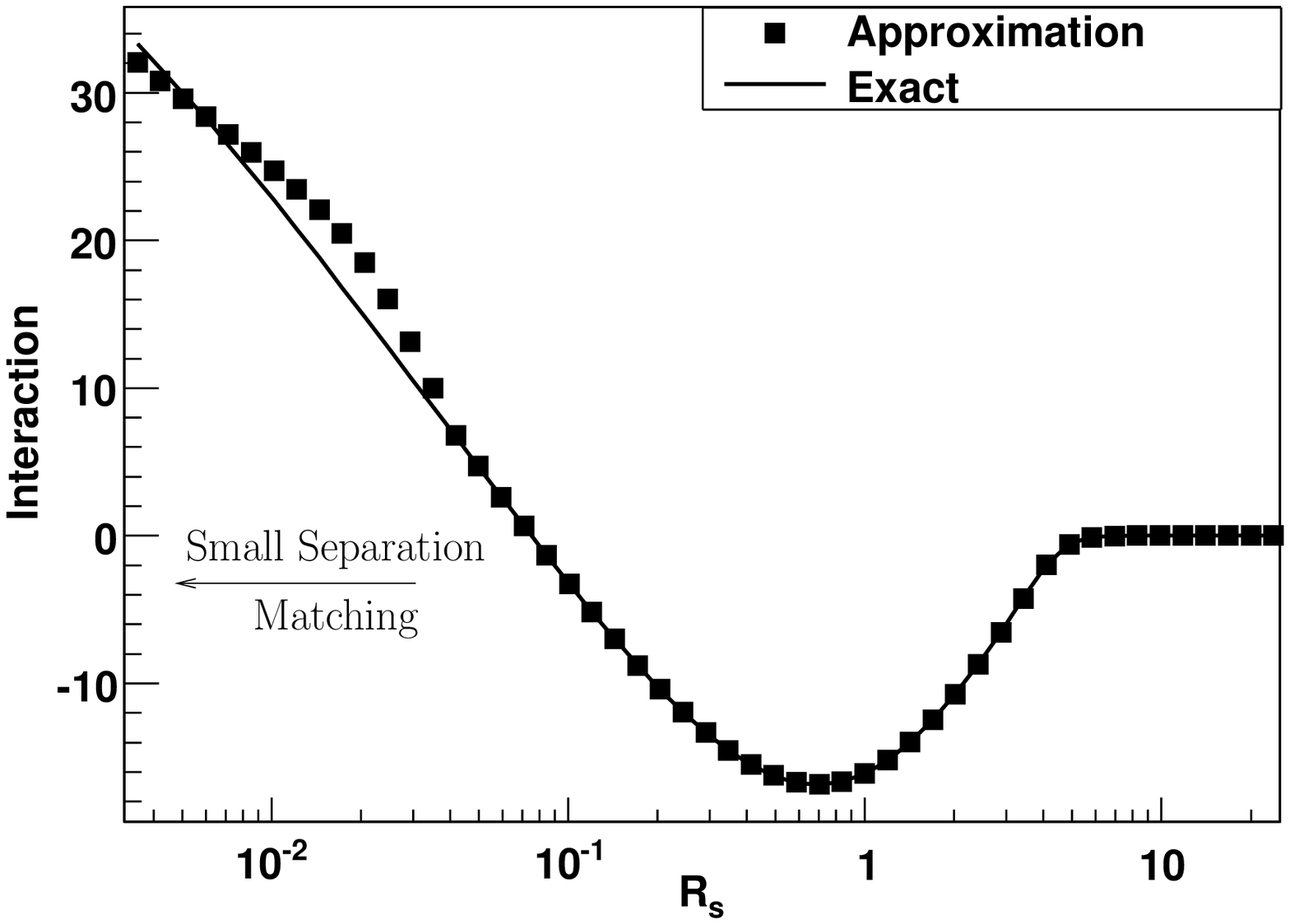}
\end{center}
 \caption{The small separation asymptotic approximation is only qualitatively correct. It overestimates the correct interaction for moderately small separations because it underestimates the quark contribution. For very small separations, where the exact quark interaction is indeed negligible, the gluonic approximation underestimates the exact result. However, the grid covers rather small separations so that the asymptotic interactions for strongly overlapping instanton--anti-instanton pairs are rarely needed. In those cases where it is needed we get a qualitatively correct repulsion. (The temporal separation $R_t$ has been chosen very small in order to see the repulsion).}\label{fig:interaction:T:approximation:total}
\end{figure}

As in the $T=0$ case, the quark overlaps in the small separation region are qualitatively incorrect. However, this region is again dominated by the gluonic repulsion. The latter is not as well approximated as in the $T=0$ case but the agreement is still on the $10\%$ level and, most importantly, qualitatively correct, see \reffig{fig:interaction:T:approximation:total}. In contrast, the large separation asymptotic formulas work well.

\subsection{Biased Monte Carlo}
\label{sec:numerical:implementation:T:mc}

The IILM is defined by its partition function,
\begin{equation}
 Z = \sum_{N_I, N_A}^\infty \frac{1}{N_I !} \frac{1}{N_A !} \prod_i^{N_I} d(\rho_i) \prod_j^{N_A} d(\rho_j) \exp\left(-S_\mathrm{int}\right)\,,\label{eq:iilm}\\
\end{equation}
with $S_\mathrm{int}$ defined in (\ref{eq:interaction:total}) and the single instanton density given by $d(\rho) = d_{0}(\rho)  d_T(\rho)$. The zero temperature contribution $d_{0}$ is given in \cite{wantz:iilm:1}. The finite temperature term describes the screening in the plasma of coherent field excitations exceeding the inverse temperature scale $\beta$, and is given by \cite{gross:pisarski:yaffe:instantons:finite:T} \cite{schaefer:shuryak:iilm},
\begin{multline}
 d_T(\rho) = \exp \left[-\frac{1}{3}(2N_c + N_f)(\pi \rho T)^2 - \left(1+\frac{N_c}{6}-\frac{N_f}{6}\right) \right.\\
 \left.\left(-\ln(1+(\pi \rho T)^2/3) + \frac{0.15}{(1+0.15 (\pi \rho T)^{-3/2})^8}\right)\right]\,. \label{eq:plasma:screening}
\end{multline}

Apart from book-keeping related technicalities, the major challenge is to adequately simulate instanton--anti-instanton molecules, the structures responsible for the chiral symmetry restoration within the IILM \cite{ilgenfritz:shuryak:chiral:symmetry:restoration:iilm}, \cite{ilgenfritz:shuryak:quark:correlations:chiral:transition}, \cite{schaefer:shuryak:verbaarschot:chiral:phase:transition:molecules}. Detailed numerical studies at finite temperature have been performed in \cite{schaefer:shuryak:iilm}. For the mass parameters used in these latter simulations no technical problems were encountered. However, for the small quark masses that we determined in the $T=0$ simulations, we found that standard Monte Carlo techniques faced severe problems with the strong attraction between instanton--anti-instanton pairs. In \cite{wantz:iilm:2} we have argued that this technical problem will occur quite generally for the semi-classical expansion with small quark masses. In that paper we have set up the framework to deal with these algorithmic issues by adapting techniques from chemical engineering and computational chemistry and developed for the study of strongly associating fluids.

The instanton--anti-instanton pair interactions become ever stronger and more localised for small quark masses, and random sampling methods will generically miss these configurations. We therefore need to preferentially sample the attraction centres. This can be achieved by resorting to biased Monte Carlo schemes. They exploit the large redundancy in devising transition probabilities $P_{ij}$ that satisfy detailed balance,
\begin{equation}
 P^\mathrm{eq}_i P_{ij} = P^\mathrm{eq}_j P_{ji}\,, \label{eq:detailed_balance}
\end{equation}
where the equilibrium distribution $P^\mathrm{eq}$ is given by the partition function. Specifically, the degeneracy follows from the fact that the transition probability consists of two parts, the proposal probability, $\mathcal{P}_{ij}$, and the acceptance probability, $\mathcal{A}_{ij}$,
\begin{equation}
 P_{ij} = \mathcal{P}_{ij} \mathcal{A}_{ij}\,.
\end{equation}
To satisfy (\ref{eq:detailed_balance}), the acceptance probability is given by
\begin{equation}
 \mathcal{A}_{ij} = \min \left[1,\frac{P^\mathrm{eq}_j}{P^\mathrm{eq}_i} \frac{\mathcal{P}_{ji}}{\mathcal{P}_{ij}} \right]\,, \label{eq:acceptance}
\end{equation}
which is the Metropolis prescription.

Recently, efficient and general purpose algorithms have been developed to achieve this importance sampling; we will use \cite{wierzchowski:kofke:associating_fluids}, the Unbonding--Bonding algorithm (UB). This algorithm focuses not on the union of all the interaction regions, a complicated and case-specific geometrical problem, but on the individual interaction regions and all the possible routes that lead to the same final state.

\begin{figure}[tbp]
\begin{center}
 \includegraphics[width=\figwidth,clip=true,trim=0mm 0mm 0mm 0mm]{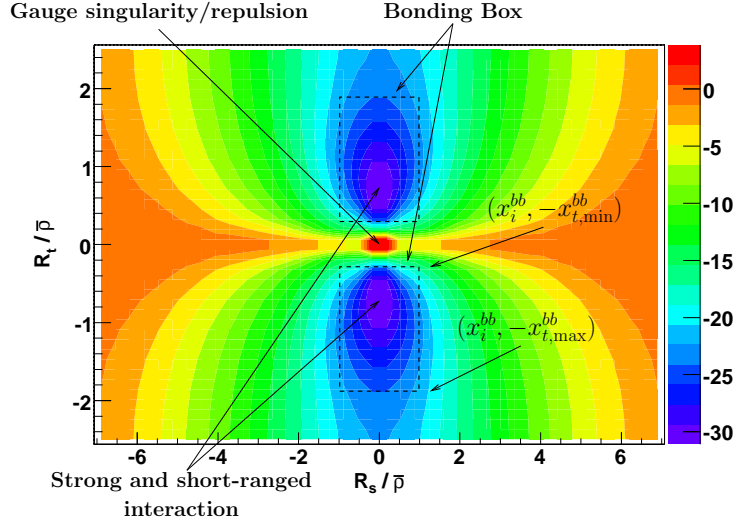}
\end{center}
 \caption{The strongest interaction is achieved for the colour orientation $U=\mathbb{I}$ and located in the two troughs. For simplicity we will choose the bonding box to consist of the two hypercubes displayed by dashed lines.
}\label{fig:interaction:T}
\end{figure}

For the IILM, we choose the bonding region to consist of the two disconnected valleys of strong interaction, see \reffig{fig:interaction:T}, and take them to be hypercubes for simplicity. The precise size of the box is a free parameter and will be chosen to achieve the fastest convergence. We will check that the results do not depend on the bonding box parameters, within statistical uncertainties. We will now present the transition probabilities for the different types of updates; details can be found in \cite{wantz:iilm:2}.

The forward and backward transition probabilities for the UB moves of instanton $i$ are given by
\begin{align}
 \mathcal{P}_{ii'} & = \sum_j^{N'^B_I(i)} \mathcal{P}^B_{i(i',j)} + \delta^B_i \mathcal{P}^U_{ii'}\,,\\
 \mathcal{P}_{i'i} & = \sum_j^{N^B_I(i)} \mathcal{P}^B_{i'(i,j)} + \delta^B_{i'} \mathcal{P}^U_{i'i}\,,\\
\end{align}
with $\delta^B_i=1$ if $i$ is bonded and $\delta^B_i=0$ otherwise. We denote by $N^B_I$ ($N^B_A$) the number of bonded instantons (anti-instantons) and $N^B_I(i)$ is the number of anti-instantons that instanton $i$ is bonded to, and analogously for $N^B_B(i)$. Unprimed quantities are evaluated before the move whereas primed ones denote the same quantity after the move. The individual bonding and unbonding transition probabilities are given by 
\begin{align}
 \mathcal{P}^B_{i(i',j)} & = \frac{1}{N_I}\frac{1}{N_A} \frac{1}{V_j}\,,\\
 \mathcal{P}^U_{ii'} & = \frac{1}{N^B_I} \frac{1}{V}\,.
\end{align}
The bonding move consists of choosing uniformly an instanton and an anti-instanton, and to place the instanton $i$ uniformly in the bonding region $V_j$ of the anti-instanton $j$. The unbonding move consists of choosing uniformly one of the bonded instantons and to place it uniformly in the simulation box.

Insertion and deletion will be constructed along the lines of the UB algorithm by either placing the instanton $i$ into the bonding region of an anti-instanton or removing the bonded instanton $i$. Adding up all possible paths, including the unbiased one, we get
\begin{align}
 \mathcal{P}_{N_I N'_I} & = p_b \sum_j^{N'^B_I(i)} \frac{1}{N_A} \frac{1}{V_j} + (1-p_b)\frac{1}{V}\,,\\
 \mathcal{P}_{N'_I N_I} & = p_b \frac{\delta^B_{i}}{N'^B_I} + (1-p_b)\frac{1}{N'_I}\,.
\end{align}
Here $p_b$ is the a-priori-probability to perform biased moves.

It has also been argued that cluster moves are important to achieve good mixing of the Markov chain \cite{orkoulas:panagiotopoulos:associating_fluids}. We assume that $I-A$ pairs are the dominant clusters that form. The pair displacements consist of translating the pair as a whole or to displace one of the constituents within the bonding region of its partner, and the resulting transition probabilities are obvious. The pair insertions and deletions are given by
\begin{align}
 \mathcal{P}_{N N'} & = \frac{\delta^P_{i_I i_A}}{4} \frac{1}{V} \left( \frac{1}{V_{i_I}} + \frac{1}{V_{i_A}} \right) + \frac{1}{2} \frac{1}{V} \frac{1}{V}\,,\\
 \mathcal{P}_{N' N} & = \frac{1}{2} \frac{\delta^P_{i_I i_A}}{N'_P} + \frac{1}{2} \frac{1}{N'_I}\frac{1}{N'_A}\,,
\end{align}
where $N_P$ is the number of pairs. The last terms in each line follow from randomly and independently inserting/deleting an instanton and an anti-instanton. We found that inclusion of these unbiased moves enhances the acceptance rates at high temperature when we approach the dilute gas limit.

So far we have only discussed the spatial arrangements of the instantons. However, the interaction also depends on the colour degrees of freedom \cite{frenkel:smit:understanting_molecular_simulation}. This further technicality can be dealt with by adapting techniques for orientation dependent forces in molecular dynamics simulations. To this end we define the measure in colour space by the Boltzmann factor of the pair interaction. This ensures that $U$ is chosen so as to increase the interaction. More precisely, given the position of the instanton within the bonding box of its partner anti-instanton, we want to sample the following equilibrium distribution:
\begin{equation}
 \mathcal{P}(U_I|x_I,x_A,U_A) = \frac{\exp \left(-S^\mathrm{pair}_\mathrm{int}(U_I,U_A,x_I,x_A) \right)}{\int dU_I \exp \left(-S^\mathrm{pair}_\mathrm{int}(U_I,U_A,x_I,x_A)\right)}\,, \label{eq:color_orientation}
\end{equation}
where $dU_I$ is the Haar measure over $SU(3)$. Note that we neglect the influence of the neighbouring instantons and anti-instantons; we could include them but since the fermionic interaction involves a determinant the computation would become rather costly.

We cannot sample this distribution analytically, but we can sample it exactly within a Monte Carlo scheme \cite{frenkel:smit:understanting_molecular_simulation}. Using the fact that the Haar measure is invariant under group composition and also that the interactions (\ref{eq:glue_interaction}) and (\ref{eq:quark_interaction}) depend on $U=U^\dagger_I U_A$, it is natural to work with $U$. To sample (\ref{eq:color_orientation}), we choose a set of $N$ different colour matrices uniformly over $SU(3)$, i.e.\ according to the Haar measure. We select one among those $N$ matrices according to the probability
\begin{equation}
 p_i = \frac{e^{-S^\mathrm{pair}_i}}{\sum_j^N e^{-S^\mathrm{pair}_j} }\,, \label{eq:color_bias_0}
\end{equation}
with $S^\mathrm{pair}_i=S^\mathrm{pair}_\mathrm{int}(U_i,x_I,x_A)$. Now, remember that the transition probability is given by a proposal and acceptance probability, and that the former includes any a-priori-probabilities for generating the proposed state. Thus, the orientational bias includes the generation probability of the set of matrices, and is given by
\begin{equation}
 \mathcal{P}_i=\left(\frac{1}{V_{SU(3)}}\right)^{2N-1} N p_i\,. \label{eq:color_bias_1}
\end{equation}
The selection probability and the first $N$ volume factors are obvious. The additional $N-1$ volume factors need further explanation. Given the form (\ref{eq:acceptance}) for the acceptance probability, apart from the $N$ matrices used in (\ref{eq:color_bias_0}), we need to generate another $N-1$ trial matrices that combine with the matrix selected in the forward move to form the set of $N$ matrices necessary to evaluate the orientational bias, (\ref{eq:color_bias_0}), in the backward move. Finally, the extra enhancement of $N$ is necessary because we are not so much interested in $U_i$ being generated at the $i\mathrm{th}$ trial but rather that the value $U_i$ is selected. Thus we must marginalise over the label $i$: adding up all the permutations of the set of trial matrices modulo the permutations of the subset of the $N-1$ matrices that are not selected produces the extra $N$ factor. Of the $2N-1$ volume factors $2N-2$ will cancel in general\footnote{One might think that all $2N-1$ volume factors drop out, but in general we might add new moves that select $U$ with a rule different from (\ref{eq:color_bias_0}); in such a case we only need $2N-2$ trial matrices.} and so we can simplify (\ref{eq:color_bias_1}) to
\begin{equation}
 \mathcal{P}_i= \frac{e^{-H_i}}{\frac{V_{SU(3)}}{N}\sum_j^N e^{-H_j}}\,, \label{eq:color_bias_2}
\end{equation}
which we recognise as an approximation to (\ref{eq:color_orientation}); note, however, that within the MC scheme is exact, i.e.\ the results do not suffer from any discretisation errors. Including the orientation bias amounts to multiplying the bonding box volume factors $V_j$ by (\ref{eq:color_bias_2}).

We found that the interaction are rather self-similar when expressed in units of $\bar{\rho}=\sqrt{\rho_I^2+\rho^2_A}$. Thus, the bonding region depends on both particles that make up the pair through a function of their sizes. We choose the functional form of the size dependence to be given by
\begin{equation}
 V_j \to V_{ij} = V_\mathrm{UB} \left( \rho_i^2 + \rho_j^2 \right)^2\,.
\end{equation}
By construction, $V_\mathrm{UB}$ is independent of the specific pair, and we will tune it to achieve fast convergence.

\begin{figure}[tbp]
\begin{center}
 \includegraphics[width=\figwidth,clip=true,trim=0mm 0mm 15mm 10mm]{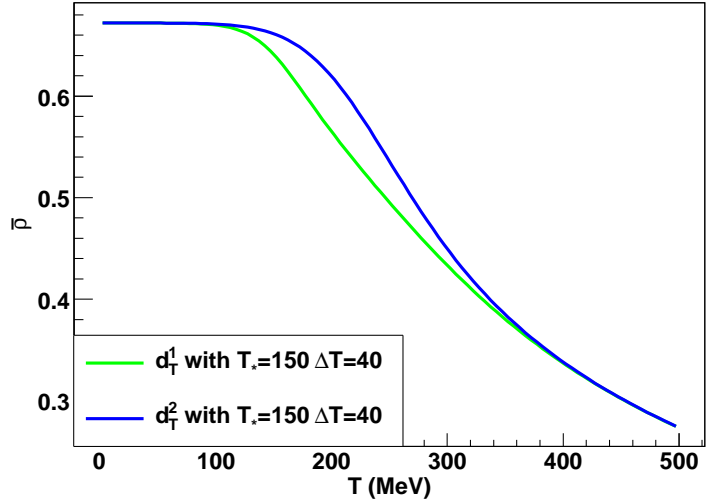}
\end{center}
 \caption{The screening factors (\ref{eq:plasma:screening:transition}) lead to constant mean instanton sizes in the confined phase, below $T_*$, as has been found from lattice investigations \cite{chu:schramm:instanton:content}. As expected, the factor $d_2$ leads to a slower and more symmetrical switch-on of the screening.}\label{fig:rho:T}
\end{figure}

The IILM does not lead to confinement; from a heuristic point of view this follows from the simple fact that the Harrington--Shepard calorons have a vanishing Polyakov loop, the order parameter of the phase transition. It has been in argued in \cite{shuryak:velkovsky:density} that at low temperatures the instanton density should only depend very weakly on temperature and that, in particular, the result (\ref{eq:plasma:screening}) is only applicable in the plasma, i.e.\ deconfined, phase. In the hadronic phase the fundamental degrees of freedom are not screened and the heat bath consists mainly of pions. We will follow the prescription put forward in \cite{schaefer:shuryak:iilm} and include in (\ref{eq:plasma:screening}) a phenomenological term that mimics the transition from the confined to the deconfined phase. We have decided to investigate this with the following two choices, see \reffig{fig:rho:T}:
\begin{align}
 d_T^1(\rho) &= d_0(\rho) \, d_T(\rho)^{\alpha_T}\,,  \nonumber\\
 d_T^2(\rho) &= d_0(\rho) \left( 1 + \alpha_T \left( d_T(\rho)-1 \right) \right)\,, \nonumber\\
 \alpha_T &= \frac{1}{2} \left( 1 + \tanh \frac{T-T_*}{\Delta T} \right)\,. \label{eq:plasma:screening:transition}
\end{align}
The functional form $d^1_T$ for the transition between hadronic and plasma phase has been used in \cite{schaefer:shuryak:iilm}. The functional form $d^2_T$ is chosen on the basis that, intuitively, it should lead to a more symmetrical behaviour around $T_*$. With these two choices we will get a rough idea on the systematics due to these phenomenological terms. We've introduced two new parameters into the IILM; they will be fixed by comparing our results to available lattice data. 

So far these modified screening factors are purely phenomenological. However, the more general non-trivial holonomy calorons \cite{kraan:baal:caloron:I,kraan:baal:caloron:II,kraan:baal:caloron:monopole} \cite{lee:lu:caloron}, that might play a crucial role in driving the confinement/deconfinement phase transition, have been shown to lead to mean instanton sizes that depend very weakly on temperature in the confined phase \cite{gerhold:ilgenfritz:mueller_preussker:kvbll:gas:confinement}. Given the assumptions of that paper, the change of the Polyakov loop with temperature is responsible for the different screening behaviour of instantons in the confined and deconfined phase: in the former, the equilibrium state corresponds to an ensemble of maximally non-trivial holonomy calorons which are not screened, whereas the latter consists of trivial holonomy Harrington--Shepard calorons which are strongly screened. We can therefore interpret the modified screening factors (\ref{eq:plasma:screening:transition}) as an effective description of the change of the Polyakov loop as the system evolves through the phase transition. Generalising the IILM to include these calorons in the future, we might be able to describe the plasma screening effects self-consistently.

\begin{figure}[tbp]
\begin{center}
 \includegraphics[width=\figwidth,clip=true,trim=0mm 0mm 15mm 10mm]{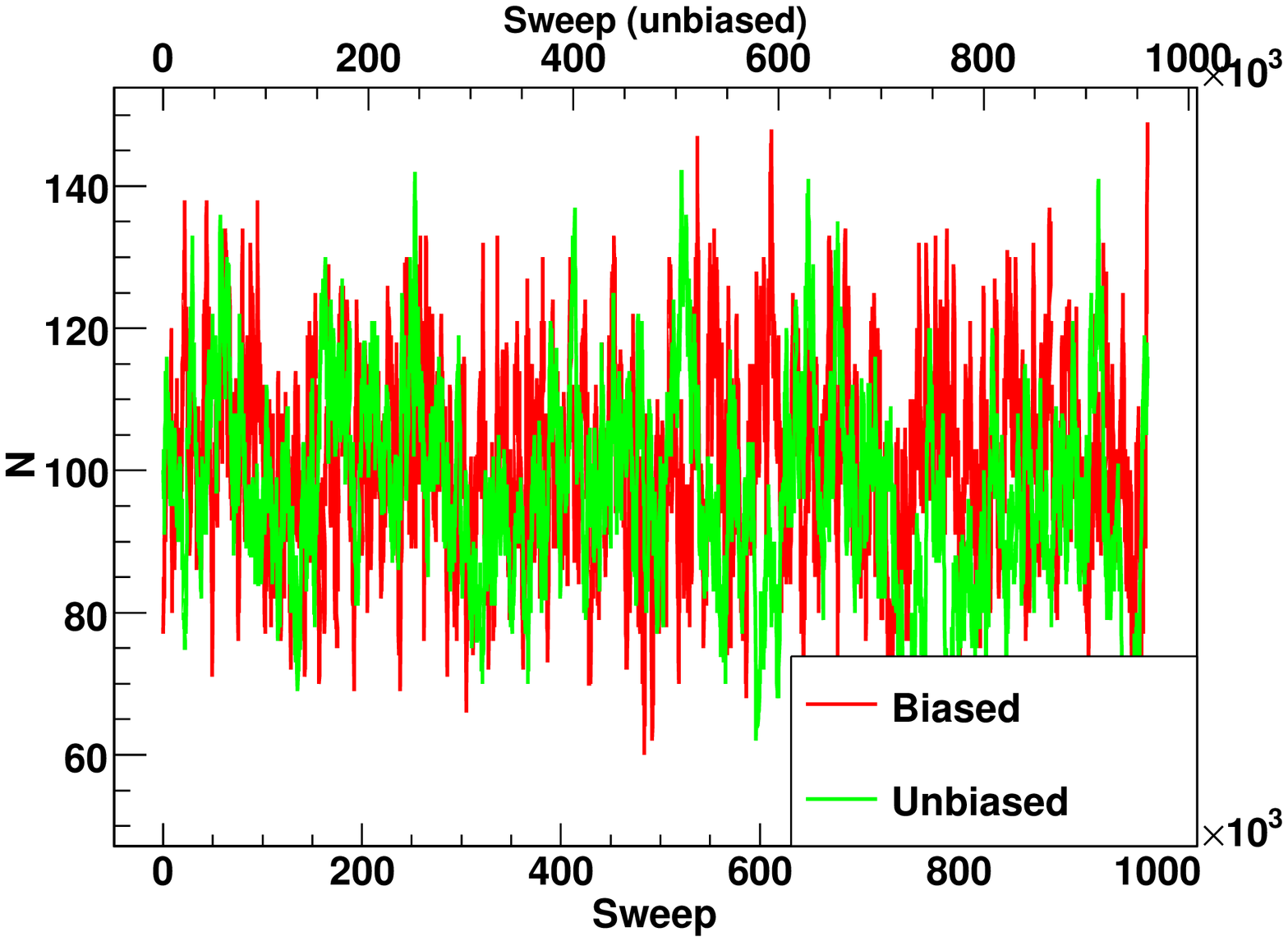}
 \includegraphics[width=\figwidth,clip=true,trim=0mm 0mm 15mm 0mm]{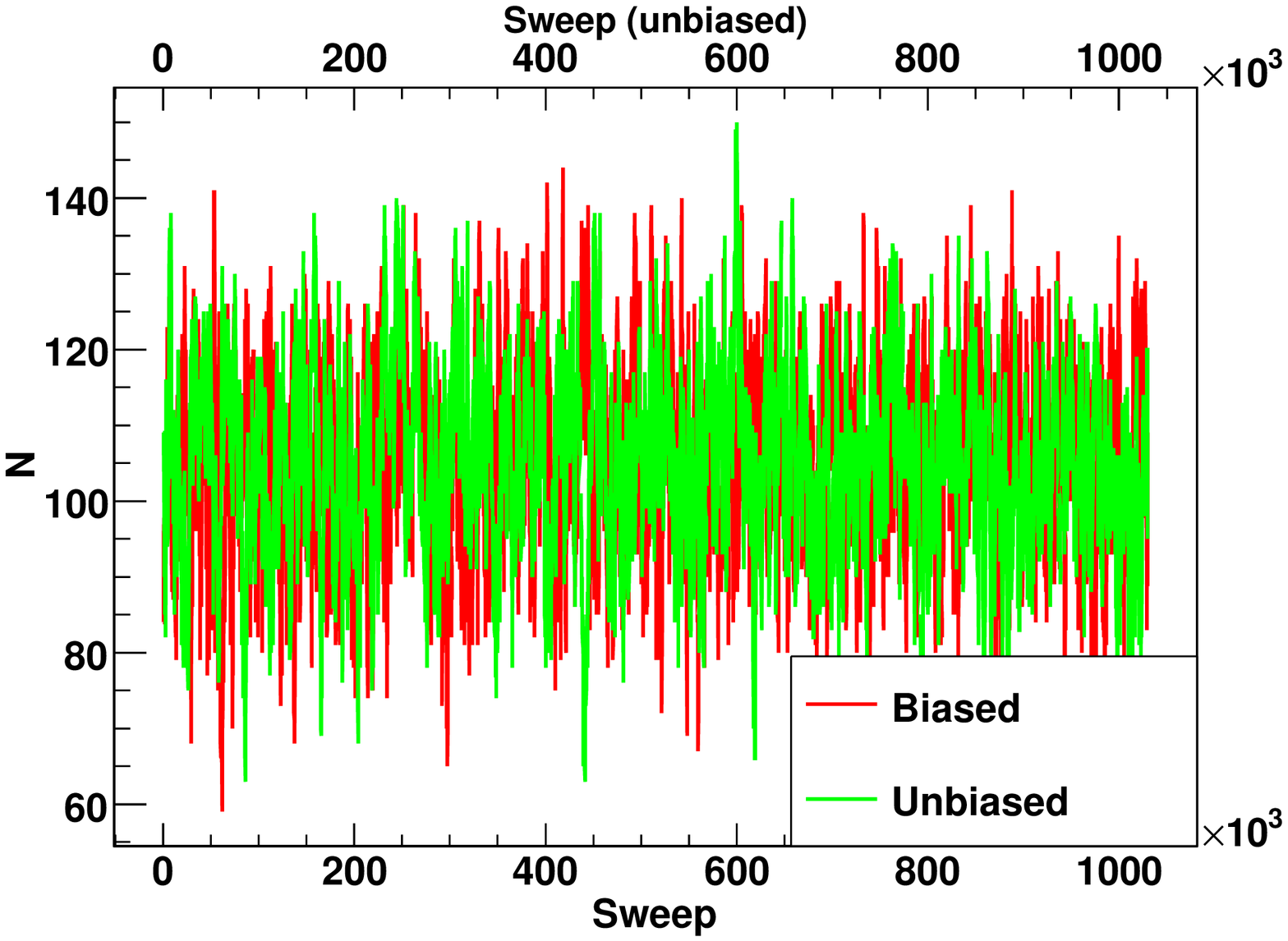}
\end{center}
\caption{In the hadronic phase, i.e.\ without the plasma screening factor (\ref{eq:plasma:screening}), biased and unbiased simulations agree very well. This is to be expected because we know from the $T=0$ simulations, see \cite{wantz:iilm:1}, that ordinary Monte Carlo leads to good sampling. These low-$T$ simulations serve thus to check the biased Monte Carlo scheme.}\label{fig:thermalization:tl}
\end{figure}

Note that for low temperatures the IILM should not be very different from the $T=0$ case, which was well dealt with by ordinary Monte Carlo. We can use this regime to test the more advanced Monte Carlo scheme against unbiased simulations. In \reffig{fig:thermalization:tl} we plot the thermalisation history at such low temperatures for two different set of quark masses; the latter are given by the physical masses determined in \cite{wantz:iilm:1} and the second set has the same dimensionless strange quark mass and identical light quarks with a value of ten times the physical up quark. We can clearly see that the histories for the biased and unbiased simulations are very similar.

\begin{figure}[tbp]
\begin{center}
 \includegraphics[width=\figwidth,clip=true,trim=0mm 0mm 15mm 10mm]{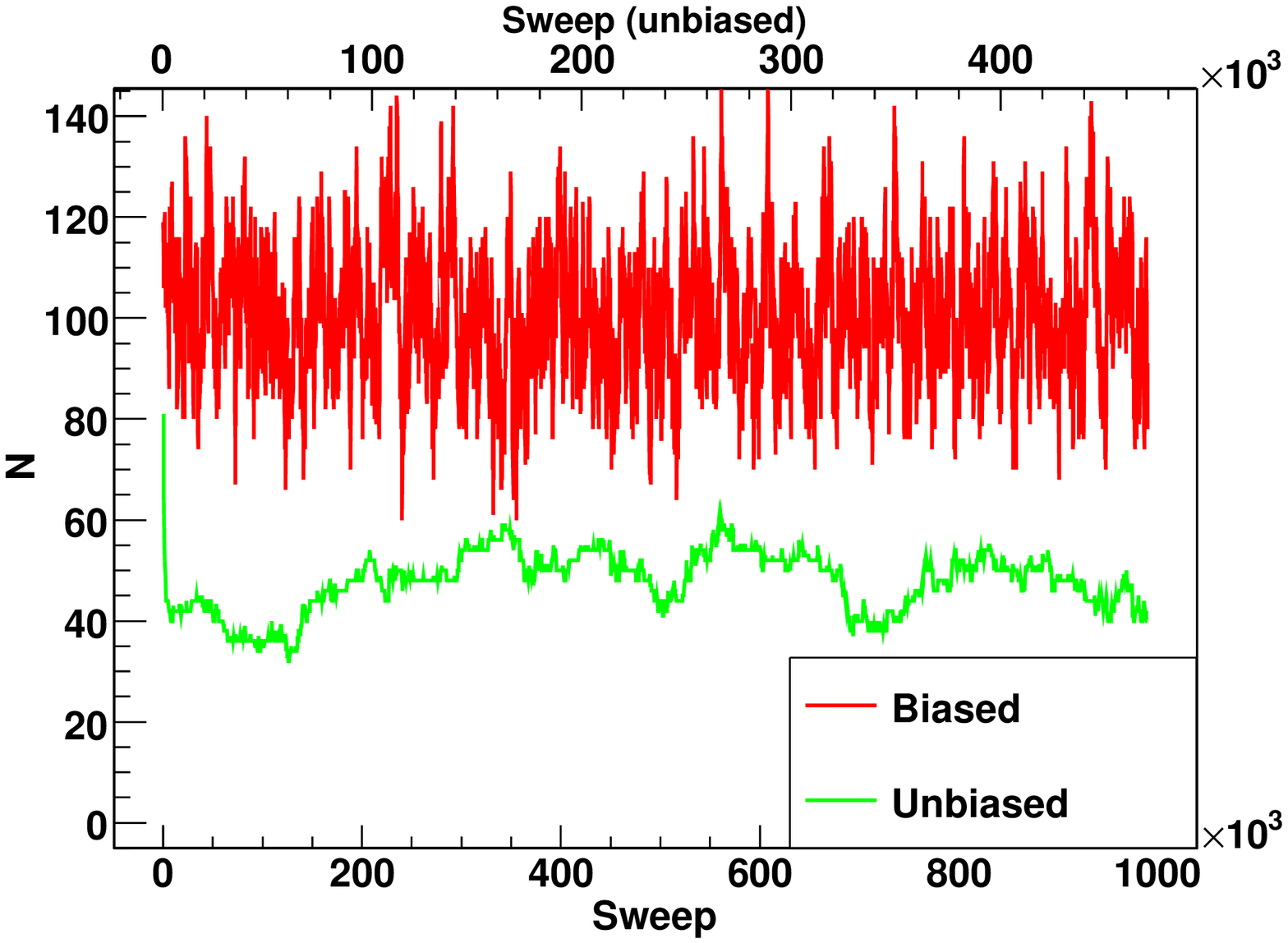}
 \includegraphics[width=\figwidth,clip=true,trim=0mm 0mm 15mm 0mm]{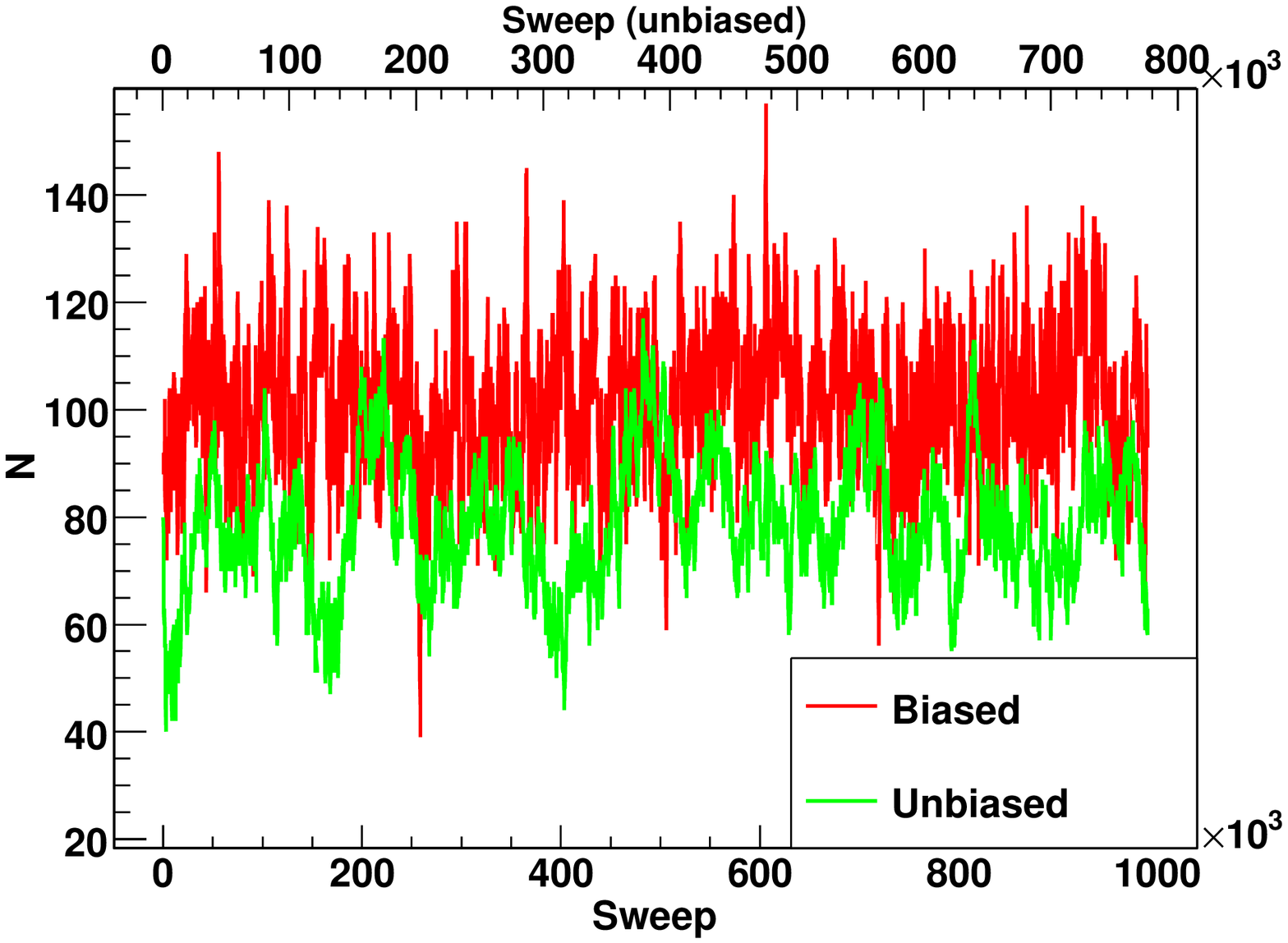}
\end{center}
\caption{Around the confinement/deconfinement transition, modeled by (\ref{eq:plasma:screening:transition}), the formation of instanton--anti-instanton pairs is important as it drives the chiral symmetry restoration in the IILM. We can clearly see that for the small, physical quark masses (top) ordinary Monte Carlo fails to sample the ensemble correctly. For the large quark masses the effect is much less pronounced but there is a clear systematic difference in the mean instanton number. It can be attributed again to fewer pairs.}\label{fig:thermalization:tc}
\end{figure}

As the temperature is increased, ordinary Monte Carlo fails for the set of physical quark masses. For intermediate temperatures, around $T_*$, there is a substantial difference in the results from biased and unbiased simulations. Even for the set of larger masses there are differences visible to the naked eye, see \reffig{fig:thermalization:tc}. For these temperatures instanton--anti-instanton pair formation, the mechanism that drives the chiral symmetry restoration in the IILM, is supposed to be very important. We find that the number of pairs is indeed quite different with $\langle N_P \rangle \approx 20$ for biased and $\langle N_P \rangle \approx 7$ for unbiased simulations and physical quarks. For the larger quark masses we find that the difference in the number of pairs between biased, $\langle N_P \rangle \approx 16$, and unbiased, $\langle N_P \rangle \approx 11$, simulations is less pronounced, as expected.

\begin{figure}[tbp]
\begin{center}
 \includegraphics[width=\figwidth,clip=true,trim=0mm 0mm 15mm 10mm]{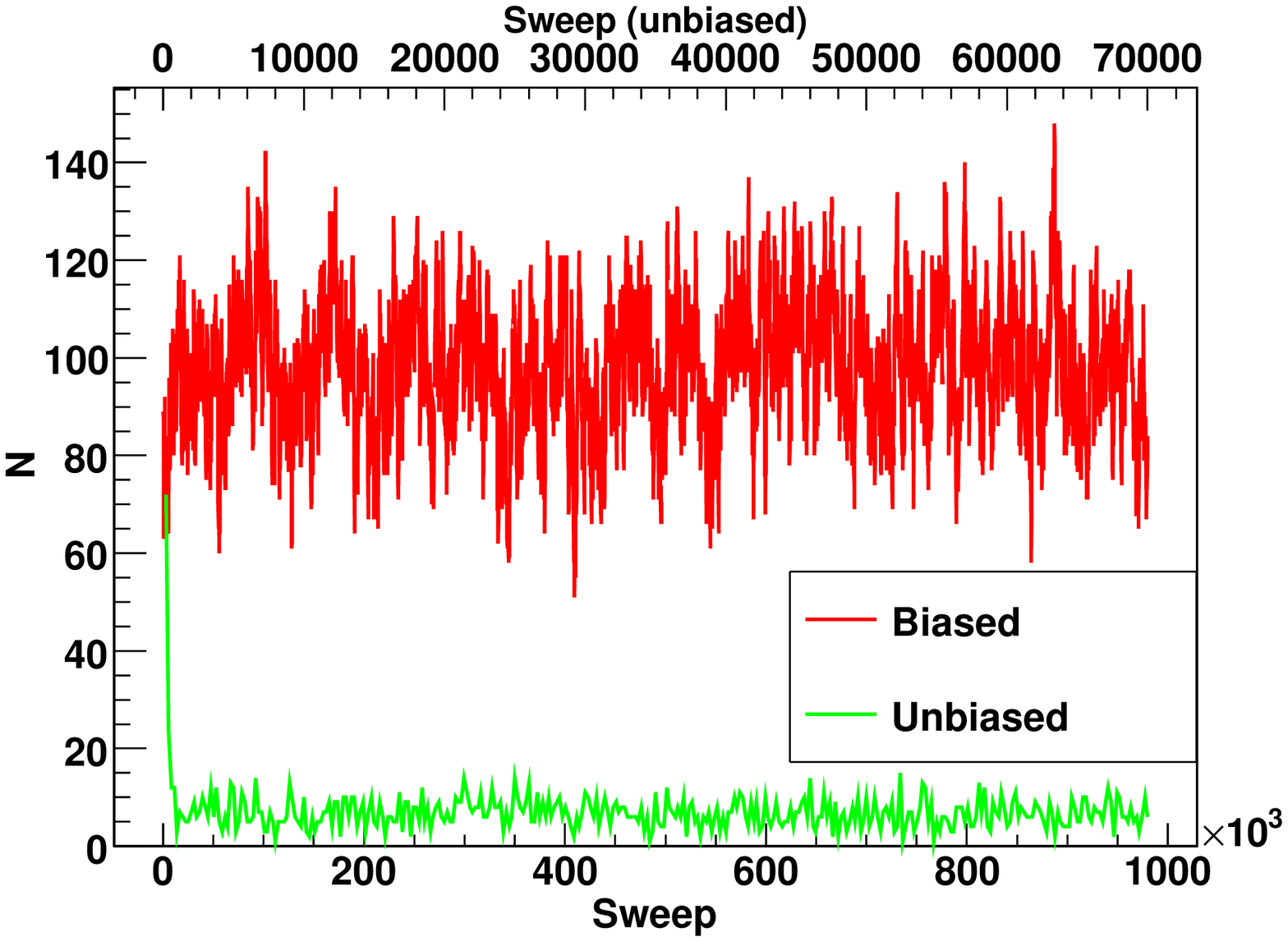}
 \includegraphics[width=\figwidth,clip=true,trim=0mm 0mm 15mm 0mm]{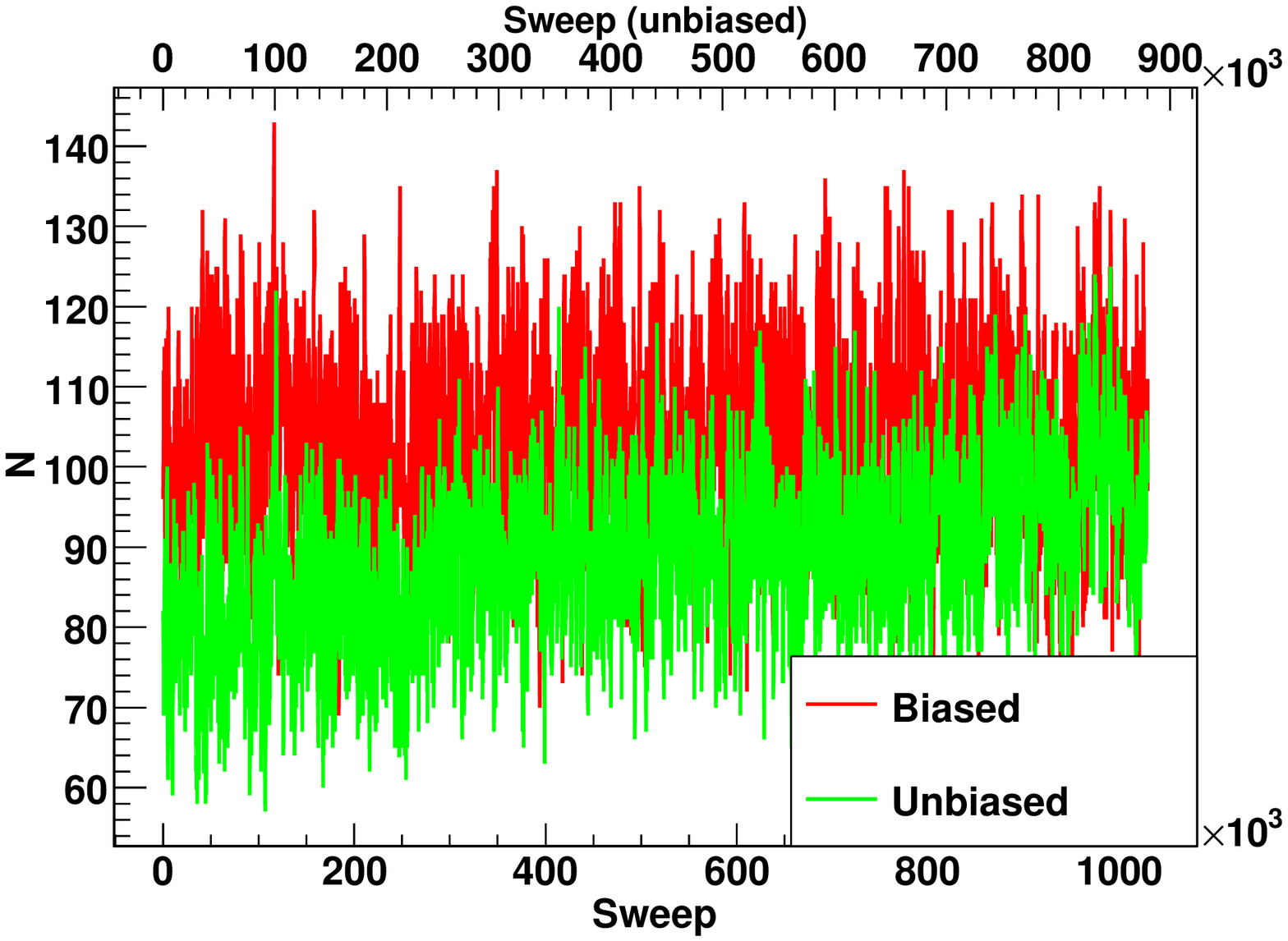}
\end{center}
\caption{In the plasma phase, ordinary Monte Carlo fails drastically for the physical quark masses (top). This trend will, however, not persist to higher temperatures. Depending on the quark masses the energy-dominated pair configurations will be outweighed in the partition function by entropy-dominated configurations at sufficiently high temperatures. For larger masses this happens at lower temperatures (bottom).}\label{fig:thermalization:th}
\end{figure}

At temperatures well inside the plasma phase, ordinary Monte Carlo fails completely to describe the system accurately for physical quark masses, see \reffig{fig:thermalization:th}. In this case the number of pairs is very different, $\langle N^\mathrm{biased}_P \rangle \approx 18$ and $\langle N^\mathrm{unbiased}_P \rangle \approx 1$. Note, however, that the number of pairs has decreased for the higher temperatures.

Simulations at larger masses converge (slowly) to the correct equilibrium distribution. Actually, ordinary Monte Carlo perform better at high temperature than around $T_*$ for these larger quark masses. The reason is that the ensemble equilibrates in a very dilute state, so that there are only a few instanton--anti-instanton pairs, $\langle N^\mathrm{biased}_P \rangle \approx 4.5$ and $\langle N^\mathrm{unbiased}_P \rangle \approx 1.5$. Note, again, that the pair concentration has dropped with respect to temperatures around $T_*$. It turns out that, quite generally, the molecule concentration drops with increasing temperature, because the energy-dominated pair configurations have a smaller and smaller entropy and are outweighed by the entropy-dominated configurations; the latter correspond to a truly random ensemble of instantons at a high enough temperature. This will happen for physical quark masses as well, albeit at higher temperatures. 

To be a bit more precise, consider the contribution to the partition function of an instanton--anti-instanton pair, which we approximate by its dilute gas limit and by the contribution from the bonding box. The latter dominates, and hence biased simulations might be needed, if
\begin{equation}
 1 < \left(\int d(\rho)\right) \Delta V \overline{\exp\left(-S_\mathrm{int}\right)}\,.
\end{equation}
The contribution from the two instanton measures have cancelled on both sides, and we used $V=N/2\int d$ for the dilute gas. The overbar is meant to indicate the average over the bonding box, size and colour orientations.

\begin{figure}[tbp]
\begin{center}
 \includegraphics[width=0.3\textheight]{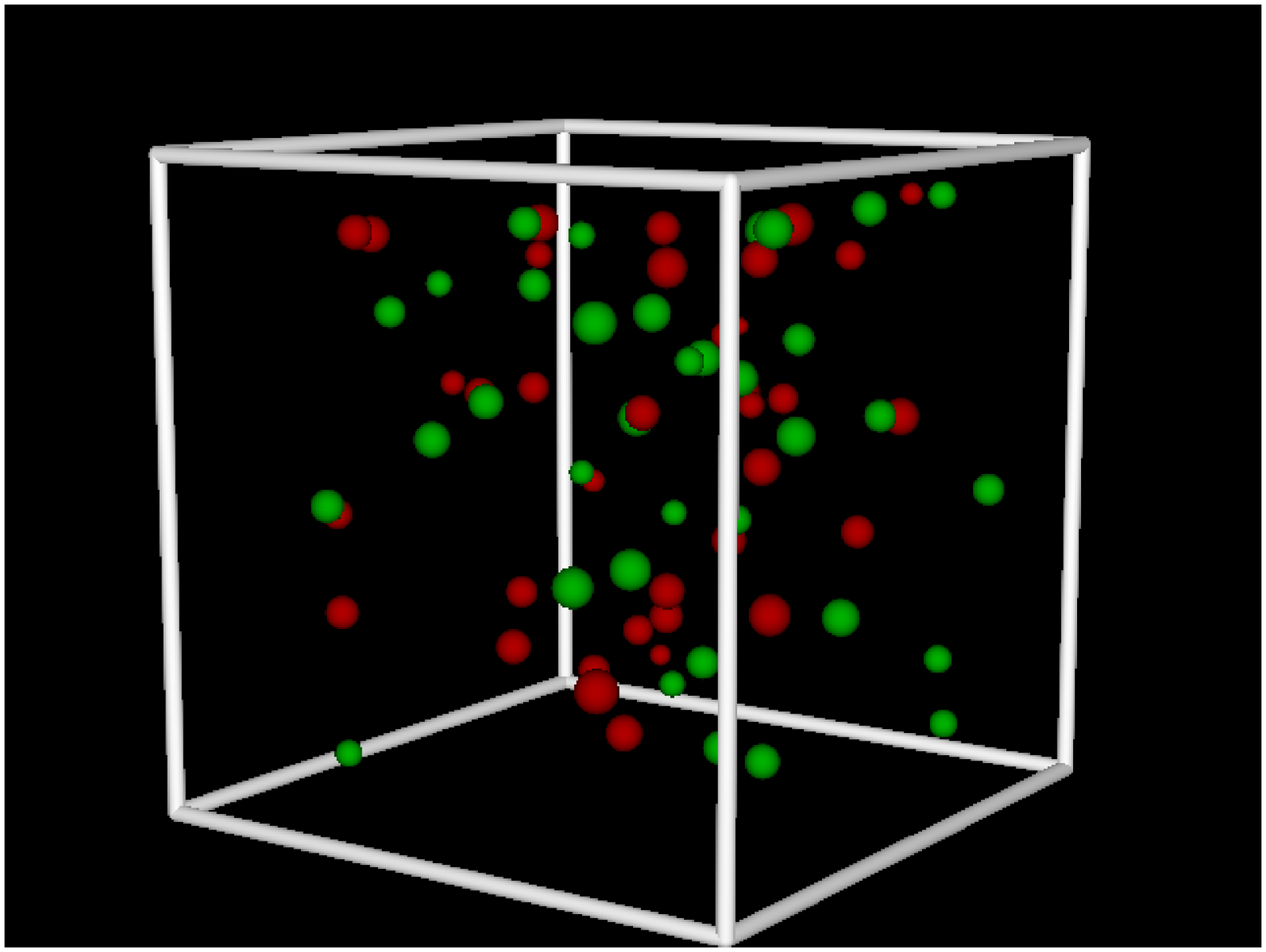}\\
 \includegraphics[width=0.3\textheight]{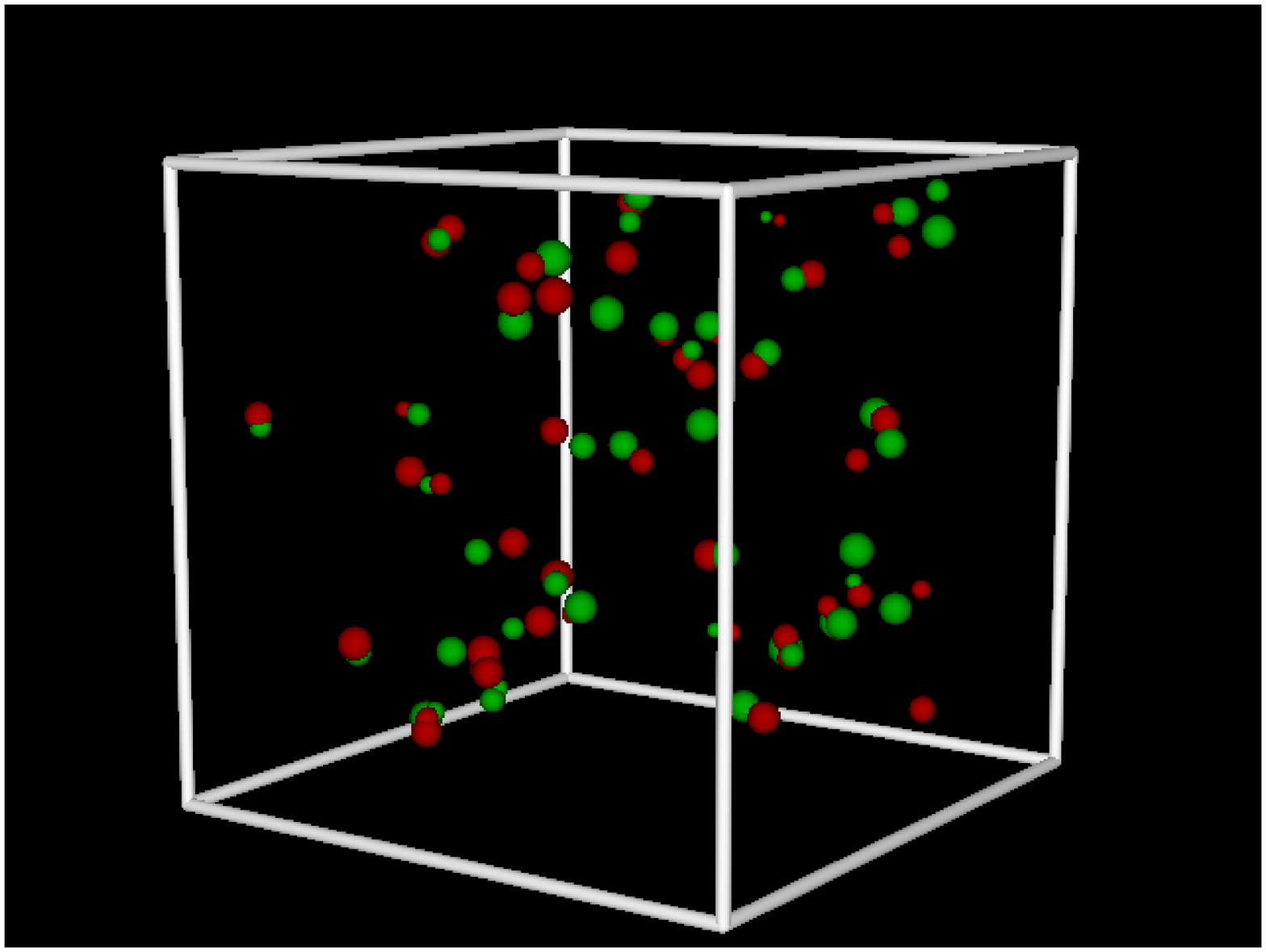}\\
 \includegraphics[width=0.3\textheight]{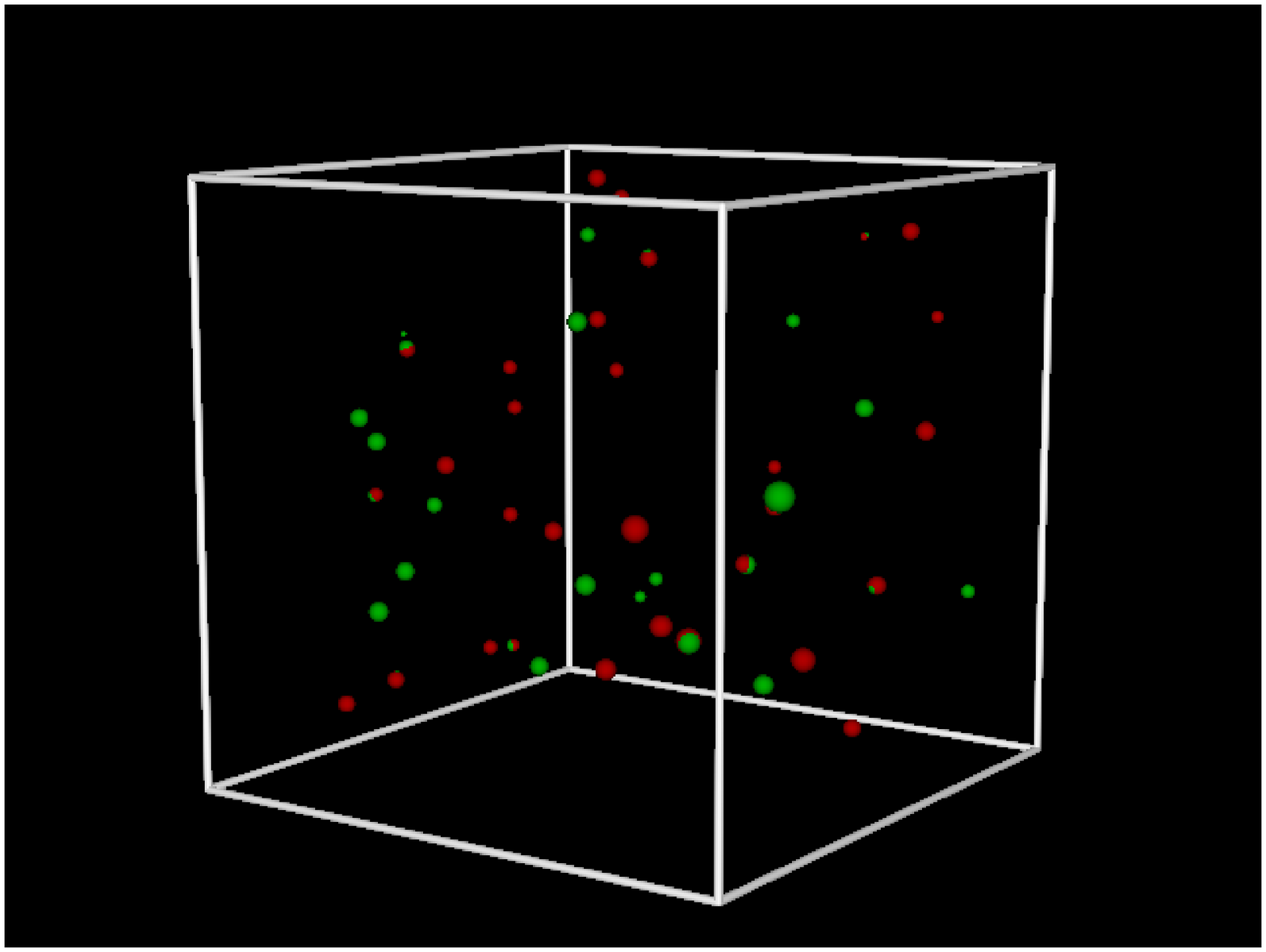}
\end{center}
\caption{We display a (spatial) snap-shot of a typical configuration. For low temperature (top) spatial correlations are not very pronounced and the system equilibrates in a `random' state. In the region $T\approx T_*$ (middle) a higher concentration of instanton--anti-instanton pairs can be seen, signalling the restoration of chiral symmetry in the IILM. At higher temperatures (bottom) the ensemble becomes dilute: the energy gain in pair-formation is outweighed by the large entropy gain of a random distribution of instantons throughout the box.}\label{fig:povray}
\end{figure}

Using the 1-loop formula for the instanton measure and choosing one specific size with $T\rho=\mathrm{const}$, this leads to
\begin{equation}
 0 < \left[-\left(7+\frac{N_f}{3}\right) + \left(11-\frac{2}{3} N_f \right) (-V_{12}) + 2N_f -4 \right] \log T +\mathrm{const}\,.
\end{equation}
We only display explicitly the temperature dependence for the one-loop instanton measure (first term), the gluonic interaction (second term), the quark interaction (third term) and the scaling with temperature of the bonding box (fourth term). It turns out that the gluonic contribution is bounded by $5$ for $N_f=3$. Hence the $\log T$ term is negative and the dilute gas approximation becomes better and better as the temperature increases, see also \reffig{fig:povray}. For higher temperatures, when more quark flavours become active, the trend seems to be reversed. Note, however, that this conclusion completely neglects the high-frequency quantum interaction, which should be investigated in the future to study this issue further.

The UB algorithm is only useful if the results do not depend strongly on the precise implementation, i.e.\ the actual definition of the bonding box. We checked this by using different bonding boxes for three different temperatures, i.e.\ above, around and below $T_*$, see \reftable{table:bondingbox}. We found that the results agree very well for temperatures below $T_*$. This corroborates our expectations that biased Monte Carlo is not essential in that regime. For a modest sample of $200$ independent configurations\footnote{This relates to $\langle N \rangle$; the autocorrelation time for $\langle Q^2 \rangle$ was always smaller, effectively leading to a larger sample size. The better agreement on the latter quantity corroborates the expectation that the differences among bonding boxes vanish for infinite sample size.}, we also find rather good agreement for the higher temperatures where random sampling fails. We have also checked the dependence on the a-priori-probabilities\footnote{The data in \reftable{table:bondingbox} corresponds to one particular choice for the a-priori-probabilities.}; we find again that the results only depend weakly on these choices for the small sample size we have used. Autocorrelation times, however, depend much more strongly on the bonding box; this allows us to fine-tune the parameters to achieve efficient sampling. The bonding box $B_4$ will be used for the final simulation.

\begin{table*}[tbp]
\begin{center}
\begin{tabular}{c|c|c|c|c|c|c}
  & $B_1$ & $B_2$ & $B_3$ & $B_4$ & $B_5$ & $B_6$ \\ \hline\hline
 $\langle N \rangle $ & $99(1)$ & $100(1)$ & $97(1)$ & $102(1)$ & $98(1)$ & $98(1)$\\ \hline
 $\langle Q^2 \rangle $ & $0.81(1)$ & $0.81(1)$ & $0.81(1)$ & $0.78(1)$ & $0.81(1)$ &  $0.779(9)$\\ \hline
 $\langle \bar{u}u \rangle $ & $230(9)$ & $219(8)$ & $222(9)$ & $217(9)$ & $211(9)$ & $225(8)$ \\ \hline\hline
 $\langle N \rangle $ & $105(1)$ & $101(1)$ & $96.7(9)$ &$99.3(8)$ & $98.6(9)$ & $95.3(9)$\\ \hline
 $\langle Q^2 \rangle $ & $0.36(1)$ & $0.31(2)$ & $0.39(2)$ & $0.30(2)$ & $0.347(8)$ & $0.349(6)$\\ \hline
 $\langle \bar{u}u \rangle $ & $84(8)$ & $68(6)$ & $80(8)$ & $69(7)$ & $75(7)$ & $72(7)$\\ \hline\hline
 $\langle N \rangle $ & $102.0(9)$ & $106(1)$ & $99.2(9)$ & $106(1)$ & $108(1)$ & $102.2(9)$\\ \hline
 $\langle Q^2 \rangle $ & $6.9(3)$ & $7.0(3)$ & $7.0(2)$ & $6.7(3)$ & $6.8(2)$ & $6.9(1)$\\ \hline
 $\langle \bar{u}u \rangle $ & $1320(30)$ & $1300(30)$ & $1300(40)$ & $1200(30)$ & $1310(40)$ & $1210(30)$
\end{tabular}
\\[1cm]
\begin{tabular}{c|c|c|c|c|c|c}
  & $B_1$ & $B_2$ & $B_3$ & $B_4$ & $B_5$ & $B_6$ \\ \hline\hline
 $x_i^{bb} $ & $0.4$ & $0.4$ & $0.4$ & $0.6$ & $0.23$ & $0.17$\\ \hline
 $x_{t,\mathrm{min}}^{bb}$ & $0.2$ & $0.25$ & $0.3$ & $0.2$ & $0.35$ & $0.4$\\ \hline
 $x_{t,\mathrm{max}}^{bb}$ & $2$ & $1.8$ & $1.6$ & $2$ & $1.2$ & $1$ \\ 
\end{tabular}
\end{center}
\caption{The temperature increases from top to bottom. We find rather good agreement between the different bonding boxes. The sample has a modest size of $200$ independent configurations for $\langle N \rangle $. The topological susceptibility has smaller autocorrelation times and a correspondingly larger sample size. The better agreement for $\langle Q^2 \rangle$ can be used to argue that the differences between the bonding boxes vanish with increasing sample size, as expected. The bottom table gives the dimensions, defined in \reffig{fig:interaction:T}, for the different bonding boxes $B_i$.}
\label{table:bondingbox}
\end{table*}

\subsection{Fermionic determinant}

We argued that for sufficiently high temperatures, the IILM goes over to a dilute ensemble. It turns out that for high dilution the numerical manipulation of the quark overlaps becomes unstable because the overlap matrix becomes nearly degenerate. To render the evaluation of the fermionic interaction stable we need to decompose the overlap matrix into smaller non-degenerate blocks.

To achieve this, we truncate the interactions to zero above a certain cutoff; it is a function of the pair sizes and determines which instantons interact with each other. Given this cutoff, we can built clusters of pairwise interacting instantons. The overlap matrix $T_{IA}$ is then decomposed into a direct sum according to these clusters. Not only does this render the numerical manipulations stable, but it also leads to a dramatic speed increase. Incidentally, we also use a cutoff for the gluonic interactions to get a better scaling with the number of instantons.

\section{Quenched simulation}
\label{sec:quenched:T}

The main purpose of this section is to investigate how well the phenomenological screening factor (\ref{eq:plasma:screening:transition}) can model the confinement/deconfinement phase transition. To this end, we compare the topological susceptibility, for various values of the free parameters, to lattice data \cite{alles:elia:giacomo:top_susc:quenched:T}. Normalising to the $T=0$ result, we get $\Lambda=206(8) \mathrm{MeV}$.

\begin{figure}[tbp]
\begin{center}
 \includegraphics[width=\figwidth,clip=true,trim=0mm 0mm 15mm 10mm]{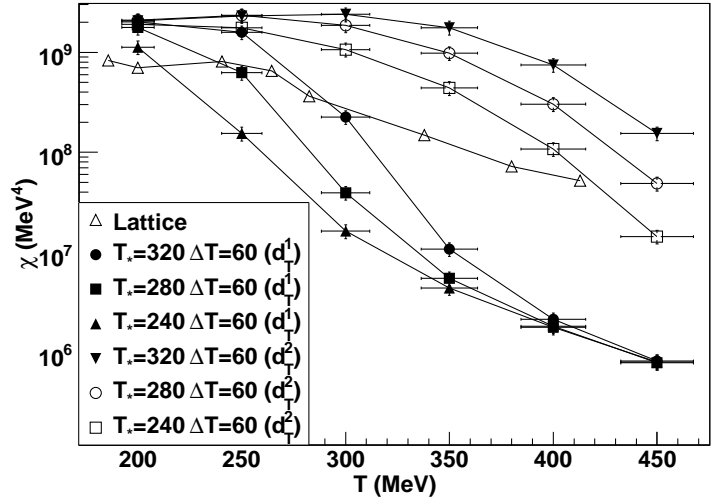}
\end{center}
\caption{A striking feature of the quenched simulation is that the topological susceptibility decays generically too fast, as compared to lattice data. At low temperatures the IILM predicts a rising topological susceptibility, in contrast to the lattice data, which is roughly constant. The screening factor $d^1_T$ underestimates the lattice result, whereas $d^2_T$ tends to overestimate it at temperatures around $T_*$. Once the plasma screening is fully effective, both lead to too fast decays: the dilute gas approximation gives an approximate power-law of $\chi \propto T^{-8}$ which is not compatible with the available lattice data.}\label{fig:chi:quenched:T}
\end{figure}

We performed many simulations with different sets $\{T_*,\Delta T\}$. We report a few of these in \reffig{fig:chi:quenched:T}, which is representative of the general findings. Namely, that the screening factors do not capture the lattice data well. In particular, we found that the IILM result for the topological susceptibility decays too fast: the power-law-like behaviour, e.g.\ $\chi \propto T^{-8}$ from the dilute gas approximation, at temperatures above $T_*$ is not compatible with the available lattice data. The screening factor $d^1_T$ generically underestimates the lattice result for rather moderate temperatures. Only higher $T_*$ could remedy this; however, for such values the topological susceptibility is too high in the low temperature regime. This can also be seen in \reffig{fig:chi:quenched:T} for the second screening factor $d^2_T$: it has a gentler switch-on of the plasma screening effects, and also leads to rather large values for the topological susceptibility at low temperature.

We actually found that without any plasma screening, and therefore also in the low temperature region, the IILM produces a susceptibility that rises with temperature. This is not what the lattice predicts, namely an almost constant $\chi$. The quenched IILM at finite temperature does not seem to give a good description of the pure Yang-Mills sector: that it fails around the phase transition could be expected since the crude phenomenological screening terms can only be expected to be a rough effective description; however, we find that the IILM fails to qualitatively describe the pure gauge dynamics at low and high temperature.

Given the fact that the dilute gas of non-trivial holonomy calorons \cite{gerhold:ilgenfritz:mueller_preussker:kvbll:gas:confinement} gave rather encouraging results, it might be that these are essential to model QCD in the quenched case. In particular, it will be interesting to find out whether these degrees of freedom can capture the smaller decay at high temperature or whether other degrees of freedom are dominant.

\section{Chiral symmetry restoration}
\label{sec:chiral}

Compared to the quenched sector, full QCD has the additional property of (softly broken) chiral symmetry. Based on it, powerful analytical approaches like chiral perturbation theory have been developed. A strong point of the IILM is its rather accurate description of the chiral properties of QCD. As compared to the quenched sector, this additional structure can be expected to improve the agreement between the IILM and the lattice at finite temperature.

We will repeat the analysis of the previous section for the unquenched IILM. Instead of the topological susceptibility, which we want to compute after all, we will use the chiral susceptibility \cite{aoki:fodor:katz:szabo:transition:temperature} to estimate the parameters $T_*$ and $\Delta T$. Note that there is a controversy with regard to the lattice results for the critical temperature \cite{detar:gupta:tc,karsch:recent:partII,aoki:et_al:transition:temperature:II} with differences on the order of $20 \units{MeV}$.

The chiral susceptibility is defined by
\begin{equation}
 \chi_{\bar{q}q} = \partial_{m}^2 \ln Z_{\mathrm{QCD}} = \int\int \langle \bar{\psi}\psi(x) \bar{\psi}\psi(y) \rangle - \left ( \int \bar{\psi}\psi(x) \right)^2\,,
\end{equation}
with $Z_{\mathrm{QCD}}$ the QCD partition function. It is actually composed of a connected and a disconnected part, and the latter is most sensitive to the chiral transition \cite{karsch:laermann:susceptibilities}. In terms of propagators both parts are given by
\begin{align}
 \chi_{\bar{q}q}^d & = \langle \left( \int \Tr S_A(x,x) \right)^2 \rangle_A - \left( \langle \int \Tr S_A(x,x)  \rangle_A \right)^2,\\
 \chi_{\bar{q}q}^c & = -\langle \int \Tr S_A(x,y) S_A(y,x) \rangle_A + \chi_{\bar{q}q}^d\,,
\end{align}
and the expectation values are over the gluon fields. These formulas make it clear that the disconnected part is the main order parameter of the transition. Within the IILM, the propagator is the sum of a low and a high frequency part; the latter is usually just approximated by the free massive propagator. As such it gives an infinite contribution to the disconnected part and has to be subtracted at $T=0$. A finite contribution remains at $T \neq 0$; it can be derived from the free partition function \cite{kapusta:finite:temperature:field:theory} and turns out to be small compared to the low-frequency contribution, and we will ignore it. This is expected, as the chiral transition is driven by the dynamics of the interacting instanton ensemble, and so the contribution from the free propagator should be negligible. By the same reasoning, we will ignore the cross terms between the low frequency and the free propagator. It is then straightforward to compute the chiral susceptibility in the IILM. The propagator, truncated to the finite dimensional low frequency part, is given by
\begin{equation}
 S_A(x,y) = (\slashed{D}+m)_{\mathrm{low}}^{-1} = \sum_n \frac{1}{m-i\lambda_n}\xi_n(x) \xi_n^\dagger(y)\,,
\end{equation}
where $\lambda$ are the eigenvalues of the matrix defined in (\ref{eq:quark_interaction}); note that we used the fact that the set $\{\xi_n\}$ of zero modes is assumed orthonormal. The susceptibilities in the IILM are then given by
\begin{align}
 \chi_{\bar{q}q}^d & = \langle \left( \sum_n \frac{1}{m-i\lambda_n} \right)^2 \rangle - \left( \langle \sum_n \frac{1}{m-i\lambda_n} \rangle \right)^2\,,\\
 \chi_{\bar{q}q}^c & = - \langle \sum_n \left( \frac{1}{m-i\lambda_n} \right)^2 \rangle + \chi_{\bar{q}q}^d\,.
\end{align}
For notational simplicity we have omitted the subscript $A$ on the expectation value, but the average is again over the gauge fields. Remembering that the non-zero eigenvalues are paired due to chiral transformations, we find that
\begin{align}
 \chi_{\bar{q}q}^d & = \langle \left( \bar{q}q - \langle \bar{q}{q} \rangle \right)^2 \rangle,\; \bar{q}q = \frac{|\nu|}{m} + \sum_n \frac{2m}{m^2+\lambda_n^2}\,,\\
 \chi_{\bar{q}q}^c & = - \langle \frac{|\nu|}{m^2} + \sum_n \frac{2(m^2-\lambda^2_n)}{(m^2+\lambda^2_n)^2} \rangle + \chi_{\bar{q}q}^d\,,
\end{align}
where $|\nu|$ is the number of zero modes and is related to the topological charge $Q$ through the index theorem, $\nu=Q$.

Now, the chiral susceptibility in \cite{aoki:fodor:katz:szabo:transition:temperature} is isospin symmetric, whereas we have broken it explicitly, i.e.\ $m_u \neq m_d$. In order to obtain effectively isospin symmetric results, we will compute
\begin{equation}
 \chi_{\bar{q}q} = \frac{1}{2} \frac{\partial^2 Z}{\partial \bar{m}^2}\,,
\end{equation}
with $\bar{m}=\frac{1}{2} (m_u + m_d)$ the mean quark mass and $m_u / m_d = \mathrm{const}$. This leads to
\begin{equation}
 \bar{m}^2 \chi_{\bar{q}q} = \frac{1}{2} m^2_u \chi_{\bar{u}u}^c + \frac{1}{2} m^2_d \chi_{\bar{d}d}^c + m_u m_d \langle ( \bar{u}u - \langle \bar{u}u \rangle ) ( \bar{d}d - \langle \bar{d}d \rangle ) \rangle \,.
\end{equation}

It turns out that the chiral susceptibility, with all plasma screening effects removed, peaks around $T_c \approx 120 \units{MeV}$, and decays very slowly afterwards. Screening effects in the IILM tend to increase the dilution of the ensemble, and thus they will be responsible for the faster decay at higher temperature. Incidentally, with screening effects included down to very low temperatures, below $T_c$, the peak in the chiral susceptibility disappears completely; this gives a first constraint on $T_*$.

The lattice data shows a peak at $T_c \approx 160 \units{MeV}$\footnote{The newer analysis in \cite{aoki:et_al:transition:temperature:II} results in a shift towards lower temperatures by about $5 \units{MeV}$. This is not a significant change with regard to the IILM prediction of $T_c \approx 120 \units{MeV}$ and we will ignore it in what follows.} and no tuning of the parameters $\{T_*,\Delta T\}$ will shift the chiral phase transition in the IILM towards the lattice result.

\begin{figure}[tbp]
\begin{center}
 \includegraphics[width=\figwidth,clip=true,trim=0mm 0mm 15mm 10mm]{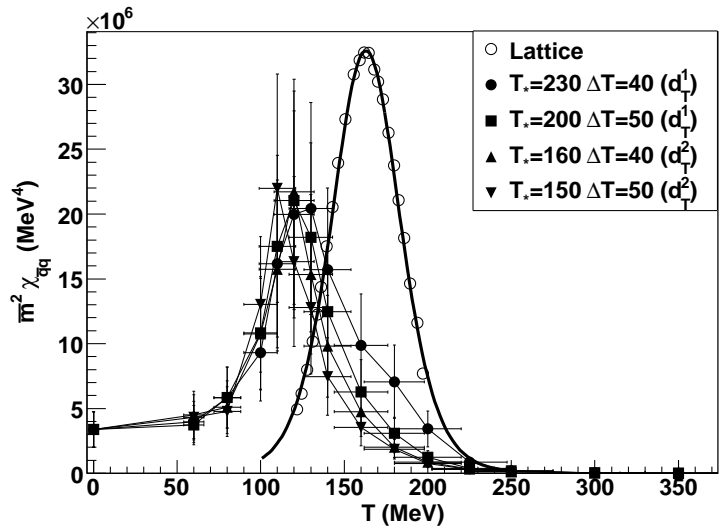}
\end{center}
\caption{The chiral susceptibility in the IILM predicts a chiral phase transition at $T_c\approx 120 \units{MeV}$. Given the constraint that $T_* > 120 \units{MeV}$ this is a robust result. The different parameter sets were chosen such that the plasma screening effects are effectively fully enhanced at about $160 \units{MeV}$, the transition predicted by the lattice result (note that we use the continuum extrapolation of \cite{aoki:fodor:katz:szabo:transition:temperature}). The first curve (solid circles) falls slightly short of this expectation. Taking the errors at face value (including those for the lattice result) the IILM is off by roughly $3\sigma$. Considering the newer analysis \cite{aoki:et_al:transition:temperature:II} that shifts the lattice result to lower temperatures, the agreement improves slightly.}\label{fig:chiralsusc}
\end{figure}

To fix the free parameters, we simply demand that the IILM result for the chiral susceptibility does not exceed the lattice result at high temperature, i.e.\ that the rapid decay due to screening effects should be around $T \approx 160 \units{MeV}$, and that its peak remains intact. We've computed the chiral susceptibility for four sets of parameters to estimate the systematics, see \reffig{fig:chiralsusc}. The large errors are due to the large systematics in $\Lambda$ that is used to set dimensions. By construction, the data below $T_c \approx 120 \units{MeV}$ is very robust as the screening effects are almost completely negligible, and the location of the peak remains fairly unaffected. Above $T_c$ the results do not depend too strongly on the parameters $\{T_*,\Delta T\}$, apart for the first curve (solid circles) which does seem to have too big a $T_*$-value. Given the phenomenological terms, see (\ref{eq:plasma:screening:transition}), and the constraint $T_*> 120 \units{MeV}$, the IILM predicts a chiral phase transition at $T_c \approx 120 \units{MeV}$.

One of the successful predictions of the IILM has been chiral symmetry restoration based on the formation of instanton--anti-instanton molecules \cite{ilgenfritz:shuryak:chiral:symmetry:restoration:iilm}, \cite{ilgenfritz:shuryak:quark:correlations:chiral:transition}, \cite{schaefer:shuryak:verbaarschot:chiral:phase:transition:molecules}: through pairing up of instantons and anti-instantons, and the strongly localised quark wavefunctions, the Dirac operator spectrum develops a gap and the Casher-Banks relation tells us that the quark condensate vanishes, i.e.\ chiral symmetry is restored.

\begin{figure}[tbp]
\begin{center}
 \includegraphics[width=\figwidth,clip=true,trim=0mm 0mm 15mm 10mm]{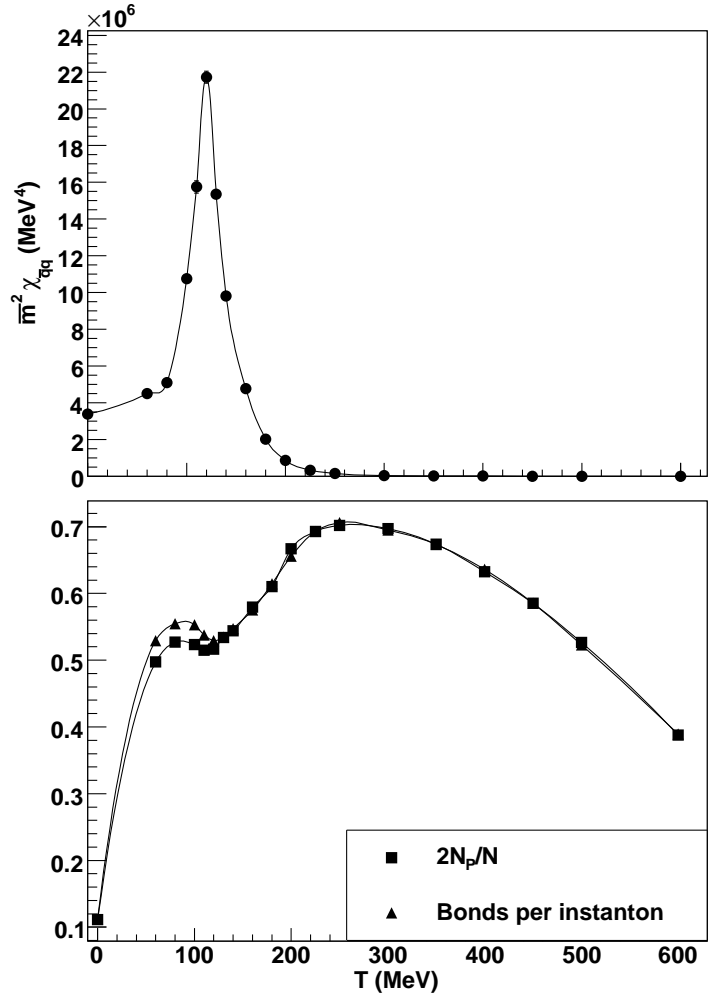}
\end{center}
\caption{Contrary to expectation, the pair concentration does not peak at the chiral phase transition, although it is large with $50\%$, but, interestingly dips at the phase transition. Also, it keeps growing to $70\%$ at roughly $T \approx 300 \units{MeV}$; beyond that point it decays. }\label{fig:chiral:molecules}
\end{figure}

\begin{figure}[tbp]
\begin{center}
 \includegraphics[width=\figwidth,clip=true,trim=0mm 0mm 15mm 10mm]{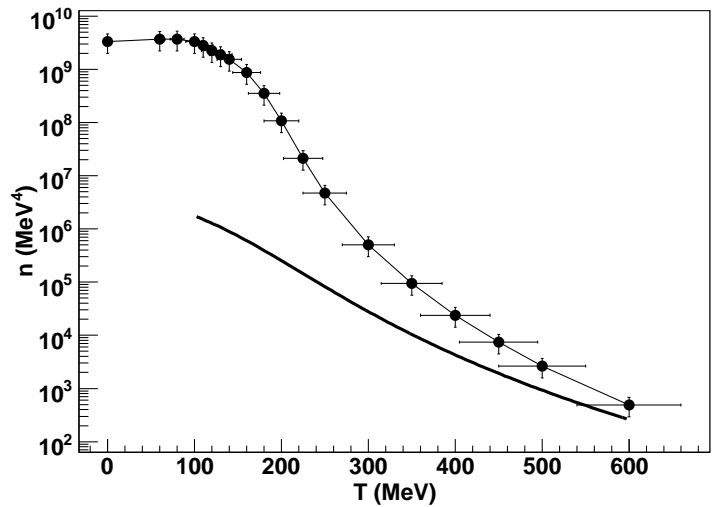}
\end{center}
\caption{The instanton density compared to the dilute gas result. The data clearly shows that even for temperatures as high as $T=400 \units{MeV}$ the ensemble is not compatible with a dilute gas. The higher density in the IILM is due to the energy stored in pairs.}\label{fig:n:unquenched:T}
\end{figure}

Naively we would expect that the pair concentration is highest around the phase transition. This is, however, not what we find and, interestingly, the number of pairs seems to drop at the phase transition. In fact, the concentration of molecules keeps growing beyond $T_c$, see \reffig{fig:chiral:molecules}, and at $T_c$ only half of the instantons are paired up into molecules. The maximum concentration of $70 \%$ is reached at $T \approx 300 \units{MeV}$, beyond which it starts to decay rather slowly. Note that even at $T = 400 \units{MeV}$ the system is still far from a dilute random gas of individual instantons, see \reffig{fig:n:unquenched:T}. It is, however, important to note that the identification of pairs is correlated to our specific definition of a bonding box, and even though the Monte Carlo results did not depend on the definition of the box, the number of instanton--anti-instantons does certainly depend on it.

\begin{figure}[tbp]
\begin{center}
 \includegraphics[width=\figwidth,clip=true,trim=0mm 0mm 15mm 10mm]{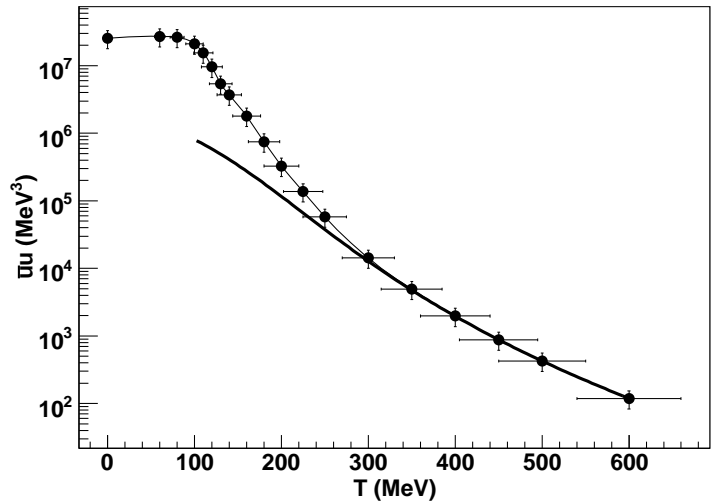}
\end{center}
\caption{Although at $T\approx 300 \units{MeV}$ the pair concentration is high, the quark condensate already tracks the dilute gas approximation (solid line).}\label{fig:qq}
\end{figure}

\begin{figure}[tbp]
\begin{center}
 \includegraphics[width=\figwidth,clip=true,trim=0mm 0mm 15mm 10mm]{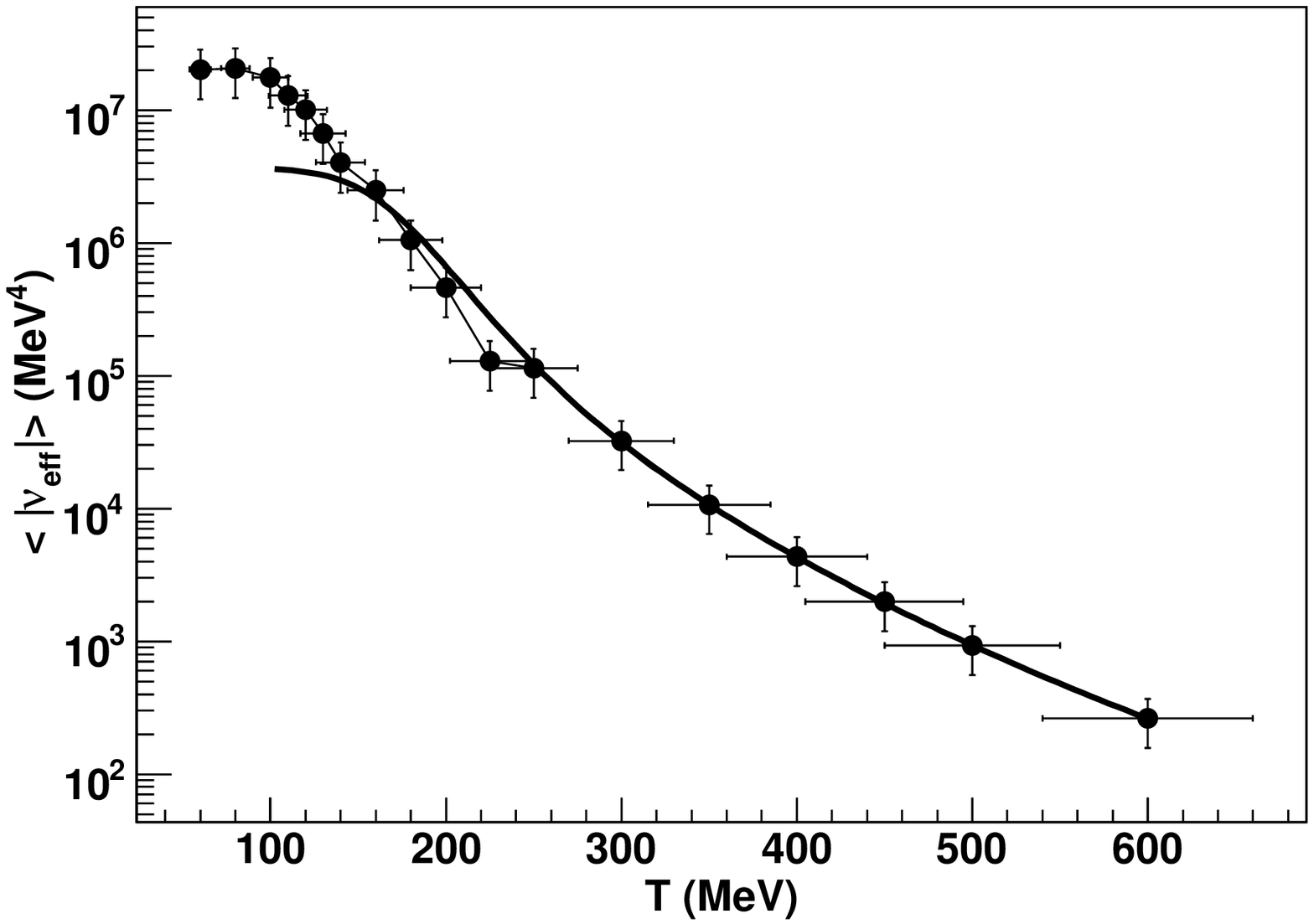}
 \includegraphics[width=\figwidth,clip=true,trim=0mm 0mm 15mm 10mm]{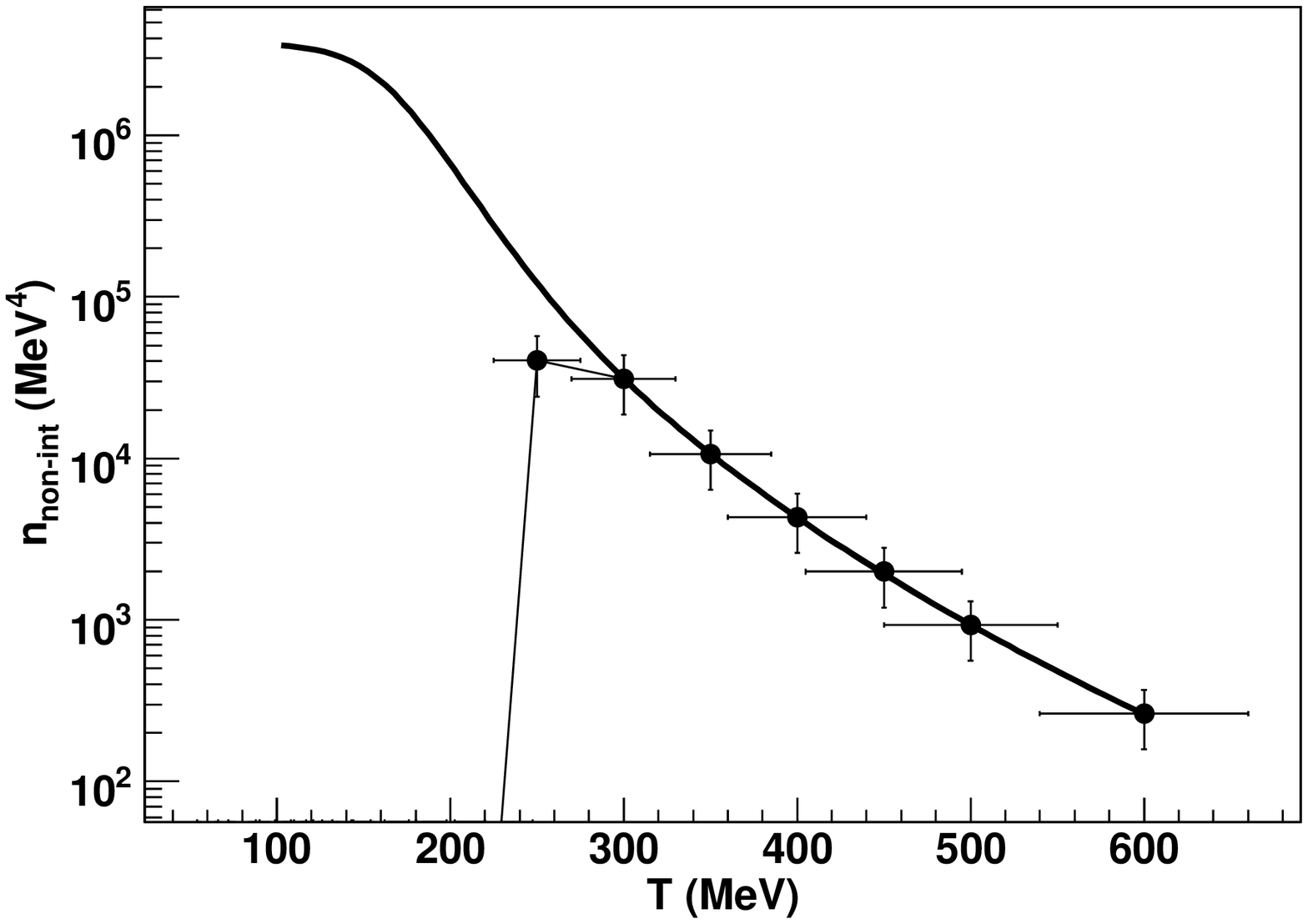}
\end{center}
\caption{It turns out that the effective number of zero modes, $|\nu_\mathrm{eff}|$, follows the temperature dependence of the dilute gas approximation for the density (solid line). Together with the observation that the condensate is dominated by $\nu_\mathrm{eff}$ at high temperatures, see text, this explains why the condensate has a temperature dependence that follows closely the dilute gas result. At temperatures around $T\approx 300\units{MeV}$ we can attribute this effect to a sub-population of non-interacting instantons.}\label{fig:nu:T}
\end{figure}

Despite the fact that the ensemble is distinct from a random gas, the quark condensate tracks the dilute gas approximation already for moderately high temperatures, see \reffig{fig:qq}. The reason is that the Dirac eigenvalues for pairs are large compared to the quark masses; this in turn follows from the fact that pairs have lined up along the time direction and thus their separation becomes smaller as the temperature increases. Therefore they are negligible compared to the zero modes. Bearing in mind the cluster decomposition of the quark overlap matrix $T_{IA}$, the quark condensate is given by
\begin{equation}
 \langle \bar{q}q \rangle \approx \frac{1}{m_q} \sum_\mathrm{clusters}\langle|\nu|\rangle \equiv \frac{\langle |\nu_\mathrm{eff}| \rangle}{m_q}\,.
\end{equation}
Note that this has almost the form of the dilute gas result, for which $T_{IA}=0$ and hence $\lambda_n=0$,
\begin{equation}
 \left. \langle \bar{q}q \rangle \right|_{\mathrm{DGA}} = \frac{\langle N \rangle/V}{m_q}\,.
\end{equation}
It turns out that $\langle |\nu_\mathrm{eff}| \rangle$ has a temperature dependence that follows the dilute gas approximation for $\langle N \rangle$, see \reffig{fig:nu:T}. The straightforward explanation that $|\nu_\mathrm{eff}|$ is approximately equal to the number of unbonded instantons did not hold up to scrutiny, especially $\langle \# \mathrm{unbonded} \rangle$ does not behave as $\left. \langle N/V \rangle \right|_{\mathrm{DGA}}$. Thus the population of unbonded instantons is still too large. But the clustering is defined through cutoffs in the interactions; these cutoffs can therefore also be used to identify a truly non-interacting sub-ensemble in the IILM. By definition, this population behaves according to the dilute gas approximation. We find that its density becomes important at $T\approx 300 \units{MeV}$ which explains the behaviour of the quark condensate.

We can conclude that, quite generally, the quark condensate is rather insensitive to the details of the molecule population. Presumably, only its concentration determines the effective number of zero modes $\nu_\mathrm{eff}$, respectively the density of the non-interacting instantons.

\section{Topological susceptibility and axion mass}
\label{sec:top:susc:axion}

\subsection{Topological susceptibility}

\begin{figure}[tbp]
\begin{center}
 \includegraphics[width=\figwidth,clip=true,trim=0mm 0mm 15mm 10mm]{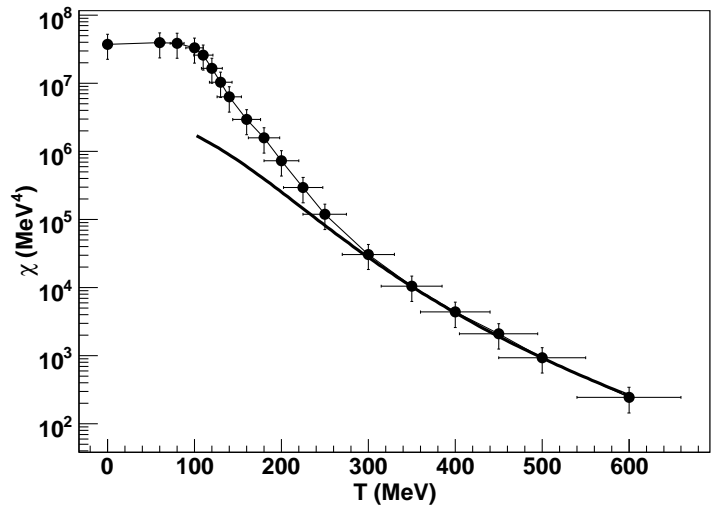}
\end{center}
\caption{The IILM reproduces the topological susceptibility in the dilute gas approximation for temperatures where the ensemble is still far from dilute. We found that this phenomenon cannot be attributed to a population of unbonded instantons. Rather, it can be explained by the population of non-interacting instantons, \reffig{fig:nu:T}, which is smaller than that of the unbonded instantons. At higher temperatures the distinction between unbonded and non-interacting instantons becomes ever more unimportant.}\label{fig:chi:unquenched:single:T}
\end{figure}

The topological susceptibility approaches the dilute gas approximation for temperatures above approximately $250 \units{MeV}$, and, like the quark condensate, does not seem to be sensitive to the high concentration of instanton--anti-instanton molecules, see \reffig{fig:chi:unquenched:single:T}.

\begin{figure}[tbp]
\begin{center}
 \includegraphics[width=\figwidth,clip=true,trim=0mm 0mm 15mm 10mm]{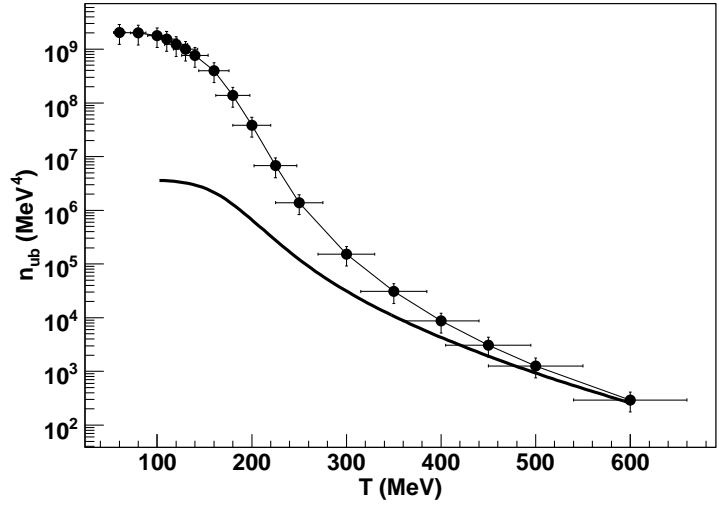}
\end{center}
\caption{For moderately high temperatures, the population of unbonded instantons cannot be responsible for the topological susceptibility (solid line); we have checked that $\langle N^2_{ub} \rangle - \langle N_{ub} \rangle^2 \approx \langle N_{ub} \rangle$. The unbonded instantons still interact too strongly with the highly correlated instanton--anti-instanton molecules. For higher temperatures, the unbonded instantons become indistinguishable from the non-interacting instantons that can account for the topological susceptibility. The latter instantons play the role of the random sub-ensemble along side the `molecular' ensemble, following the ideas of \cite{ilgenfritz:shuryak:quark:correlations:chiral:transition}.}\label{fig:unbonded}
\end{figure}

Guided by the ideas of \cite{ilgenfritz:shuryak:quark:correlations:chiral:transition} that the IILM leads to a mixture of a highly correlated and a random component, it is natural to identify the unbonded instantons as the dominant contribution to the topological susceptibility; the molecules, at least at zeroth order, do not lead to charge fluctuations and thus cannot account for the topological susceptibility. For a random ensemble $N_r$ it follows that
\begin{equation}
 \chi \propto \langle N^2_r \rangle - \langle N_r \rangle^2 \propto \langle N_r \rangle \,.
\end{equation}
We found, however, that the population of unbonded instantons is far too large to be responsible for the small topological susceptibility, under the assumption that $N_{ub}$ is a completely random ensemble, see \reffig{fig:unbonded}. Again, as in the case of the quark condensate, it is the non-interacting instantons, not the unbonded instantons, that fit the bill. Note that for higher temperatures, $T\approx 600 \units{MeV}$, where the pair concentration is still quite high, the concentration of unbonded and non-interacting instantons becomes equal and there is no more distinction between the two. For lower temperatures a coupling between unbonded and bonded instantons still exists.

Note that, in the chiral limit, the topological susceptibility is related to the quark condensate through chiral perturbation theory. It is therefore consistent that the condensate and the topological susceptibility behave rather similarly.

\begin{figure}[tbp]
\begin{center}
 \includegraphics[width=\figwidth,clip=true,trim=0mm 0mm 15mm 10mm]{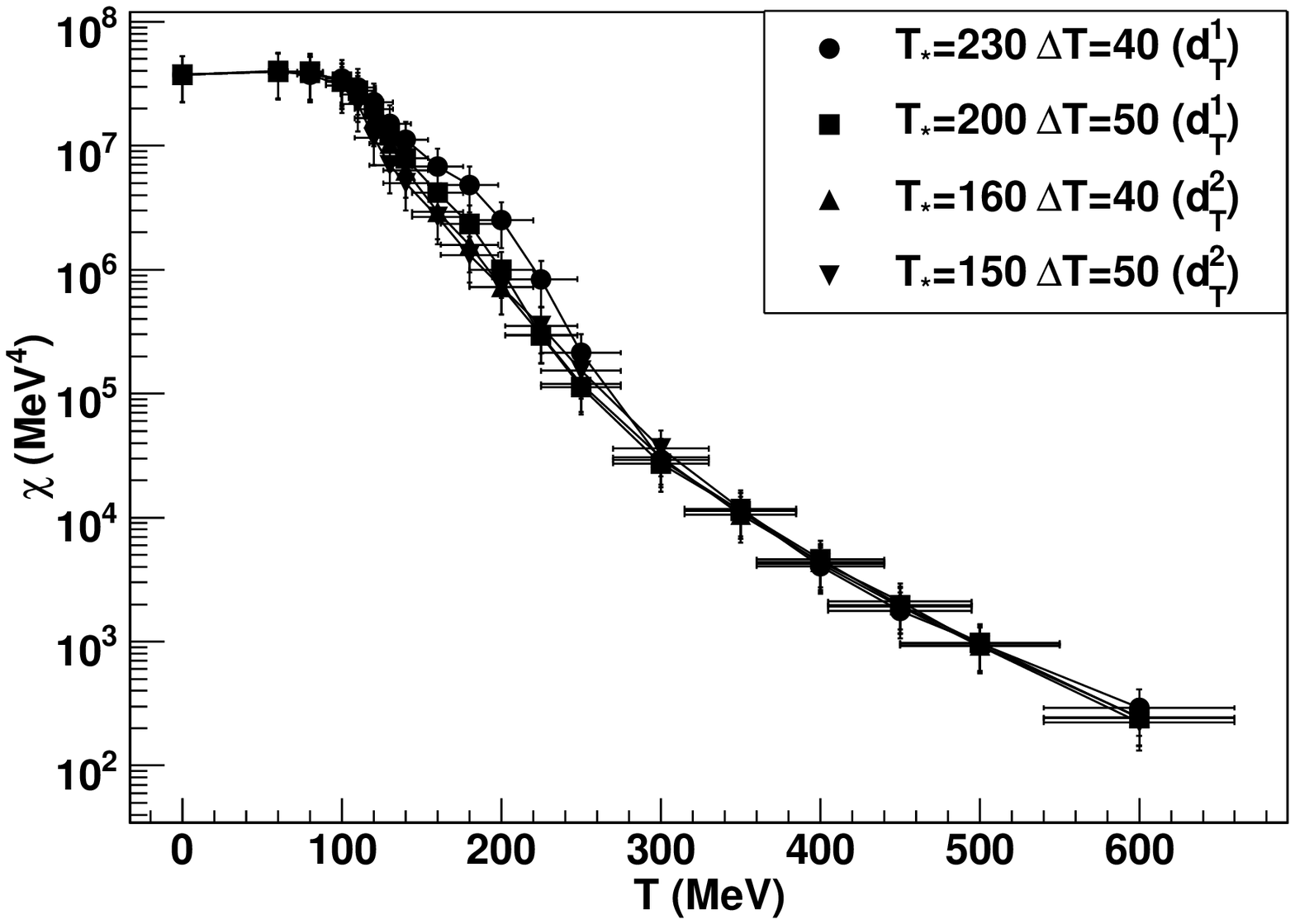}
 \includegraphics[width=\figwidth,clip=true,trim=0mm 0mm 15mm 10mm]{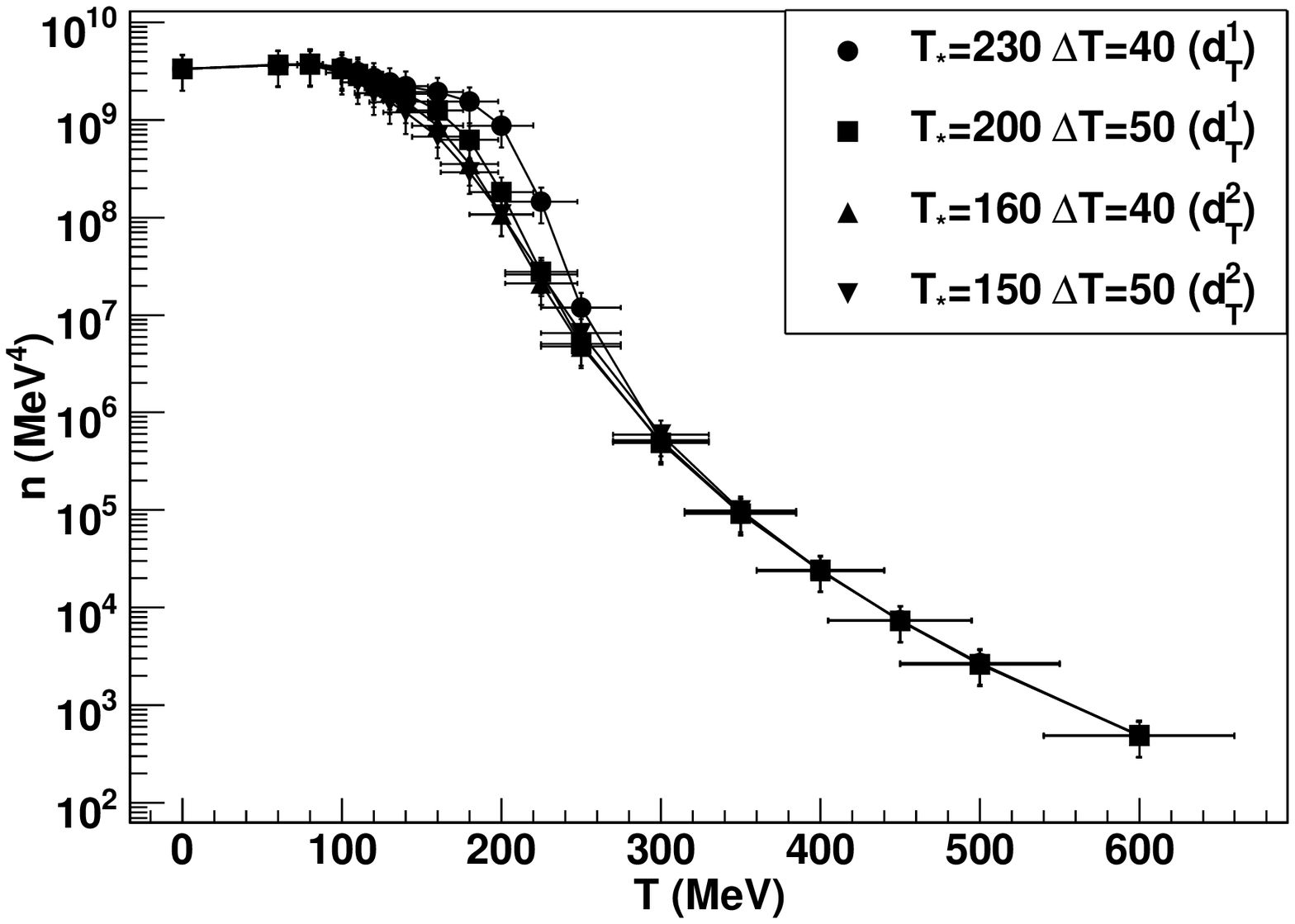}
\end{center}
\caption{Both the topological susceptibility and the density are fairly constant for low temperatures, except for a very gentle growth from $T=0$ towards a maximum at around $T=100 \units{MeV}$. In contrast to the quenched case, the IILM is not as sensitive to the free parameters, $\{T_*,\Delta T\}$, that effectively describe the phase transition. The first curve (solid circles) could be argued to be unphysical as it delays the decay in a rather unnatural fashion.}\label{fig:chi:unquenched:T}
\end{figure}

Finally, it is worth noting that the topological susceptibility, and also the quark condensate, are not overly sensitive to the free parameters $T_*$ and $\Delta T$, see \reffig{fig:chi:unquenched:T}.

\subsection{Temperature-dependent axion mass}

In order to solve the strong CP problem, Peccei and Quinn introduced a new field into the Standard Model of particle physics \cite{peccei:quinn:cp1,peccei:quinn:cp2}. It was soon realised that this field gives rise to a new light particle, a pseudo-Goldstone boson, the axion \cite{wilczek:axion,weinberg:axion}. At the time it seemed natural to tie the axion to the electro-weak scale but laboratory experiments have ruled out such an axion. Leaving the scale of the PQ field $f_a$ free, and large to evade the previously mentioned constraints, the so-called `invisible' axions were born, e.g.\ \cite{dine:fischler:srednicki:axion,zhitnitsky:axion,kim:axion,shifman:vainshtein:zakharov:cp}.

Through their weak couplings to ordinary matter, invisible axions can play the role of a dark matter candidate. They have a rich phenomenology, and can be produced through a variety of production channels: the thermal scenario (similar to WIMP production), cosmic string decay (axions are the the Goldstone boson of a spontaneously broken $U_\mathrm{PQ}(1)$ symmetry) and the so-called misalignment mechanism. The latter is essentially the PQ mechanism: due to the anomalous $U_\mathrm{PQ}(1)$ symmetry an axion mass term is generated through the coupling of the axion to the topological charge
\begin{equation}
 \mathcal{L}_{a-g} \propto \frac{\phi}{f_a} \frac{g^2}{32 \pi^2} F^a_{\mu\nu}\tilde{F}^a_{\mu\nu}\,.
\end{equation}
This term combines with the vacuum angle $\theta$ and the axion field can be shifted to get rid of $\theta$, i.e.\ the vacuum angle becomes a dynamical field, the axion angle $\theta_a$. This is the key insight because now we can evoke the principle of least action to argue that $\theta_a \to 0$ dynamically to solve the strong CP problem.

Indeed, integrating out the gluons, we can determine the axion mass from the effective action,
\begin{equation}
 \exp(-V V_\mathrm{eff}(\phi)) = \int [dA] \det(\slashed{D}+M) \exp(-S_g - S_{a-g})\,,
\end{equation}
according to
\begin{equation}
 m^2_\phi = \left.\frac{\partial^2 V_\mathrm{eff}}{\partial \phi^2}\right|_{\phi=0} = \frac{\left.\chi\right|_{\theta=0}}{f_a^2}\,,
\end{equation}
and the axion eventually evolves to its minimum at $\theta_a=0$ \cite{vafa:witten:parity}. It also demonstrates that the mass for the QCD axion is set by the topological susceptibility.

Of all the production channels, the misalignment mechanism is most sensitive to the axion mass. From above it is clear that the axion mass is inherently a non-perturbative problem. At high temperatures, the dilute gas approximation to the instanton ensemble can be used. However, at lower temperatures it breaks down. From our determination of the topological susceptibility we can for the first time give a well-motivated axion mass that covers all temperatures down to $T=0$.

\begin{figure}[tbp]
\begin{center}
 \includegraphics[width=\figwidth,clip=true,trim=0mm 0mm 15mm 10mm]{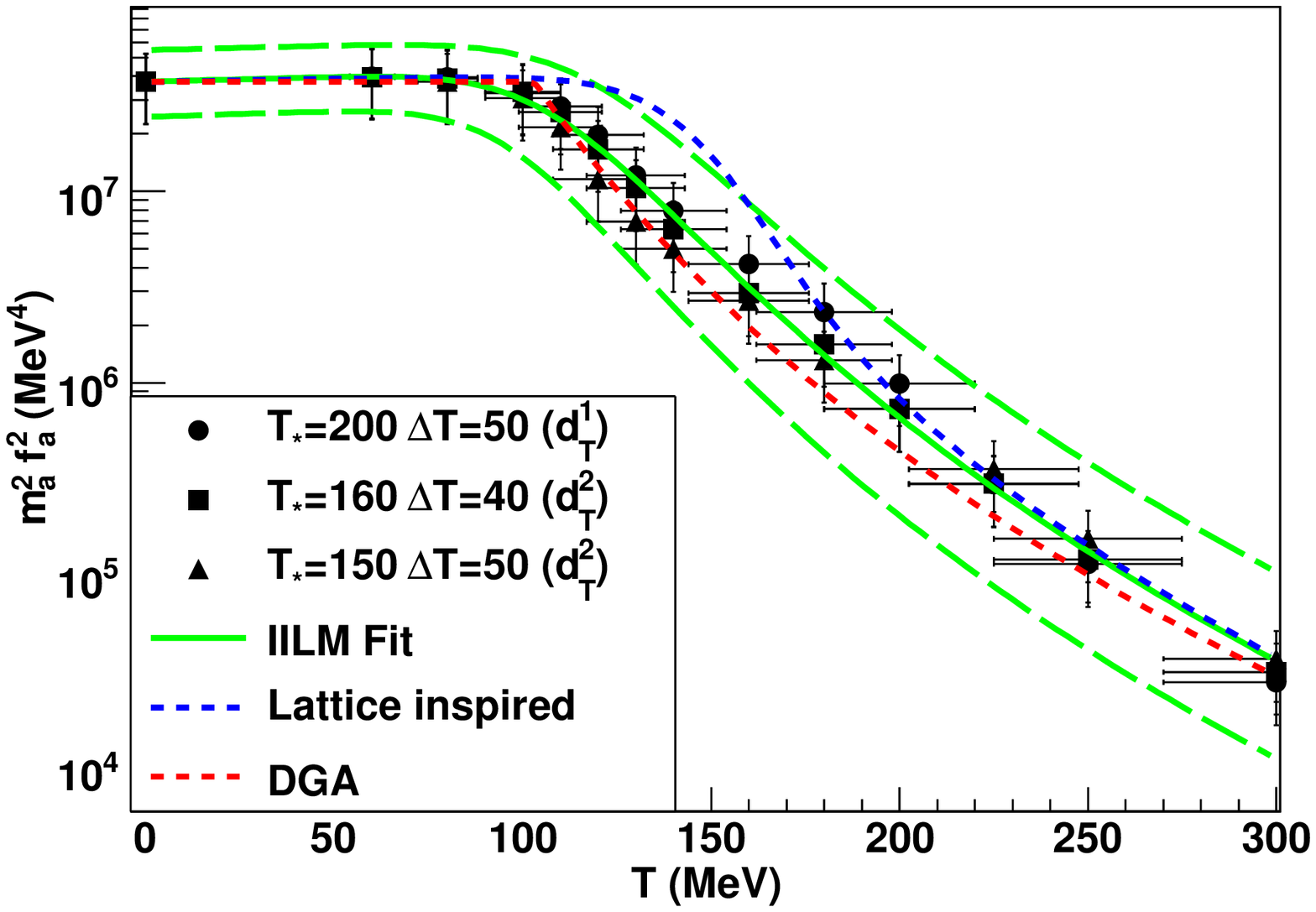}
 \includegraphics[width=\figwidth,clip=true,trim=0mm 0mm 15mm 10mm]{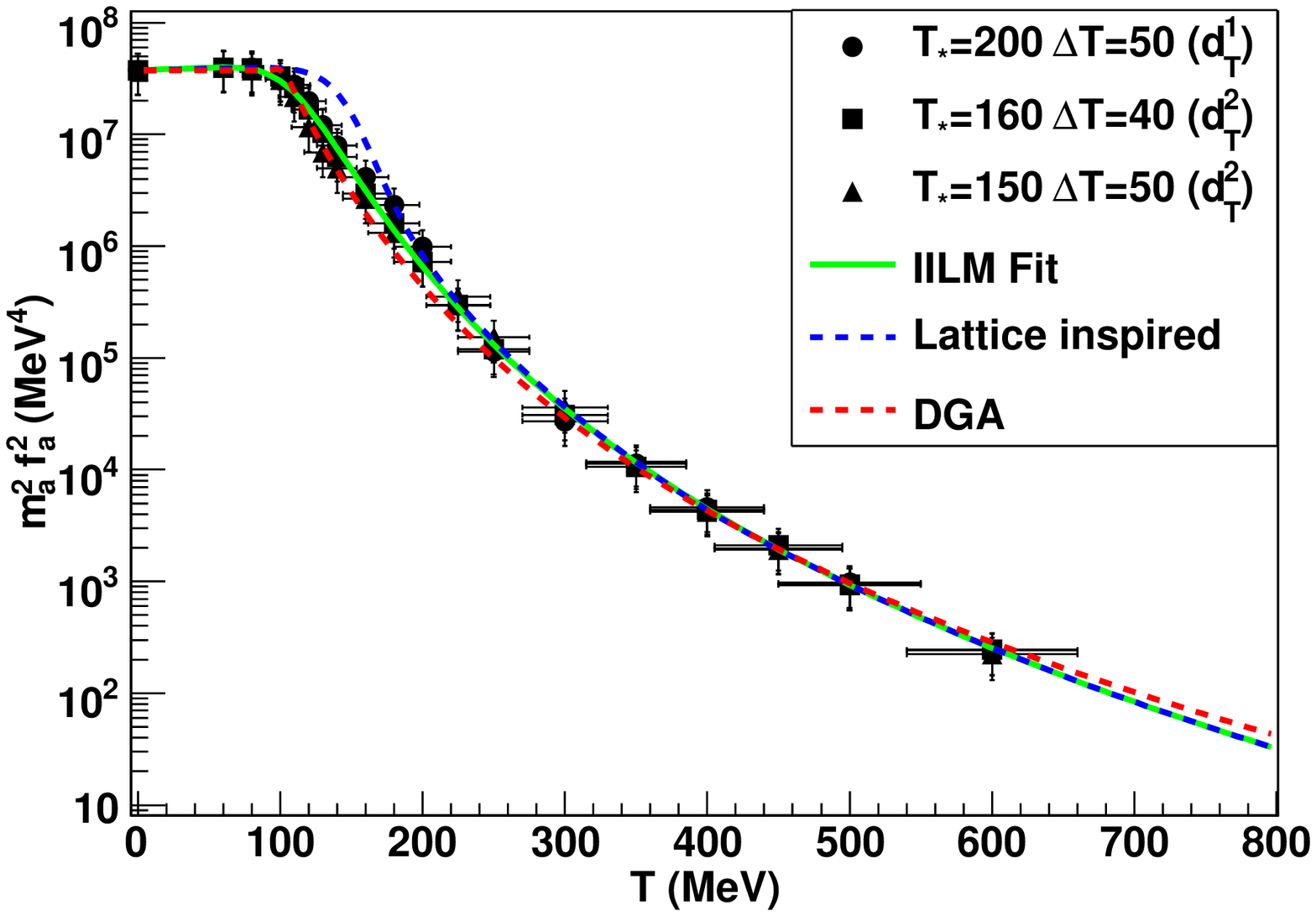}
\end{center}
\caption{The mass for the QCD axion follows from the topological susceptibility, $m^2_a f^2_a = \chi$. The fit goes over to the dilute gas approximation for moderately high temperatures, in accordance to the IILM data. Note that the large errors are mostly due to the large uncertainties in the determination of $\Lambda$, used to set dimensions. We also include a similar fit that is slightly shifted towards higher temperatures to mimic the phase transition as seen on the lattice, and a simple power-law approximation to the dilute gas limit (DGA), see (\ref{eq:dga}). We see that such a simple power-law approximation to the full result certainly has its merit as it is fairly accurate given the analytic simplicity.}\label{fig:axion:mass}
\end{figure}

In \reffig{fig:axion:mass} we display the axion mass together with a fit and its error range mainly due to the error in $\Lambda$. The data suggests that the axion mass turns into the dilute gas approximation rather quickly; around $T\approx 300 \units{MeV}$ the differences are negligible, and the fit takes this into account. We have seen in the previous sections that the IILM predicts a phase transition that is slightly too low. Using the lattice data for the phase transition temperature $T_c$, we also include a tentative fit to what the ultimate lattice data might look like, and again we impose the dilute gas limit at moderately high temperatures. It is worth noting that we currently do not have lattice data available that could corroborate such a result; remember that in the quenched case the lattice data did not behave according to the dilute gas result at high temperatures! Given the state of lattice calculations it should not take too much longer before a comprehensive lattice study with physical quark masses and across a wide range of temperatures will be available to give the exact axion mass.

Within the IILM some progress can be made by including the non-trivial holonomy calorons; a dilute gas study in the quenched sector has given encouraging results \cite{gerhold:ilgenfritz:mueller_preussker:kvbll:gas:confinement} that these degrees of freedom might play a role in the confinement/deconfinement transition. It would be very interesting to investigate their role in the unquenched sector, where we might expect less dramatic qualitative changes for chiral quantities, such as the topological susceptibility, because chiral properties are reasonably well modeled by the IILM. However, we certainly expect a closer agreement with the lattice results if the QCD vacuum is indeed dominated by non-perturbative, topological fluctuations.

A second improvement with regard to finite temperature effects is the implication of quantum interactions on the plasma screening effects. In particular, we have found within a toy-model that the topological susceptibility can change qualitatively if the screening effects become subdominant to the quark zero mode interactions. This would favour a higher concentration of molecules and therefore reduce the non-interacting instanton density that sets the axion mass. Specifically, it would lead to a faster decay of the axion mass.

The IILM result has a slight rise at low temperatures. We do not have enough low-$T$ data to determine the shape of this rise; however, in the gauge sector we found that the topological susceptibility had a roughly linear dependence on $T$. We will therefore constrain the fit at low temperatures to be a first order polynomial in temperature. It is given by
\begin{equation}
m^2_a f^2_a = 1.46\;10^{-3}\Lambda^4 \frac{1+0.50\,T/\Lambda}{\displaystyle 1+\left(3.53\, T/\Lambda\right)^{7.48}},\,  0 <T< 0.45 \,,
\end{equation}
where $\Lambda=400\units{MeV}$, and the errors will be mainly due to the uncertainties in $\Lambda$.

Inspired by the lattice result for the chiral phase transition \cite{aoki:fodor:katz:szabo:transition:temperature}, we include a fit that delays the decay of the topological susceptibility until $T_c \approx 160\units{MeV}$. As mentioned above, there is a rather large disagreement between different lattice collaborations, and the chiral phase transition could occur at higher temperatures still. We assume again that the dilute gas limit will be recovered rather quickly above the phase transition. This assumption is hard to justify quantitatively given the lack of lattice data in that regime; the dilute gas result is really the best estimate we have at the moment. For the lattice inspired fit to smoothly connect with the dilute gas limit at around $T\approx 200\units{MeV}$, we patch together two different rational functions with the help of a washed-out step function, $\alpha(T)=\frac{1}{2}\left(1-\tan\frac{T/\Lambda-0.40}{0.075}\right)$. The result is 
\begin{equation}
m^2_a f^2_a = 1.46\;10^{-3}\Lambda^4 \left( \frac{\displaystyle 1+0.3\, T/\Lambda}{\displaystyle 1+\left(2.5\,T/\Lambda\right)^{8.3}}\right)^{\alpha(T)} \left(1+\left(3.4\,T/\Lambda\right)^{7.4}\right)^{\alpha(T)-1}\,,
\end{equation}
for the temperature range $0 <T< 0.41 \units{GeV}$.

These fits cover the low temperature regime, bounded by what we call $T_\mathrm{DGA}$. Given our aim to improve on the current axion mass computations, we will also be more systematic for high temperatures and take the effects of quark thresholds into account. To that end, remember that we really are using the language of effective field theory: the results are given as a function of $g^{(3)}$, the strong coupling for $N_f=3$ active flavours. It is well known that in order for S-matrix elements to be smooth, the `free' parameters, such as masses and coupling constants, become `discontinuous' \cite{weinberg:effective:gauge:theories,ovrut:schnitzer:decoupling:ms,ovrut:schnitzer:effective:field:theory:background,rodrigo:santamaria:matching:threshold}; actually, the parameters are not discontinuous but belong to different theories that are matched at quark mass thresholds, i.e.\ $g^{(3)}=g^{(4)}+O((g^{(4)})^2)$. Since part of the quantum effects in the instanton density are two-loop improved, we expect that such `discontinuities' do arise. In the present case the coupling in the exponential is the only place that displays two-loop improvements, and, at the threshold, the dilute gas approximations are related to each other by
\begin{equation}
 \chi^{(4)}=\chi^{(3)} \exp(\gamma_{3,4})\,.
\end{equation}
The factor $\gamma_{3,4}$ is essentially given by the Dirac determinant of the newly active quark flavour. In practice we just determine $\gamma_{3,4}$ such that the dilute gas approximations are smooth across the quark threshold.

At high temperatures the dilute gas result is almost given by a power-law. We found that the corrections could be straightforwardly taken into account by a higher order polynomial in log-log space; therefore the fit takes the form of an exponential of a polynomial in logarithms. We find
\begin{equation}
m^2_a f^2_a = \Lambda^4
\left\{
 \begin{array}{l@{,\;}l}
 \exp\left[ d^{(3)}_0 + d^{(3)}_1 l + d^{(3)}_2 l^2 + d^{(3)}_3 l^3 \right] & T_\mathrm{DGA} < T < 1.200\\
 \exp\left[ d^{(4)}_0 + d^{(4)}_1 l + d^{(4)}_2 l^2 \right] & 1.200 < T < 4.200\\
\exp\left[ d^{(5)}_0 + d^{(5)}_1 l + d^{(5)}_2 l^2 \right] & 4.200 < T < 100\\
\end{array}
\right. \,,
\end{equation}
with $l=\ln\frac{T}{\Lambda}$, and the different parameters are given by
\begin{equation}
\begin{array}{c|c|c|c|c}
N_f & d^{(N_f)}_0 & d^{(N_f)}_1 & d^{(N_f)}_2 & d^{(N_f)}_3 \\\hline\hline
3 & -15.6 & -6.68 & -0.947 & 0.555 \\\hline
4 & 15.4 & -7.04 & -0.139 & \\\hline
5 & -14.8 & -7.47 & -0.0757 &\\
\end{array}\,. \label{eq:dga:full}
\end{equation}
The $\gamma_{N_f-1,N_f}$-factors have already been absorbed into $d^{(N_f)}_0$; they are $\gamma_{3,4}=0.444$ and $ \gamma_{4,5}=1.54$.

We conclude by giving a very simple approximation to the dilute gas result in the form of a power-law, as in earlier work \cite{turner:axion:cosmology,bae:huh:kim:axion},
\begin{equation}
 m^2_a = \frac{\alpha_a \Lambda^4}{f_a^2 (T/\Lambda)^n}\,,\label{eq:dga}
\end{equation}
where $n=6.68$ and $\alpha=1.68 \,10^{-7}$, from (\ref{eq:dga:full}), and compare well with the more recent study \cite{bae:huh:kim:axion}; it compares well with \cite{bae:huh:kim:axion}. We believe it is a rather lucky coincidence that such a simple fit, based solely on the high temperature regime, still gives such a good overall approximation to the much more elaborate result of the IILM simulations, see \reffig{fig:axion:mass}. The small qualitative differences are that the power-law approximation: overshoots the IILM result at high temperatures, due to the wrong running of the QCD $\beta$-function, and underestimates the axion mass at low temperature where it is by construction constant whereas it reaches the $T=0$ limit from above in the IILM. Given its analytic simplicity and its unexpectedly good agreement, such a simple power-law has certainly its merit. Whether such a conclusion pertains to an improved IILM including the more general calorons, and ultimately to the lattice, remains an open question.

\section{Conclusion}

We have been able to improve on the finite temperature interactions in the IILM. The numerical framework we set up in the first paper \cite{wantz:iilm:1} could successfully be implemented at finite temperature as well, and well-defined interactions that lead to a consistent thermodynamic limit have been derived.

Using these improved interactions, we investigated the IILM at finite temperature with light, physical quark masses. For these small quark masses we have found that the usual `random' Monte Carlo sampling is very inefficient and can even break down. Using the results form \cite{wantz:iilm:2}, where we introduced biased Monte Carlo techniques and, in particular, adapted the Unbonding--Bonding algorithm to the grand canonical ensemble, including a biasing scheme to deal with the orientation-dependent interaction of the IILM, we could run efficient simulations at finite temperature. We have found that the screening factors, from single instanton quantum fluctuations, will lead to a dilute, random ensemble at high enough temperatures. There is, however, a possibility that this trend could be reversed when more quarks become active; this does seem rather unnatural though. We want to point out that the high-frequency quantum interactions from overlapping instantons might be important to settle this question.

In the Yang-Mills sector we found rather poor overlap of the IILM and the lattice data. Most strikingly, we found that the IILM failed to reproduce the lattice data, even in a qualitative manner, in the regions where the phenomenological factors that mimic the phase transition in the IILM are unimportant, namely at low and high temperatures. To model the pure gauge sector, the IILM might have to be generalised to include the non-trivial holonomy calorons. Ultimately this is needed in any case because, as for the zero temperature IILM, confinement is still lacking and these more general degrees of freedom, the KvBLL calorons \cite{kraan:baal:caloron:I,kraan:baal:caloron:II,kraan:baal:caloron:monopole,lee:lu:caloron}, might play a crucial\footnote{Recent lattice studies see evidence of the lumpy structure characteristic for an ensemble of these new caloron solutions \cite{ilgenfritz:mueller_preussker:peschka:caloron:su_3:lattice:evidence}.}.

In the unquenched sector, we investigated the chiral susceptibility and the quark condensate to gauge the free parameters introduced by the phenomenological screening factors; we found that the IILM is not overly sensitive to these, given some mild restrictions. The IILM unambiguously leads to chiral symmetry restoration at $T_c=120 \units{MeV}$, slightly too low as compared to lattice data.

We investigated the population of instanton--anti-instanton molecules and found that, rather surprisingly, the maximum concentration does not occur at $T_c$ but rather at higher temperatures. A large population of instanton pairs prevailed to fairly high temperatures, indicating that the dilute gas limit of the instanton ensemble is only reached far beyond the critical temperature. However, the quark condensate and the topological susceptibility behave according to the dilute gas result much earlier. We could attribute this behaviour to a sub-ensemble of non-interacting instantons, distinct from the unbonded instantons that still interact considerably with the instanton--anti-instanton molecule population. At higher temperature, the distinction between unbonded and non-interacting instantons becomes irrelevant.

Given the topological susceptibility, we have presented a fit to the axion mass. We paid due attention to extrapolate the axion mass to higher temperature by including threshold effects due to heavier quarks that become active as the temperature rises. The main improvement, however, is a real computation of the low temperature axion mass that matches smoothly to high temperatures. Considerations of the anthropic axion for which $\theta$ and $f_a$ are considered free parameters, one `environmental' and the other fundamental, need knowledge of the axion mass for all temperatures, which this work provides. Comparison with lattice data leaves open the intriguing possibility that the high temperature axion mass does not behave according to the dilute gas result based on Harrington--Shepard calorons. A considerably different fall-off behaviour of the axion mass will change the cosmological bounds decidedly: the classic axion misalignment scenario, where $\theta_a$ is set by its rms fluctuations at the time of symmetry breaking, would get a weaker upper bound. This conclusion relies on our findings from the pure gauge simulations, where the lattice topological susceptibility did not fall off as quickly as predicted by the IILM. The unquenched case is, however, significantly different and chiral symmetry, which has proved powerful at zero temperature, might constrain the discrepancies between the lattice and the IILM. Especially, a higher instanton--anti-instanton molecule density, possibly due to the non-trivial holonomy calorons or weaker screening effects for the strongly overlapping pairs (the effective size of such a pair will be smaller), would lead to stronger upper bounds and could possibly rule out the classic axion window. This was one motivation for the present work, but within the present IILM this expectation could not be corroborated. Further investigations are necessary to settle this issue. It might turn out, however, that the non-interacting sub-population that sets the axion mass in the IILM will not change considerably; such robustness within the IILM could be interpreted as evidence that the axion mass is rather insensitive to the details of the QCD phase transition. Given the advance of lattice QCD simulations, the ultimate axion mass determination might be available in the near future.

\section*{Acknowledgements}

We are very grateful for many informative discussions with P. Faccioli, R. Horgan and M. Wingate. Simulations were performed on the COSMOS supercomputer (an Altix 4700) which is funded by STFC, HEFCE and SGI. OW was supported by STFC grant PPA/S/S2004/03793 and an Isaac Newton Trust European Research Studentship. E.P.S.\ S. was supported by STFC grant ST/F002998/1 and the Centre for Theoretical Cosmology.

\appendix
\section{Gluonic Interactions}
\label{app:interaction:T:gluonic}

The ratio ansatz for an instanton--anti-instanton pair is defined by 
\begin{eqnarray}
 A^a_\mu &=& - \frac{\bar{\eta}^a_{\mu\nu}\partial_\nu \Pi_1(x,\{x_1,\rho_1\}) + O^{ab} \eta^b_{\mu\nu} \partial_\nu \Pi_2(x,\{x_2,\rho_2\})}{1+\Pi_1(x,\{x_1,\rho_1\})+\Pi_2(x,\{x_2,\rho_2\})}\,,\\
\Pi(x,\{y,\rho\}) &=& \frac{\pi \rho^2}{\beta r} \frac{\sinh\frac{2\pi r}{\beta}}{\cosh\frac{2\pi r}{\beta}-\cos\frac{2\pi t}{\beta}}\,,
\end{eqnarray}
where $O=O_1^t O_2$, with $O_i$ the respective colour embeddings. A global colour rotation has been performed to bring the gauge potential into this form. Since we are ultimately interested in the action, we do not need to bother about it since the action is gauge invariant. Instanton--instanton and anti-instanton--anti-instanton pairs differ by having either only $\bar{\eta}$ or $\eta$ in the above formula. A brute force computation then gives 

\begin{multline}
 F^a_{\mu\nu} F^a_{\mu\nu} = I + (\Tr O^tO + (\bar{\eta} O \eta)_{\mu\nu\mu\nu}) J + (\bar{\eta} O \eta)_{\rho\mu\rho\nu} I_{\mu\nu} \\
+ (\bar{\eta} O \eta)_{\mu\rho\nu\sigma} I_{\mu\rho\nu\sigma} + (\eta O^tO \eta)_{\mu\rho\nu\sigma} J_{\mu\rho\nu\sigma} + (\bar{\eta} O \eta)_{\alpha\mu\alpha\rho} (\bar{\eta} O \eta)_{\beta\nu\beta\sigma} K_{\mu\rho\nu\sigma}\,.
\end{multline}

The different terms have the following form
\begin{multline}
I = \frac{4}{(1+\Pi_1+\Pi_2)^2} \left[ (\partial_\mu \partial_\nu \Pi_1) (\partial_\mu \partial_\nu \Pi_1) + (\partial_\mu \partial_\nu \Pi_2) (\partial_\mu \partial_\nu \Pi_2) \right]\\
\shoveleft{- \frac{8}{(1+\Pi_1+\Pi_2)^3} \left[ (\partial_\mu \partial_\nu \Pi_1) (\partial_\mu \Pi_1)(\partial_\nu \Pi_2) + (\partial_\mu \partial_\nu \Pi_2) (\partial_\mu \Pi_2)(\partial_\nu \Pi_1) \right.}\\
\shoveright{\left. + 2(\partial_\mu \partial_\nu \Pi_1) (\partial_\mu \Pi_1)(\partial_\nu \Pi_1) + 2(\partial_\mu \partial_\nu \Pi_2) (\partial_\mu \Pi_2)(\partial_\nu \Pi_2) \right]}\\
\shoveleft{+ \frac{4}{(1+\Pi_1+\Pi_2)^4} \left[3(\partial_\mu \Pi_1 \partial_\mu \Pi_1)(\partial_\mu \Pi_1 \partial_\mu \Pi_2) +  3(\partial_\mu \Pi_2 \partial_\mu \Pi_2)(\partial_\mu \Pi_2 \partial_\mu \Pi_1)\right.}\\
\shoveright{\left .+ 3(\partial_\mu \Pi_1 \partial_\mu \Pi_1)^2 + 3(\partial_\mu \Pi_2 \partial_\mu \Pi_2)^2 + 2(\partial_\mu \Pi_1 \partial_\mu \Pi_1)(\partial_\mu \Pi_2 \partial_\mu \Pi_2) \right.}\\
\left. + (\partial_\mu \Pi_1 \partial_\mu \Pi_2)^2  \right]\,.
\end{multline}

\begin{multline}
\shoveright{J = \frac{2}{(1+\Pi_1+\Pi_2)^4} (\partial_\mu \Pi_1 \partial_\mu \Pi_1)(\partial_\mu \Pi_2 \partial_\mu \Pi_2)\,.}
\end{multline}

\begin{multline}
I_{\mu\nu} = \frac{4}{(1+\Pi_1+\Pi_2)^2} (\partial_\mu \partial_\sigma \Pi_1) (\partial_\mu \partial_\sigma \Pi_2)\\
\shoveleft{+ \frac{4}{(1+\Pi_1+\Pi_2)^3} \left[ (\partial_\mu \partial_\nu \Pi_1)(\partial_\sigma \Pi_2 \partial_\sigma \Pi_2) + (\partial\mu \partial_\nu \Pi_2)(\partial_\sigma \Pi_1 \partial_\sigma \Pi_1) \right.}\\
\shoveright{\left. - 2(\partial_\mu \partial_\sigma \Pi_1)(\partial_\nu \Pi_2)(\partial_\sigma \Pi_2) - 2(\partial_\mu \Pi_1)(\partial_\sigma \Pi_1)(\partial_\nu \partial_\sigma \Pi_2)\right.}\\
\shoveright{\left. - 2(\partial_\mu \partial_\sigma \Pi_1)(\partial_\sigma \Pi_1)(\partial_\nu \Pi_2) - 2(\partial_\mu \Pi_1)(\partial_\nu \partial_\sigma \Pi_2)(\partial_\sigma \Pi_2) \right]}\\
\shoveleft{+ \frac{4}{(1+\Pi_1+\Pi_2)^4} \left[ - (\partial_\mu \Pi_1)(\partial_\nu \Pi_1)(\partial_\sigma \Pi_2 \partial_\sigma \Pi_2) \right.}\\
\shoveright{\left. - (\partial_\mu \Pi_2)(\partial_\nu \Pi_2)(\partial_\sigma \Pi_1 \partial_\sigma \Pi_1) + 3(\partial_\mu \Pi_1)(\partial_\nu \Pi_2)(\partial_\sigma \Pi_1 \partial_\sigma \Pi_1)\right.}\\
\left. + 3(\partial_\mu \Pi_1)(\partial_\nu \Pi_2)(\partial_\sigma \Pi_2 \partial_\sigma \Pi_2) + 3(\partial_\mu \Pi_1)(\partial_\nu \Pi_2)(\partial_\sigma \Pi_1 \partial_\sigma \Pi_2) \right]\,.
\end{multline}

\begin{multline}
I_{\mu\rho\nu\sigma} = \frac{4}{(1+\Pi_1+\Pi_2)^2} (\partial_\mu \partial_\nu \Pi_1)(\partial_\rho \partial_\sigma \Pi_2)\\
\shoveleft{+ \frac{8}{(1+\Pi_1+\Pi_2)^3} \left[ (\partial_\mu \Pi_2)(\partial_\rho \partial_\nu \Pi_1)(\partial_\sigma \Pi_2) + (\partial_\mu \Pi_1)(\partial_\rho \partial_\nu \Pi_2)(\partial_\sigma \Pi_1) \right]}\\
\shoveleft{+ \frac{8}{(1+\Pi_1+\Pi_2)^4} (\partial_\mu \Pi_1)(\partial_\rho \Pi_2)(\partial_\nu \Pi_1)(\partial_\sigma \Pi_2)\,.}\hfill
\end{multline}

\begin{multline}
\shoveright{J_{\mu\rho\nu\sigma} = \frac{2}{(1+\Pi_1+\Pi_2)^4}(\partial_\mu \Pi_1)(\partial_\rho \Pi_2)(\partial_\nu \Pi_1)(\partial_\sigma \Pi_2)\,.}
\end{multline}

\begin{multline}
\shoveright{K_{\mu\rho\nu\sigma} = \frac{2}{(1+\Pi_1+\Pi_2)^4}(\partial_\mu \Pi_1)(\partial_\rho \Pi_2)(\partial_\nu \Pi_1)(\partial_\sigma \Pi_2)\,.}
\end{multline}

\subsection{Exact Interactions}
\label{app:interaction:T:gluonic:exact}

When computing the look-up tables we use global translations and rotations in $\mathbb{R}^3$, and periodicity in $S^1$, to place one instanton at the origin and the partner at $y'_3=R_s=\sqrt{R_i R_i}=|y^{I_1}_s-y^{I_2}_s|$ and $y'_4=R_t=|y^{I_1}_t-y^{I_2}_t|$; $y^i$ are the instanton centres. The rotation will reemerge in contractions of $R_i$ with the colour structure as we will now see. The relation between the spatial position vector $R_i$ and $R_i' \equiv (0,0,R_s)$ is given by the following $O(3)$ rotation matrix
\begin{eqnarray}
 R_{i}' &=&  \mathcal{O}^t_{ij} R_j\,,\\
 \mathcal{O}_{i 3} &=& \frac{R_i}{R_s}\,,
\end{eqnarray}
and the other components of the rotation matrix are irrelevant.

Note that with the choice of $R_i'$ the integrands are $O(2)$ symmetric in the subspace orthogonal to the $3$-direction. Denoting the spatial arguments of the 't Hooft potentials, $\Pi(x,\{y,\rho\})$, by $x_i$ and $\tilde{x}_i \equiv x_i-R_i$, we can extract the $R_i$ from the integrands with help of the following formulas, which we order according to the tensor structure in the $x_i$-dependence of the integrand

\begin{equation}
 \int x_i = \mathcal{O}_{i3} \int x_3'\,.
\end{equation}

\begin{equation}
 \int x_i x_j = \delta_{i j} \int x'^2_1+ \mathcal{O}_{i 3} \mathcal{O}_{j 3} \int (x'^2_3-x'^2_1)\,.
\end{equation}

\begin{eqnarray}
 \int x_i x_j x_k &=&  (\delta_{ij} \mathcal{O}_{k 3} + \delta_{ki} \mathcal{O}_{j 3} + \delta_{jk} \mathcal{O}_{i 3})\int x'^2_1 x'_3\\
 &+& \mathcal{O}_{i 3} \mathcal{O}_{j 3} \mathcal{O}_{k 3} \int (x'^3_3-3x'^2_1 x'_3)\,.
\end{eqnarray}

\begin{eqnarray}
 \int x_i x_j x_k x_h &=&  (\delta_{ij}\delta_{kh} + \delta_{ik}\delta_{jh} + \delta_{ih}\delta_{kj})\int x'^2_1 x'^2_2\\
 &+& (\delta_{ij}\mathcal{O}_{k 3} \mathcal{O}_{h 3} + \mathrm{perm.} ) \int (x'^2_3 x'^2_1-x'^2_1 x'^2_2)\\
&+& \mathcal{O}_{i 3} \mathcal{O}_{j 3} \mathcal{O}_{k 3} \mathcal{O}_{h 3} \int (x'^4_3-6x'^2_1 x'^2_3 + 3 x'^2_1 x'^2_2)\,.
\end{eqnarray}

Insertion of $\tilde{x}$ can be constructed from these. Incidentally splitting the different integrands according to the above formulas is the most stable procedure numerically. Taking into account the antisymmetry of the 't Hooft symbols, we end up with the following integrands

\begin{multline}
I = \frac{4}{(1+\Pi_1+\Pi_2)^2} \left[ (\Pi''_1)^2 + 2(\Pi'_1/r)^2 + 2(\dot{\Pi}_1')^2 + (\ddot{\Pi}_1)^2 \right.\\
\shoveright{\left. + (\Pi''_2)^2 +2(\Pi'_2/\tilde{r})^2 + 2(\dot{\Pi}_2')^2 + (\ddot{\Pi}_2)^2 \right]}\\
\shoveleft{- \frac{8}{(1+\Pi_1+\Pi_2)^3} \left[ 2\Pi''_1(\Pi'_1)^2 + 4\dot{\Pi}_1' \dot{\Pi}_1
 \Pi_1' + 2\ddot{\Pi}_1 (\dot{\Pi}_1)^2 \right.}\\
\left. + 2\Pi''_2(\Pi'_2)^2 + 4\dot{\Pi}_2' \dot{\Pi}_2 \Pi_2' + 2\ddot{\Pi}_2 (\dot{\Pi}_2)^2 \right.\\
\left. + \dot{\Pi}_1' \Pi_1' \dot{\Pi}_2 + \dot{\Pi}_1 \dot{\Pi}_2' \Pi_2' + \ddot{\Pi}_1 \dot{\Pi}_1 \dot{\Pi}_2 + \dot{\Pi}_1 \ddot{\Pi}_2 \dot{\Pi}_2 \right.\\
 + \shoveright{\left. \frac{x\tilde{x}}{r\tilde{r}} \left( \Pi''_1 \Pi'_1 \Pi'_2 + \Pi'_1 \Pi''_2 \Pi'_2 + \dot{\Pi}_1' \dot{\Pi}_1 \Pi_2' + \Pi_1' \dot{\Pi}_2' \dot{\Pi}_2 \right) \right]}\\
\shoveleft{+ \frac{4}{(1+\Pi_1+\Pi_2)^4} \left[12 ((\Pi'_1)^2 + (\dot{\Pi}_1)^2)^2 + 12((\Pi'_2)^2 + (\dot{\Pi}_2)^2)\right.}\\
\left. + 8 ((\Pi'_1)^2 + (\dot{\Pi}_1)^2)((\Pi'_2)^2 + (\dot{\Pi}_2)^2) + 4 (\frac{x\tilde{x}}{r\tilde{r}} \Pi'_1 \Pi'_2 + \dot{\Pi}_1 \dot{\Pi}_2)^2\right.\\
\left. + 12((\Pi'_1)^2 + (\dot{\Pi}_1)^2 + (\Pi'_2)^2 + (\dot{\Pi}_2)^2)(\frac{x\tilde{x}}{r\tilde{r}} \Pi'_1 \Pi'_2 + \dot{\Pi}_1 \dot{\Pi}_2) \right]\,.
\end{multline}

Note that to achieve good numerical precision, we need to subtract the one-instanton integrands from the above before performing the numerical integration

\begin{multline}
\shoveright{J = \frac{2}{(1+\Pi_1+\Pi_2)^4} ((\Pi'_1)^2 + (\dot{\Pi}_1)^2)((\Pi'_2)^2 + (\dot{\Pi}_2)^2)\,.}
\end{multline}

\begin{multline}
\shoveright{I_{\mu\nu} = \delta_{ij} \tilde{I}_{ii} + \frac{R_i R_j}{R^2_s} \tilde{I}_{ij} + \frac{R_i}{R_s} \tilde{I}_{it} + \frac{R_j}{R_s}\tilde{I}_{tj} + \tilde{I}_{tt}\,.}
\end{multline}

\begin{multline}
 \tilde{I}_{ii} = \frac{4}{(1+\Pi_1+\Pi_2)^2} \left[ \frac{x'^2_1}{r^2} ( \Pi''_1- (\Pi'_1/r))(\Pi'_2/\tilde{r}) \right.\\
\left.  + \frac{x'^2_1}{\tilde{r}^2} (\Pi'_1/r)( \Pi''_2-(\Pi'_2/\tilde{r})) + (\Pi'_1/r)(\Pi'_2/\tilde{r}) \right.\\
 \left.  + \frac{x'^2_1}{r\tilde{r}}(\frac{x\tilde{x}}{r\tilde{r}}(\Pi''_1 \Pi''_2 -\Pi''_1 (\Pi'_2/\tilde{r})-(\Pi'_1/r)\Pi''_2 + (\Pi'_1/r)(\Pi'_2/\tilde{r})) + \dot{\Pi}_1' \dot{\Pi}_2') \right]\\
\shoveleft{+ \frac{1}{(1+\Pi_1+\Pi_2)^3}\left[ 4 ((\Pi'_1/r)((\Pi'_2)^2 + (\dot{\Pi}_1)^2) + ((\Pi'_1)^2 + (\dot{\Pi}_2)^2)(\Pi'_2/\tilde{r})) \right.}\\
\left. + \frac{x'^2_1}{r^2}( 4(\Pi''_1-(\Pi'_1/r))((\Pi'_2)^2 + (\dot{\Pi}_2)^2)-8 (\Pi'_1)^2(\Pi'_2/\tilde{r})) \right.\\
\left. +\frac{x'^2_1}{\tilde{r}^2} (4((\Pi'_1)^2 + (\dot{\Pi}_1)^2) ( \Pi''_2-(\Pi'_2/\tilde{r})) - 8 (\Pi'_1/r)(\Pi'_2)^2) \right.\\
\left. + \frac{x'^2_1}{r\tilde{r}}(-8\frac{x\tilde{x}}{r\tilde{r}} ((\Pi''_1-(\Pi'_1/r))(\Pi'_2)^2+(\Pi'_1)^2(\Pi''_2-(\Pi'_2/\tilde{r}))) \right.\\
\shoveright{\left. -8\dot{\Pi}_1' \dot{\Pi}_2 \Pi_2' - 8 \dot{\Pi}_1 \Pi_1' \dot{\Pi}_2' -8(\Pi_1'' \Pi_1' + \dot{\Pi}_1' \dot{\Pi}_1) \Pi_2' - 8 \Pi_1' (\Pi_2'' \Pi_2' + \dot{\Pi}_2' \dot{\Pi}_2')  \right]}\\
\shoveleft{+\frac{1}{(1+\Pi_1+\Pi_2)^4} \left[ -4 \frac{x'^2_1}{r^2} (\Pi'_1)^2 ((\Pi'_2)^2 + (\dot{\Pi}_2)^2) -4 \frac{x'^2_1}{\tilde{r}^2} ((\Pi'_1)^2 + (\dot{\Pi}_1)^2)(\Pi'_2)^2 \right.}\\
\left. + 12 \frac{x'^2_1}{r\tilde{r}} \Pi'_1 \Pi'_2 ( (\Pi'_1)^2 + (\dot{\Pi}_1)^2 + (\Pi'_2)^2 + (\dot{\Pi}_2)^2 +  \frac{x\tilde{x}}{r\tilde{r}} \Pi'_1 \Pi'_2 + \dot{\Pi}_1 \dot{\Pi}_2) \right]\,.
\end{multline}

\begin{multline}
 \tilde{I}_{ij} = \frac{4}{(1+\Pi_1+\Pi_2)^2} \left[ \frac{x'^2_3-x'^2_1}{r^2} ( \Pi''_1- (\Pi'_1/r))(\Pi'_2/\tilde{r}) \right.\\
\left.  + \frac{(x'_3-R_s)^2-x'^2_1}{\tilde{r}^2} (\Pi'_1/r)( \Pi''_2-(\Pi'_2/\tilde{r})) \right.\\
 \left.  + \frac{x'_3(x'_3-R_s)-x'^2_1}{r\tilde{r}}(\frac{x\tilde{x}}{r\tilde{r}}(\Pi''_1 \Pi''_2 -\Pi''_1 (\Pi'_2/\tilde{r})-(\Pi'_1/r)\Pi''_2 + (\Pi'_1/r)(\Pi'_2/\tilde{r}))\right.\\
\shoveright{\left. + \dot{\Pi}_1' \dot{\Pi}_2') \right]}\\
\shoveleft{+ \frac{1}{(1+\Pi_1+\Pi_2)^3}\left[ \frac{x'^2_3-x'^2_1}{r^2}( 4(\Pi''_1-(\Pi'_1/r))((\Pi'_2)^2 +(\dot{\Pi}_2)^2)\right.}\\
\left. -8 (\Pi'_1)^2(\Pi'_2/\tilde{r})) \right.\\
\left. +\frac{(x'_3-R_s)^2-x'^2_1}{\tilde{r}^2} (4((\Pi'_1)^2 + (\dot{\Pi}_1)^2)( \Pi''_2-(\Pi'_2/\tilde{r})) - 8(\Pi'_1/r)(\Pi'_2)^2) \right.\\
\left. + \frac{x'_3(x'_3-R_s)-x'^2_1}{r\tilde{r}}(-8\frac{x\tilde{x}}{r\tilde{r}} ((\Pi''_1-(\Pi'_1/r))(\Pi'_2)^2+(\Pi'_1)^2(\Pi''_2-(\Pi'_2/\tilde{r}))) \right.\\
\shoveright{\left. -8 \dot{\Pi}_1' \dot{\Pi}_2 \Pi_2' -8\dot{\Pi}_1 \Pi_1' \dot{\Pi}_2' -8 (\Pi''_1 \Pi'_1 + \dot{\Pi}_1' \dot{\Pi}_1) \Pi'_2 - 8 \Pi'_1 (\Pi''_2 \Pi'_2 + \dot{\Pi}_2' \dot{\Pi}_2)\right]}\\
\shoveleft{+\frac{1}{(1+\Pi_1+\Pi_2)^4} \left[ -4 \frac{x'^2_3-x'^2_1}{r^2} (\Pi'_1)^2 ((\Pi'_2)^2 + (\dot{\Pi}_2)^2) \right.}\\
\left. -4 \frac{(x'_3-R_s)^2-x'^2_1}{\tilde{r}^2} ((\Pi'_1)^2 +(\dot{\Pi}_1)^2)(\Pi'_2)^2 \right.\\
\left. + 12 \frac{x'_3(x'_3-R_s)-x'^2_1}{r\tilde{r}} \Pi'_1 \Pi'_2 ( (\Pi'_1)^2 + (\dot{\Pi}_1)^2 + (\Pi'_2)^2 + (\dot{\Pi}_2)^2 \right.\\
\left. + \frac{x\tilde{x}}{r\tilde{r}} \Pi'_1 \Pi'_2 + \dot{\Pi}_1 \dot{\Pi}_2) \right]\,.
\end{multline}

\begin{multline}
 \tilde{I}_{it} = \frac{4}{(1+\Pi_1+\Pi_2)^2} \left[ \frac{x'_3}{r} (\frac{x\tilde{x}}{r\tilde{r}} (\Pi_1'' - (\Pi_1'/r)) \dot{\Pi}_2' + \dot{\Pi}_1' \ddot{\Pi}_2) \right.\\
\shoveright{\left. \frac{x'_3-R_s}{\tilde{r}} (\Pi_1'/r)\dot{\Pi}_2'\right] }\\
\shoveleft{+ \frac{1}{(1+\Pi_1+\Pi_2)^3}\left[ \frac{x_3'}{r} \left( 4 \dot{\Pi}_1'((\Pi_2')^2 + (\dot{\Pi}_2)^2)\right.\right.}\\
\left.\left. - 8 \frac{x\tilde{x}}{r\tilde{r}}(\Pi_1'' - (\Pi_1'/r)) \dot{\Pi}_2 \Pi_2' -8 \dot{\Pi}_1' (\dot{\Pi}_2)^2 -8 \Pi_1'(\frac{x\tilde{x}}{r\tilde{r}} \Pi_1' \dot{\Pi}_2' + \dot{\Pi}_1 \ddot{\Pi}_2)\right.\right.\\
\left.\left. -8 (\Pi_1'' \Pi_1' + \dot{\Pi}_1' \dot{\Pi}_1) \dot{\Pi}_2 -8 \Pi_1' ( \dot{\Pi}_2' \Pi_2' + \ddot{\Pi}_2 \dot{\Pi}_2) \right)\right.\\
\shoveright{\left. + \frac{x_3'-R_s}{\tilde{r}} ( 4 ((\Pi_1')^2 + (\dot{\Pi}_1)^2) \dot{\Pi}_2' - 8 (\Pi_1'/r) \dot{\Pi}_2 \Pi_2') \right]}\\
\shoveleft{+\frac{1}{(1+\Pi_1+\Pi_2)^4} \left[ \frac{x'_3}{r}\left(-4\dot{\Pi}_1\Pi_1' ( (\Pi_2')^2 + (\dot{\Pi}_2)^2) \right.\right.}\\
\left.\left. +12\Pi_1' \dot{\Pi}_1 ((\Pi_1')^2 + (\dot{\Pi}_1)^2 + (\Pi_2')^2 + (\dot{\Pi}_2)^2 +\frac{x\tilde{x}}{r\tilde{r}}\Pi_1'\Pi_2' + \dot{\Pi}_1 \dot{\Pi}_2 ) \right)\right.\\
\left. -4 \frac{x_3'-R_s}{\tilde{r}} ((\Pi_1')^2 + (\dot{\Pi}_1)^2)\dot{\Pi}_2\Pi_2'\right]\,.
\end{multline}

\begin{multline}
 \tilde{I}_{ti} = \frac{4}{(1+\Pi_1+\Pi_2)^2} \left[ \frac{x'_3}{r} \dot{\Pi}_1'(\Pi_2'/\tilde{r}) \right.\\
\shoveright{\left. \frac{x'_3-R_s}{\tilde{r}} (\frac{x\tilde{x}}{r\tilde{r}} \dot{\Pi}_1'(\Pi_2'' - (\Pi_2'/\tilde{r})) + \ddot{\Pi}_1 \dot{\Pi}_2')\right] }\\
\shoveleft{+ \frac{1}{(1+\Pi_1+\Pi_2)^3}\left[ \frac{x_3'}{r} ( 4 \dot{\Pi}_1'((\Pi_2')^2 + (\dot{\Pi}_2)^2) - 8 \dot{\Pi}_1 \Pi_1' (\Pi_2'/\tilde{r}))\right.}\\
\left. + \frac{x_3'-R_s}{\tilde{r}} \left( 4 ((\Pi_1')^2 + (\dot{\Pi}_1)^2)\dot{\Pi}_2' -8 (\frac{x\tilde{x}}{r\tilde{r}} \dot{\Pi}_1' \Pi_1' +\ddot{\Pi}_1 \dot{\Pi}_1) \Pi_2' \right.\right.\\
\left.\left. - 8 \frac{x\tilde{x}}{r\tilde{r}}\dot{\Pi}_1 \Pi_1'(\Pi_2'' - (\Pi_2'/\tilde{r})) -8 (\dot{\Pi}_1)^2 \dot{\Pi}_2'\right.\right.\\
\shoveright{\left.\left. -8 (\dot{\Pi}_1' \Pi_1' + \ddot{\Pi}_1 \dot{\Pi}_1) \Pi_2' -8 \dot{\Pi}_1 ( \Pi_2'' \Pi_2' + \dot{\Pi}_2' \dot{\Pi}_2) \right)\right]}\\
\shoveleft{+\frac{1}{(1+\Pi_1+\Pi_2)^4} \left[ -4\frac{x'_3}{r}\dot{\Pi}_1\Pi_1' ( (\Pi_2')^2 + (\dot{\Pi}_2)^2) \right.}\\
\left. \frac{x_3'-R_s}{\tilde{r}} \left( -4 ((\Pi_1')^2 + (\dot{\Pi}_1)^2)\dot{\Pi}_2\Pi_2' \right.\right.\\
\left.\left. + 12\dot{\Pi}_1 \Pi_2' ((\Pi_1')^2 + (\dot{\Pi}_1)^2 + (\Pi_2')^2 + (\dot{\Pi}_2)^2 +\frac{x\tilde{x}}{r\tilde{r}}\Pi_1'\Pi_2' + \dot{\Pi}_1 \dot{\Pi}_2 \right)\right]\,.
\end{multline}

\begin{multline}
 \tilde{I}_{tt} = \frac{4}{(1+\Pi_1+\Pi_2)^2} \left[ \frac{x\tilde{x}}{r\tilde{r}} \dot{\Pi}_1' \dot{\Pi}_2' + \ddot{\Pi}_1 \ddot{\Pi}_2 \right]\\
\shoveleft{+ \frac{1}{(1+\Pi_1+\Pi_2)^3}\left[ 4 \ddot{\Pi}_1 ((\Pi_2')^2 + (\dot{\Pi}_2)^2) + 4 ((\Pi_1')^2 + (\dot{\Pi}_1)^2) \ddot{\Pi}_2 \right.}\\
\left. -8(\frac{x\tilde{x}}{r\tilde{r}} \dot{\Pi}_1'\dot{\Pi}_2 \Pi_2' + \ddot{\Pi}_1 (\dot{\Pi}_2)^2)  -8(\frac{x\tilde{x}}{r\tilde{r}} \dot{\Pi}_1 \Pi_1' \dot{\Pi}_2' + (\dot{\Pi}_1)^2 \ddot{\Pi}_2) \right.\\
\shoveright{\left.-8(\dot{\Pi}_1' \Pi_1' \dot{\Pi}_2 + \ddot{\Pi}_1 \dot{\Pi}_1 \dot{\Pi}_2) -8(\dot{\Pi}_1 \dot{\Pi}_2' \Pi_2' + \dot{\Pi}_1 \ddot{\Pi}_1 \dot{\Pi}_2) \right]}\\
\shoveleft{+\frac{1}{(1+\Pi_1+\Pi_2)^4} \left[ -4 (\dot{\Pi}_1)^2 ((\Pi_2')^2 +(\dot{\Pi}_2)^2) -4 ((\Pi_1')^2 +(\dot{\Pi}_1)^2) (\dot{\Pi}_2)^2 \right.}\\
\left. +12 \dot{\Pi}_1 \dot{\Pi}_2 ((\Pi_1')^2 +(\dot{\Pi}_1)^2 + (\Pi_2')^2 +(\dot{\Pi}_2)^2 + \frac{x\tilde{x}}{r\tilde{r}} \Pi_1' \Pi_2' + \dot{\Pi}_1 \dot{\Pi}_2) \right]\,.
\end{multline}

\begin{multline}
\shoveright{I_{\mu\rho\nu\sigma} = \delta_{ij} \delta_{kh} \tilde{I}_{ijij} + \delta_{ij} \frac{R_k R_h}{R_s^2} \tilde{I}_{ikih} + \delta_{kh}(\delta_{\nu t}\frac{R_i}{R_s} \tilde{I}_{i k t k} + \delta_{\mu t}\frac{R_j}{R_s} \tilde{I}_{t k j k})}\\
+ \delta_{\mu t} \delta_{\nu t} (\delta_{kh} \tilde{I}_{tktk} + \frac{R_k R_h}{R^2_s} \tilde{I}_{tkth})\,.
\end{multline}

\begin{multline}
\shoveright{\tilde{I}_{ijij} = 0 \quad (\mathrm{analytically})\,.}
\end{multline}

\begin{multline}
\tilde{I}_{ikih} = \frac{4}{(1+\Pi_1+\Pi_2)^2} \left[ \frac{x'^2_3-x'^2_1}{r^2} ( \Pi''_1-(\Pi'_1/r)) (\Pi'_2/\tilde{r}) \right.\\
\left. + \frac{(x'_3-R_s)^2-x'^2_1}{\tilde{r}^2} (\Pi'_1/r)(\Pi''_2-(\Pi'_2/\tilde{r})) \right. \\
\shoveright{\left. \frac{x'^2_1 R_s^2}{(r\tilde{r})^2}(\Pi''_1-(\Pi'_1/r))(\Pi''_2-(\Pi'_2/\tilde{r})) \right]}\\
\shoveleft{ - \frac{8}{(1+\Pi_1+\Pi_2)^3} \left[ \frac{x'^2_3-x'^2_1}{r^2} (\Pi'_1)^2(\Pi'_2/\tilde{r}) + \frac{(x'_3-R_s)^2-x'^2_1}{\tilde{r}^2} (\Pi'_1/r)(\Pi'_2)^2 \right.}\\
\shoveright{\left. + \frac{x'^2_1 R_s^2}{(r\tilde{r})^2} ( (\Pi''_1-(\Pi'_1/r))(\Pi'_2)^2 + (\Pi'_1)^2 (\Pi''_2-(\Pi'_2/\tilde{r}))) \right]}\\
\shoveleft{ + \frac{8}{(1+\Pi_1+\Pi_2)^4}\left[ \frac{x'^2_1 R_s^2}{(r\tilde{r})^2} (\Pi'_1)^2 (\Pi'_2)^2 \right]\,.}\hfill
\end{multline}

\begin{multline}
\tilde{I}_{iktk} = \tilde{I}_{t k j k} = \frac{4}{(1+\Pi_1+\Pi_2)^2} \left[\frac{x_3'}{r} \dot{\Pi}_1' (\Pi_2' /\tilde{r}) + \frac{x_3'-R_s}{\tilde{r}} (\Pi_1' /r) \dot{\Pi}_2' \right.\\
\shoveright{\left. + \frac{x'^2_1 R_s}{r \tilde{r}^2} \dot{\Pi}_1' ( \Pi_2'' - (\Pi_2' /\tilde{r})) - \frac{x'^2_1 R_s}{r^2 \tilde{r}} ( \Pi_1'' - (\Pi_1' /\tilde{r})) \dot{\Pi}_2' \right]}\\
\shoveleft{ - \frac{8}{(1+\Pi_1+\Pi_2)^3} \left[ \frac{x'_3}{r} \dot{\Pi}_1 \Pi_1' (\Pi_2' / \tilde{r}) + \frac{x'_3-R_s}{\tilde{r}} (\Pi_1' /r) \dot{\Pi}_2 \Pi_2' \right.}\\
\left. + \frac{x'^2_1 R_s^2}{r\tilde{r}^2} ( \dot{\Pi}_1' (\Pi_2')^2 + \dot{\Pi}_1 \Pi_1'(\Pi_2'' - (\Pi_2' /\tilde{r})))\right.\\
\shoveright{\left. - \frac{x'^2_1 R_s^2}{r^2\tilde{r}} ( (\Pi_1')^2 \dot{\Pi}_2' + (\Pi_1'' - (\Pi_1' /\tilde{r}))\dot{\Pi}_2 \Pi_2')\right]}\\
\shoveleft{+ \frac{8}{(1+\Pi_1+\Pi_2)^4}\left[ \frac{x'^2_1 R_s}{r \tilde{r}^2} \dot{\Pi}_1 \Pi_1' (\Pi_2')^2 - \frac{x'^2_1 R_s}{r^2 \tilde{r}} (\Pi_1')^2 \dot{\Pi}_2 \Pi_2' \right] \,.}\hfill
\end{multline}

\begin{multline}
\tilde{I}_{tktk} = \frac{4}{(1+\Pi_1+\Pi_2)^2} \left[ \ddot{\Pi}_1 (\Pi_2'/\tilde{r}) + (\Pi_1'/r) \ddot{\Pi}_2 + \frac{x'^2_1}{r^2} (\Pi_1'' - (\Pi_1'/r))\ddot{\Pi}_2 \right.\\
\shoveright{\left. + \frac{x'^2_1}{\tilde{r}^2} \ddot{\Pi}_1 (\Pi_2'' - (\Pi_2'/\tilde{r})) - 2\frac{x'^2_1}{r\tilde{r}} \dot{\Pi}_1' \dot{\Pi}_2' \right]}\\
\shoveleft{ - \frac{8}{(1+\Pi_1+\Pi_2)^3} \left[ (\Pi_1'/r)(\dot{\Pi}_2)^2 + (\dot{\Pi}_1)^2 (\Pi_2'/\tilde{r}) \right.}\\
\left. + \frac{x'^2_1}{r^2} ((\Pi_1'' - (\Pi_1'/r)) (\dot{\Pi}_2)^2 + (\Pi_1')^2 \ddot{\Pi}_2)\right.\\
\left. + \frac{x'^2_1}{\tilde{r}^2}( (\dot{\Pi}_1)^2 (\Pi_2'' - (\Pi_2'/\tilde{r})) + \ddot{\Pi}_1 (\Pi_2')^2)  \right.\\
\shoveright{\left. -2 \frac{x'^2_1}{r\tilde{r}}( \dot{\Pi}_1' \dot{\Pi}_2 \Pi_2' + \dot{\Pi}_1 \Pi_1' \dot{\Pi}_2')\right]}\\
\shoveleft{+ \frac{8}{(1+\Pi_1+\Pi_2)^4}\left[ \frac{x'^2_1}{r^2} (\Pi_1')^2(\dot{\Pi}_2)^2 + \frac{x'^2_1}{\tilde{r}^2} (\dot{\Pi}_1)^2(\Pi_2')^2 -2\frac{x'^2_1}{r\tilde{r}} \dot{\Pi}_1 \Pi_1' \dot{\Pi}_2 \Pi_2'\right] \,.}\hfill
\end{multline}

\begin{multline}
\tilde{I}_{tkth} = \frac{4}{(1+\Pi_1+\Pi_2)^2} \left[ \frac{x'^2_3-x'^2_1}{r^2} (\Pi_1'' - (\Pi_1'/r))\ddot{\Pi}_2 \right.\\
\shoveright{\left. + \frac{(x'_3-R_s)^2-x'^2_1}{\tilde{r}^2} \ddot{\Pi}_1 (\Pi_2'' - (\Pi_2'/\tilde{r})) - 2\frac{x'_3(x'_3-R_s)-x'^2_1}{r\tilde{r}} \dot{\Pi}_1' \dot{\Pi}_2' \right]}\\
\shoveleft{ - \frac{8}{(1+\Pi_1+\Pi_2)^3} \left[ \frac{x'^2_3-x'^2_1}{r^2} ((\Pi_1'' - (\Pi_1'/r)) (\dot{\Pi}_2)^2 + (\Pi_1')^2 \ddot{\Pi}_2)\right.}\\
\left. + \frac{(x'_3-R_s)^2-x'^2_1}{\tilde{r}^2}( (\dot{\Pi}_1)^2 (\Pi_2'' - (\Pi_2'/\tilde{r})) + \ddot{\Pi}_1 (\Pi_2')^2)  \right.\\
\shoveright{\left. -2 \frac{x'_3(x'_3-R_s)-x'^2_1}{r\tilde{r}}( \dot{\Pi}_1' \dot{\Pi}_2 \Pi_2' + \dot{\Pi}_1 \Pi_1' \dot{\Pi}_2')\right]}\\
\shoveleft{+ \frac{8}{(1+\Pi_1+\Pi_2)^4}\left[ \frac{x'^2_3-x'^2_1}{r^2} (\Pi_1')^2(\dot{\Pi}_2)^2 + \frac{(x'_3-R_s)^2-x'^2_1}{\tilde{r}^2} (\dot{\Pi}_1)^2(\Pi_2')^2 \right.}\\ \left. -2\frac{x'_3(x'_3-R_s)-x'^2_1}{r\tilde{r}} \dot{\Pi}_1 \Pi_1' \dot{\Pi}_2 \Pi_2'\right]\,.
\end{multline}

\begin{multline}
J_{\mu\rho\nu\sigma} = \delta_{ij} \frac{R_k R_h}{R_s^2} \tilde{J}_{ikih} +  \delta_{kh}(\delta_{\nu t}\frac{R_i}{R_s} \tilde{J}_{i k t k} + \delta_{\mu t}\frac{R_j}{R_s} \tilde{J}_{t k j k})\\
+ \delta_{\mu t} \delta_{\nu t} (\delta_{kh} \tilde{J}_{tktk} + \frac{R_k R_h}{R^2_s} \tilde{J}_{tkth})\,.
\end{multline}

\begin{multline}
\shoveright{\tilde{J}_{ikih} = \frac{2}{(1+\Pi_1+\Pi_2)^4} \frac{x'^2_1 R_s^2}{(r\tilde{r})^2} (\Pi'_1)^2 (\Pi'_2)^2\,.}
\end{multline}

\begin{multline}
\tilde{J}_{iktk} = \tilde{J}_{tkik} = \frac{2}{(1+\Pi_1+\Pi_2)^4} \left[ \frac{x'^2_1 R_s^2}{r\tilde{r}^2} \dot{\Pi}_1 \Pi_1' (\Pi'_2)^2\right.\\
\left. - \frac{x'^2_1 R_s^2}{r^2\tilde{r}} (\Pi_1')^2 \dot{\Pi}_2 \Pi'_2 \right]\,.
\end{multline}

\begin{multline}
\tilde{J}_{tktk} = \frac{2}{(1+\Pi_1+\Pi_2)^4} \left[ \frac{x'^2_1}{r^2} (\Pi'_1)^2 (\dot{\Pi}_2)^2 + \frac{x'^2_1}{\tilde{r}^2} (\dot{\Pi}_1)^2 (\Pi_2')^2 \right.\\
\left. - 2 \frac{x'^2_1}{r\tilde{r}}\dot{\Pi}_1 \Pi_1' \dot{\Pi}_2 \Pi_2' \right]\,.
\end{multline}

\begin{multline}
\tilde{J}_{tkth} = \frac{2}{(1+\Pi_1+\Pi_2)^4} \left[ \frac{x'^2_3-x'^2_1}{r^2} (\Pi'_1)^2 (\dot{\Pi}_2)^2 \right.\\
\left. + \frac{(x'_3-R_s)^2-x'^2_1}{\tilde{r}^2} (\dot{\Pi}_1)^2 (\Pi_2')^2 - 2 \frac{x'_3(x'_3-R_s)-x'^2_1}{r\tilde{r}}\dot{\Pi}_1 \Pi_1' \dot{\Pi}_2 \Pi_2' \right]\,.
\end{multline}

\begin{multline}
K_{\mu\rho\nu\sigma} = (\delta_{i j} \delta_{k h} + \delta_{i k} \delta_{j h} + \delta_{i h} \delta_{k j }) \tilde{K}_{ikik} + \delta_{i j} \frac{R_k R_h}{R_s^2} \tilde{K}_{ikih} + \delta_{k h} \frac{R_i R_j}{R_s^2} \tilde{K}_{ikjk}\\
+ (\delta_{i k} \frac{R_j R_h}{R_s^2} + \delta_{i h} \frac{R_j R_k}{R_s^2} + \delta_{j k} \frac{R_i R_h}{R_s^2} + \delta_{j h} \frac{R_i R_k}{R_s^2} ) \tilde{K}_{iijh}\\
+ \frac{R_i R_j R_k R_h}{R_s^4} \tilde{K}_{ikjh}\\
+ \delta_{\mu t} \delta_{kh} \frac{R_j}{R_s} \tilde{K}_{tkjk} + \delta_{\mu t} \delta_{jh} \frac{R_k}{R_s} \tilde{K}_{tkjj} + \delta_{\mu t} \delta_{jk} \frac{R_h}{R_s} \tilde{K}_{tjjh} + \delta_{\mu t} \frac{R_j R_k R_h}{R^3_s} \tilde{K}_{tkjh}\\
+ \delta_{\rho t} \delta_{jh} \frac{R_i}{R_s} \tilde{K}_{itjj} + \delta_{\rho t} \delta_{ih} \frac{R_j}{R_s} \tilde{K}_{itji} + \delta_{\rho t} \delta_{ij} \frac{R_h}{R_s} \tilde{K}_{itih} + \delta_{\rho t} \frac{R_i R_j R_h}{R^3_s} \tilde{K}_{itjh}\\
\delta_{\mu t} \delta_{\rho t} \delta_{jh} \tilde{K}_{ttjj} + \delta_{\mu t} \delta_{\rho t} \frac{R_j R_h}{R^2_s} \tilde{K}_{ttjh} + \delta_{\mu t} \delta_{\sigma t} \delta_{jk} \tilde{K}_{tjjt} + \delta_{\mu t} \delta_{\sigma t} \frac{R_j R_k}{R^2_s} \tilde{K}_{tkjt}\\
\delta_{\mu t} \delta_{\nu t} \delta_{kh} \tilde{K}_{tktk} + \delta_{\mu t} \delta_{\nu t} \frac{R_k R_h}{R^2_s} \tilde{K}_{tkth} + \delta_{\rho t} \delta_{\sigma t} \delta_{ij} \tilde{K}_{itjt} + \delta_{\rho t} \delta_{\sigma t} \frac{R_i R_j}{R^2_s} \tilde{K}_{itjt}\\
\delta_{\mu t} \delta_{\nu t} \delta_{\sigma t} \frac{R_k}{R_s} \tilde{K}_{tktt} + \delta_{\rho t} \delta_{\nu t} \delta_{\sigma t} \frac{R_i}{R_s} \tilde{K}_{ittt} + \delta_{\mu t}\delta_{\rho t} \delta_{\nu t} \delta_{\sigma t} \tilde{K}_{tttt}\,.
\end{multline}

\begin{multline}
\shoveright{\tilde{K}_{ikik} = \frac{2}{(1+\Pi_1+\Pi_2)^4} \frac{x'^2_1 x'^2_2}{(r\tilde{r})^2} (\Pi'_1)^2 (\Pi'_2)^2\,.}
\end{multline}

\begin{multline}
\shoveright{\tilde{K}_{ikih} = \frac{2}{(1+\Pi_1+\Pi_2)^4} \frac{(x'_3-R_s)^2 x'^2_1- x'^2_1 x'^2_2}{(r\tilde{r})^2} (\Pi'_1)^2 (\Pi'_2)^2\,.}
\end{multline}

\begin{multline}
\shoveright{\tilde{K}_{ikjk} = \frac{2}{(1+\Pi_1+\Pi_2)^4} \frac{x'^2_3 x'^2_1 - x'^2_1 x'^2_2}{(r\tilde{r})^2} (\Pi'_1)^2 (\Pi'_2)^2\,.}
\end{multline}

\begin{multline}
\tilde{K}_{iijh} = \frac{2}{(1+\Pi_1+\Pi_2)^4} \frac{x'_3(x'_3-R_s)x'^2_1 - x'^2_1 x'^2_2}{(r\tilde{r})^2} (\Pi'_1)^2 (\Pi'_2)^2\,.\hfill
\end{multline}

\begin{multline}
\tilde{K}_{ikjh} = \frac{2}{(1+\Pi_1+\Pi_2)^4} \left(x'^2_3(x'_3-R_s)^2 + 3x'^2_1 x'^2_2 - x'^2_3 x'^2_1 \right.\\
\left. - (x'_3-R_s)^2 x'^2_1 - 4x'_3(x'_3-R_s)x'^2_1 \right)\frac{1}{(r\tilde{r})^2} (\Pi'_1)^2 (\Pi'_2)^2\,.
\end{multline}

\begin{multline}
\shoveright{\tilde{K}_{tkjk} = \frac{2}{(1+\Pi_1+\Pi_2)^4} \frac{x'_3 x'^2_1}{r\tilde{r}^2} \dot{\Pi}_1 \Pi'_1 (\Pi'_2)^2\,.}
\end{multline}

\begin{multline}
\shoveright{\tilde{K}_{tkjj} = \tilde{K}_{tjjh} = \frac{2}{(1+\Pi_1+\Pi_2)^4} \frac{(x'_3-R_s) x'^2_1}{r\tilde{r}^2} \dot{\Pi}_1 \Pi'_1 (\Pi'_2)^2\,.}
\end{multline}

\begin{multline}
\tilde{K}_{tkjh} = \frac{2}{(1+\Pi_1+\Pi_2)^4} \left(x'_3(x'_3-R_s)^2-x'_3 x'^2_1 \right.\\
\left. - 2(x'_3-R_s) x'^2_1\right)\frac{1}{r\tilde{r}^2} \dot{\Pi}_1 \Pi'_1 (\Pi'_2)^2\,.
\end{multline}

\begin{multline}
\shoveright{\tilde{K}_{itjj} = \tilde{K}_{itji} = \frac{2}{(1+\Pi_1+\Pi_2)^4} \frac{x'_3 x'^2_1}{r^2\tilde{r}} (\Pi'_1)^2 \dot{\Pi}_2 \Pi'_2\,.}
\end{multline}

\begin{multline}
\shoveright{\tilde{K}_{itih} = \frac{2}{(1+\Pi_1+\Pi_2)^4} \frac{(x'_3-R_s) x'^2_1}{r^2\tilde{r}} (\Pi'_1)^2 \dot{\Pi}_2 \Pi'_2\,.}
\end{multline}

\begin{multline}
\tilde{K}_{itjh} = \frac{2}{(1+\Pi_1+\Pi_2)^4} \left(x'_3(x'_3-R_s)^2-x'_3 x'^2_1 \right.\\
\left. - 2(x'_3-R_s) x'^2_1\right)\frac{1}{r^2\tilde{r}} (\Pi'_1)^2 \dot{\Pi}_2 \Pi'_2\,.
\end{multline}

\begin{multline}
\shoveright{\tilde{K}_{ttjj} = \tilde{K}_{tjjt} = \frac{8}{(1+\Pi_1+\Pi_2)^4} \frac{x'^2_1}{r\tilde{r}} \dot{\Pi}_1 \Pi'_1 \dot{\Pi}_2 \Pi'_2\,.}
\end{multline}

\begin{multline}
\tilde{K}_{ttjh} = \tilde{K}_{tkjt} = \frac{8}{(1+\Pi_1+\Pi_2)^4} \frac{x'_3(x'_3-R_s)-x'^2_1}{r\tilde{r}} \dot{\Pi}_1 \Pi'_1 \dot{\Pi}_2 \Pi'_2\,.\hfill
\end{multline}

\begin{multline}
\shoveright{\tilde{K}_{tktk} = \frac{2}{(1+\Pi_1+\Pi_2)^4} \frac{x'^2_1}{\tilde{r}^2} (\dot{\Pi}_1)^2 (\Pi'_2)^2\,.}
\end{multline}

\begin{multline}
\shoveright{\tilde{K}_{tkth} = \frac{2}{(1+\Pi_1+\Pi_2)^4} \frac{(x'_3-R_s)^2-x'^2_1}{\tilde{r}^2} (\dot{\Pi}_1)^2 (\Pi'_2)^2\,.}
\end{multline}

\begin{multline}
\shoveright{\tilde{K}_{itit} = \frac{2}{(1+\Pi_1+\Pi_2)^4} \frac{x'^2_1}{r^2} (\Pi_1')^2 (\dot{\Pi}_2)^2\,.}
\end{multline}

\begin{multline}
\shoveright{\tilde{K}_{itjt} = \frac{2}{(1+\Pi_1+\Pi_2)^4} \frac{x'^2_3-x'^2_1}{r^2} (\Pi_1')^2 (\dot{\Pi}_2)^2\,.}
\end{multline}

\begin{multline}
\shoveright{\tilde{K}_{tktt} = \frac{4}{(1+\Pi_1+\Pi_2)^4} \frac{x'_3-R_s}{\tilde{r}} (\dot{\Pi_1})^2 \dot{\Pi}_2 \Pi_2'\,.}
\end{multline}

\begin{multline}
\shoveright{\tilde{K}_{ittt} = \frac{4}{(1+\Pi_1+\Pi_2)^4} \frac{x'_3}{r} \dot{\Pi_1} \Pi_1' (\dot{\Pi}_2)^2\,.}
\end{multline}

\begin{multline}
\shoveright{\tilde{K}_{tttt} = \frac{2}{(1+\Pi_1+\Pi_2)^4} (\dot{\Pi_1})^2 (\dot{\Pi}_2)^2\,.}
\end{multline}

\subsection{Asymptotic Interactions}

As explained in the main text, the small separation asymptotic formulas get contributions which have the same functional form as those for the large separation asymptotics but with different explicit integration limits, see \reffig{fig:interaction:T:separation:small}. We will therefore start with the large separation formulas and not evaluate the integrals explicitly.

\subsubsection{Large Separation}
\label{app:interaction:T:gluonic:large}

As explained in \reffig{fig:integration:split}, the 't Hooft potential is approximated by the $T=0$ BPST form for small arguments up to $r_{4d} = \sqrt{x_\mu x_\mu} = \beta/2$; for larger arguments we use the $r_{3d} \to \infty$ asymptotic form of the $T \neq 0$ potential. These simple rational expressions allow us to factor out completely the 't Hooft potential of the partner instanton, whose argument is kept fixed, except for its contribution to the integration limit. For fixed $I_2$, and depending on the integration region, this leads to the two integration limits
\begin{eqnarray}
 z^2_{4d} &=& \frac{1+\Pi_2}{\rho^2_1} r^2_{4d}\,,\\
 z_{3d} &=& \frac{1+\Pi_2}{\tilde{\rho}_1} r_{3d}\,.
\end{eqnarray}
We have defined $\tilde{\rho} = \pi \rho^2 /\beta$. In the following we will encounter the integrals $\int^z \equiv \int_0^z$ and $\int_z \equiv \int_z^\infty$.

Finally, for $r_{3d} > \beta/2$ the approximate 't Hooft potential is independent of $t$. This will lead to some indices only running over the spatial set $i=\{1,2,3\}$ whenever we use roman letters.

The integral over $I$ contains terms that do not mix the 't Hooft potential $\Pi_1$ and $\Pi_2$ except for the denominators. Just as in the $T=0$ case, these can be transformed to exactly match the single instanton contributions by exploiting scale invariance and independence of the action on $\beta$. Remembering that we actually subtract the one-instanton contributions to get the interactions, we can neglect these terms altogether. We then end up with the following formulas

\begin{multline}
\int I = 72 \pi^2 \rho^2 \frac{\partial_\mu \Pi \partial_\mu \Pi}{(1+\Pi)^3} \int^{z_{4d}} \frac{s^5 ds}{(s^2+1)^4}\\
 + \left( 32 \pi \beta \tilde{\rho} \frac{\partial_\mu \Pi \partial_\mu \Pi}{(1+\Pi)^3} + \frac{16}{3} \pi \beta \tilde{\rho} \frac{\partial_i \Pi \partial_i \Pi}{(1+\Pi)^3} \right) \int_{z_{3d}} \frac{s^2 ds}{(s+1)^4} + \mathrm{sym}\,.
\end{multline}

\begin{multline}
 \int J = 16 \pi^2 \rho^2 \frac{\partial_\mu \Pi \partial_\mu \Pi}{(1+\Pi)^3} \int^{z_{4d}} \frac{s^5 ds}{(s^2+1)^4}\\
+ 8 \pi \beta \tilde{\rho} \frac{\partial_\mu \Pi \partial_\mu \Pi}{(1+\Pi)^3} \int_{z_{3d}} \frac{s^2 ds}{(s+1)^4} + \mathrm{sym}\,.
\end{multline}

\begin{multline}
 \int I_{\mu\nu} = 16 \pi^2 \rho^2 \frac{\partial_\mu \partial_\nu \Pi}{(1+\Pi)^2} \int^{z_{4d}} \frac{s^3 ds}{(s^2+1)^3}\\
- \left( 8 \pi^2 \rho^2 \delta_{\mu\nu} \frac{\partial_\sigma \Pi \partial_\sigma \Pi}{(1+\Pi)^3} + 8\pi^2 \rho^2 \frac{(\partial_\mu \Pi)(\partial_\nu \Pi)}{(1+\Pi)^3}\right) \int^{z_{4d}} \frac{s^5 ds}{(s^2+1)^4}\\
+ \left( 16\pi \beta \tilde{\rho} \frac{\partial_\mu \partial_\nu \Pi}{(1+\Pi)^2} - \frac{32}{3} \pi \beta \tilde{\rho} \frac{\partial_i \partial_\nu \Pi}{(1+\Pi)^2} \right) \int_{z_{3d}} \frac{s ds}{(s+1)^3}\\
- \left( \frac{16}{3} \pi \beta \tilde{\rho} \delta_{ij} \frac{\partial_\sigma \Pi \partial_\sigma \Pi}{(1+\Pi)^3} + 16\pi \beta \tilde{\rho} \frac{(\partial_\mu \Pi)(\partial_\nu \Pi)}{(1+\Pi)^3} \right.\\
\left. - 16\pi \beta \tilde{\rho} \frac{(\partial_i \Pi)(\partial_\nu \Pi)}{(1+\Pi)^3} \right) \int_{z_{3d}} \frac{s^2 ds}{(s+1)^4} + \mathrm{sym}\,.
\end{multline}

At zeroth order, partial integration and the antisymmetry of the 't Hooft symbols can be used to simplify
\begin{equation}
 I_{\mu\rho\nu\sigma} \to \frac{8}{(1+\Pi_1+\Pi_2)^4} (\partial_\mu \Pi_1) (\partial_\rho \Pi_2) (\partial_\nu \Pi_1) (\partial_\sigma \Pi_2)\,.
\end{equation}
With asymptotic behaviour
\begin{multline}
 \int I_{\mu\rho\nu\sigma} = 16 \pi^2 \rho^2 \delta_{\mu\nu}\frac{(\partial_\rho \Pi)(\partial_\sigma \Pi)}{(1+\Pi)^3} \int^{z_{4d}} \frac{s^5 ds}{(s^2+1)^4}\\
 + \frac{32}{3} \pi \beta \tilde{\rho} \delta_{ij} \frac{(\partial_\rho \Pi)(\partial_\sigma \Pi)}{(1+\Pi)^3} \int_{z_{3d}} \frac{s^2 ds}{(s+1)^4} + \mathrm{sym}\,.
\end{multline}

\begin{multline}
 \int J_{\mu\rho\nu\sigma} = 4 \pi^2 \rho^2 \delta_{\mu\nu}\frac{(\partial_\rho \Pi)(\partial_\sigma \Pi)}{(1+\Pi)^3} \int^{z_{4d}} \frac{s^5 ds}{(s^2+1)^4}\\
 + \frac{8}{3} \pi \beta \tilde{\rho} \delta_{ij} \frac{(\partial_\rho \Pi)(\partial_\sigma \Pi)}{(1+\Pi)^3} \int_{z_{3d}} \frac{s^2 ds}{(s+1)^4} + \mathrm{sym}\,.
\end{multline}

For $K_{\mu\rho\nu\sigma}$ no 't Hooft symbols can be used to exchange the index pairs $(\mu,\nu) \leftrightarrow (\rho,\sigma)$, and so we cannot simplify with a symmetry argument anymore

\begin{multline}
 \int K_{\mu\rho\nu\sigma} = 4 \pi^2 \rho^2_1 \delta_{\mu\nu}\frac{(\partial_\rho \Pi_2)(\partial_\sigma \Pi_2)}{(1+\Pi_2)^3} \int^{z_1^{4d}} \frac{s^5 ds}{(s^2+1)^4}\\
 + 4 \pi^2 \rho^2_2 \delta_{\rho\sigma}\frac{(\partial_\mu \Pi_1)(\partial_\nu \Pi_1)}{(1+\Pi_1)^3} \int^{z_2^{4d}} \frac{s^5 ds}{(s^2+1)^4}\\
 + \frac{8}{3} \pi \beta \tilde{\rho}_1 \delta_{ij}\frac{(\partial_\rho \Pi_2)(\partial_\sigma \Pi_2)}{(1+\Pi_2)^3} \int_{z_1^{3d}} \frac{s^2 ds}{(s+1)^4}\\
 + \frac{8}{3} \pi \beta \tilde{\rho}_2 \delta_{kh}\frac{(\partial_\mu \Pi_1)(\partial_\nu \Pi_1)}{(1+\Pi_1)^3} \int_{z_2^{3d}} \frac{s^2 ds}{(s+1)^4}\,.
\end{multline}

\subsubsection{Small Separation}
\label{app:interaction:T:gluonic:small}

As explained in \reffig{fig:interaction:T:separation:small}, the small separation asymptotic formulas get contributions from the large asymptotics. Also, in this case we have performed a global translation so that the instantons sit at $\pm R_\mu /2$. For the terms in $\int^z$ the integration limit is given by $z^2 = \frac{1+\Pi}{\rho^2} (R/2)^2$. The $T\neq 0$ terms given by the $\int_z$ integrals are to be interpreted as $\theta_H(R-\beta/2) \int_z$, i.e.\ they only contribute if the separation $R$ is bigger than $\beta/2$. In practice, these terms do not contribute because they are covered by the look-up tables. The proper small separation asymptotic formulas, that encode the repulsion through the gauge singularity, are then given by the $T=0$ formulas, which we repeat here for convenience.

Introducing another explicit upper limit $z$
\begin{equation}
 z^2 = \frac{R^2}{\rho^2_1+\rho^2_2}, \quad z^2_i = \frac{R^2}{\rho^2_i}\,,
\end{equation}
and approximating the arguments $x_\mu \pm R_\mu/2 \to x_\mu$, we arrive at

\begin{multline}
 \int I = 384 \pi^2 \left[ \frac{\rho^4_1 + \rho^4_2}{(\rho^2_1 + \rho^2_2)^2} \int_z \frac{ds}{s(s^2+1)^2} \right. \\
\left. -\left( \frac{\rho^2_1 \rho^2_2}{(\rho^2_1 + \rho^2_2)^2} + 2 \frac{\rho^6_1 + \rho^6_2}{(\rho^2_1 + \rho^2_2)^3}  \right) \int_z \frac{ds}{s(s^2+1)^3} \right.\\
\left. \frac{\rho^8_1 + \rho^8_2 + \rho^4_1 \rho^4_2 + \rho^6_1 \rho^2_2 + \rho^2_1 \rho^6_2}{(\rho^2_1 + \rho^2_2)^4} \int_z \frac{ds}{s(s^2+1)^4} \right.\\
\left. -\int_{z_1} \frac{s^4 ds}{s(s^2+1)^4} --\int_{z_2} \frac{s^4 ds}{s(s^2+1)^4} \right]\,.
\end{multline}

\begin{multline}
\shoveright{ \int J = 64 \pi^2 \frac{\rho^4_1 \rho^4_2}{(\rho^2_1 + \rho^2_2)^4} \int_z \frac{ds}{s(s^2+1)^4}\,.}
\end{multline}

\begin{multline}
 \int I_{\mu\nu} = \delta_{\mu\nu} \left[ 96 \pi^2 \frac{\rho^2_1 \rho^2_2}{(\rho^2_1 + \rho^2_2)^2} \int_z \frac{ds}{s(s^2+1)^2} \right.\\
\left. - 192 \pi^2 \frac{\rho^2_1 \rho^2_2}{(\rho^2_1 + \rho^2_2)^2} \int_z \frac{ds}{s(s^2+1)^3} \right.\\
\left. + 32 \pi^2 \frac{\rho^4_1 \rho^4_2 + 3\rho^6_1 \rho^2_2 + 3\rho^2_1 \rho^6_2}{(\rho^2_1 + \rho^2_2)^4} \int_z \frac{ds}{s(s^2+1)^4} \right]\,.
\end{multline}

\begin{multline}
 \int I_{\mu\rho\nu\sigma} = \delta_{\mu\nu} \delta_{\rho\sigma} \left[ -32 \pi^2 \frac{\rho^2_1 \rho^2_2}{(\rho^2_1 + \rho^2_2)^2} \int_z \frac{ds}{s(s^2+1)^2} \right.\\
\left. +32 \pi^2 \frac{\rho^2_1 \rho^2_2}{(\rho^2_1 + \rho^2_2)^2} \int_z \frac{ds}{s(s^2+1)^3} \right]\,.
\end{multline}

\begin{multline}
 \shoveright{\int J_{\mu\rho\nu\sigma} = 0\,.}
\end{multline}

\begin{multline}
\int K_{\mu\rho\nu\sigma} = \frac{8}{3} \pi^2 ( \delta_{\mu\nu} \delta_{\rho\sigma} + \delta_{\mu\rho} \delta_{\nu\sigma} + \delta_{\mu\sigma} \delta_{\nu\rho})\frac{\rho^4_1 \rho^4_2}{(\rho^2_1 + \rho^2_2)^4} \int_z \frac{ds}{s(s^2+1)^4}\,.\hfill
\end{multline}

\section{Fermionic Interactions}
\label{app:interaction:T:quark}

The Dirac overlap matrix elements are given by
\begin{multline}
T_{IA} = \int_{\mathbb{R}^3 \times S^1} \frac{1}{4\pi^2\rho_I \rho_A} \left(\frac{1}{2}\Tr (U \tau^{+}_{\beta}) I_\beta - \frac{i}{2} \Tr (U \tau^{+}_\beta \tau_a) \bar{\eta}^a_{\mu\alpha} J_{\beta\mu\alpha} \right.\\
\left. + \frac{i}{2} \Tr (U \tau_a \tau^{+}_\beta) \eta^a_{\mu\alpha} K_{\beta\mu\alpha}\right)\,.
\end{multline}

Some straightforward algebra leads to
\begin{multline}
I_\beta = \frac{-3}{(1+\Pi_I+\Pi_A)(1+\Pi_I)^{1/2}(1+\Pi_A)^{1/2}} \\
\left[ \frac{\Pi_A}{1+\Pi_I} \left( (\partial_\mu \Pi_A \partial_\mu \chi_A) - \frac{(\partial_\mu \Pi_I \partial_\mu \Pi_I) \chi_A}{1+\Pi_I} \right) \left( \partial_\beta \chi_A - \frac{(\partial_\beta \Pi_A) \chi_A}{1+\Pi_A} \right) \right.\\
+ \left. \left( (\partial_\mu \Pi_A \partial_\mu \chi_A) - \frac{(\partial_\mu \Pi_A \partial_\mu \Pi_A) \chi_I}{1+\Pi_A} \right) \left( \partial_\beta \chi_I - \frac{(\partial_\beta \Pi_I) \chi_I}{1+\Pi_I} \right) \right]\,.
\end{multline}

\begin{multline}
J_{\beta\mu\alpha} = \frac{1}{(1+\Pi_I+\Pi_A)(1+\Pi_I)^{1/2}(1+\Pi_A)^{1/2}} \\
\frac{\Pi_A}{1+\Pi_I} (\partial_\mu \chi_I \partial_\alpha \Pi_I) \left( \partial_\beta \chi_A - \frac{(\partial_\beta \Pi_A) \chi_A}{1+\Pi_A} \right)\,.
\end{multline}

\begin{multline}
K_{\beta\mu\alpha} = \frac{1}{(1+\Pi_I+\Pi_A)(1+\Pi_I)^{1/2}(1+\Pi_A)^{1/2}} \\
(\partial_\mu \chi_A \partial_\alpha \Pi_A) \left( \partial_\beta \chi_I - \frac{(\partial_\beta \Pi_I) \chi_I}{1+\Pi_I} \right)\,.
\end{multline}

\subsection{Exact Interactions}
\label{app:interaction:T:quark:exact}

The three contributions give rise to two different colour structures which can be combined with the help of the colour four-vector $i u_\beta \equiv \frac{1}{2}\Tr (U \tau^{+}_{\beta})$, used for instance in \cite{schaefer:shuryak:iilm}; we have that
\begin{equation}
T_{IA} = i \int_{\mathbb{R}^3 \times S^1} \left( u_b \frac{R_b}{R_s} \tilde{T}_s + u_4 \tilde{T}_t \right)\,,
\end{equation}
with
\begin{multline}
\tilde{T}_s = \frac{-1}{(1+\Pi_I+\Pi_A)(1+\Pi_I)^{1/2}(1+\Pi_A)^{1/2}} \\
\left[\frac{x'_3}{r} \left\{\frac{\Pi_A}{1+\Pi_I} \left( \dot{\chi}_A - \frac{\dot{\Pi}_A \chi_A}{1+\Pi_A} \right) \left( \dot{\chi}_I \Pi_I' - \chi_I' \dot{\Pi}_I \right) \right. \right.\\
\left.\left. + 3 \left( \chi_I' - \frac{\Pi_I' \chi_I}{1+\Pi_I} \right) \left( \Pi_A'\chi_A' + \dot{\Pi}_A \dot{\chi}_A - \frac{((\Pi_A')^2 + (\dot{\Pi}_A)^2) \chi_A}{1+\Pi_A} \right) \right\} \right.\\
\left. + \frac{x'_3-R_s}{\tilde{r}} \left\{ \left( \dot{\chi}_I - \frac{\dot{\Pi}_I \chi_I}{1+\Pi_I} \right) \left( \dot{\chi}_A \Pi_A' - \chi_A' \dot{\Pi}_A \right) \right. \right.\\
\left. \left. + 3 \frac{\Pi_A}{1+\Pi_I} \left( \chi_A' - \frac{\Pi_A' \chi_A}{1+\Pi_A} \right) \left( \Pi_I'\chi_I' + \dot{\Pi}_I \dot{\chi}_I - \frac{((\Pi_I')^2 + (\dot{\Pi}_I)^2) \chi_I}{1+\Pi_I} \right) \right\} \right]\,.
\end{multline}

\begin{multline}
\tilde{T}_t = \frac{1}{(1+\Pi_I+\Pi_A)(1+\Pi_I)^{1/2}(1+\Pi_A)^{1/2}}\\
\left[ \frac{x'_3(x'_3-r_s)+2x'^2_1}{r\tilde{r}} \left\{ \frac{\Pi_A}{1+\Pi_I} \left( \chi_A' - \frac{\Pi_A' \chi_A}{1+\Pi_A} \right) (\dot{\chi}_I \Pi_I' - \chi_I' \dot{\Pi}_I) \right. \right.\\
\left. \left. + \left( \chi_I' - \frac{\Pi_I' \chi_I}{1+\Pi_I} \right) (\dot{\chi}_A \Pi_A' - \chi_A' \dot{\Pi}_A) \right\} \right.\\
\left. - 3 \frac{\Pi_A}{1+\Pi_I} \left( \dot{\chi}_A- \frac{\dot{\Pi}_A \chi_A}{1+\Pi_A}\right) \left( \Pi_I' \chi_I' + \dot{\Pi}_I \dot{\chi}_I - \frac{((\Pi_I')^2 + (\dot{\Pi}_I)^2)\chi_I}{1+\Pi_I} \right)\right.\\
\left. - 3 \left( \dot{\chi}_I- \frac{\dot{\Pi}_I \chi_I}{1+\Pi_I}\right) \left( \Pi_A' \chi_A' + \dot{\Pi}_A \dot{\chi}_A - \frac{((\Pi_A')^2 + (\dot{\Pi}_A)^2)\chi_A}{1+\Pi_A} \right)\right]\,.
\end{multline}

\subsection{Asymptotic Interactions}

\subsubsection{Large Separation}
\label{app:interaction:T:quark:large}

At $T\neq 0$ the quark zero mode has an additional factor $\chi \propto \exp(-r/\beta)$. Therefore we completely neglect the contributions from $r>\beta/2$, and recover the $T=0$ formulas. For convenience we display the final results again here; details are given in \cite{wantz:iilm:1}. The only difference with the $T=0$ formulas is that the upper limit $z\propto \beta/2$ instead of infinity

\ifthenelse{\equal{\Qclass}{revtex4}}{
\begin{multline}
 \int I_\beta = 8 \pi^2 \rho_I^2 \frac{\Pi_A \partial_\beta \Pi_A}{(1+\Pi_A)^{3/2}} \int^{z_I} \frac{s^4 ds}{(s^2+1)^{7/2}}-8 \pi^2 \rho_A^2 \frac{\partial_\beta \Pi_I}{(1+\Pi_I)^{3/2}} \int^{z_A} \frac{s^4 ds}{(s^2+1)^{5/2}}\,.
\end{multline}
}{}
\ifthenelse{\equal{\Qclass}{elsarticle}}{
\begin{multline}
 \int I_\beta = 8 \pi^2 \rho_I^2 \frac{\Pi_A \partial_\beta \Pi_A}{(1+\Pi_A)^{3/2}} \int^{z_I} \frac{s^4 ds}{(s^2+1)^{7/2}}\\
-8 \pi^2 \rho_A^2 \frac{\partial_\beta \Pi_I}{(1+\Pi_I)^{3/2}} \int^{z_A} \frac{s^4 ds}{(s^2+1)^{5/2}}\,.
\end{multline}
}{}
At zeroth order, we have that $\int J_{\beta\mu\alpha}=\int K_{\beta\mu\alpha}=0$. For $J_{\beta\mu\alpha}$ this follows from the fact that within the $r<\beta/2$ region we approximate $\chi \to \Pi$ and so the integration over $I_I$ yields zero because of the anti-symmetry of the 't Hooft symbols. Integration over $I_A$ vanishes too because of $O(4)$ symmetry.

\subsubsection{Small Separation}
\label{app:interaction:T:quark:small}

At zeroth order, i.e.\ $x_\mu \pm R_\mu /2 \to x_\mu$, the contribution to $I_\beta$ vanishes because of $O(4)$ symmetry. It turns out the large separation asymptotics falls off too quickly as $R \to 0$. However this is not important because in this regime the gluonic interaction is dominant.



\end{document}